\documentclass[english,11pt,a4paper]{article}
\usepackage[latin9]{inputenc}
\usepackage{geometry}
\usepackage{color}
\usepackage{babel}
\usepackage{amsmath}
\usepackage{amsthm}
\usepackage{amssymb}
\usepackage[unicode=true,pdfusetitle,
 bookmarks=true,bookmarksnumbered=false,bookmarksopen=false,
 breaklinks=false,pdfborder={0 0 0},pdfborderstyle={},backref=false,colorlinks=true]
 {hyperref}
\hypersetup{
 linkcolor=vblue, citecolor=vblue}

\makeatletter

\numberwithin{equation}{section}
\theoremstyle{plain}
\newtheorem{thm}{\protect\theoremname}[section]
\theoremstyle{plain}
\newtheorem{prop}[thm]{\protect\propositionname}
\theoremstyle{plain}
\newtheorem{lem}[thm]{\protect\lemmaname}
\theoremstyle{plain}
\newtheorem{cor}[thm]{\protect\corollaryname}
\theoremstyle{plain}
\newtheorem*{prop*}{\protect\propositionname}
\theoremstyle{plain}
\newtheorem*{thm*}{\protect\theoremname}

\definecolor{vblue}{RGB}{0,0,255}

\allowdisplaybreaks

\makeatother

\providecommand{\corollaryname}{Corollary}
\providecommand{\lemmaname}{Lemma}
\providecommand{\propositionname}{Proposition}
\providecommand{\theoremname}{Theorem}

\allowdisplaybreaks

\begin{document}
\title{The Gell-Mann--Brueckner Formula for the Correlation Energy of the
Electron Gas: A Rigorous Upper Bound in the Mean-Field Regime}
\author{Martin Ravn Christiansen, Christian Hainzl, Phan Th\`anh Nam\\
\\
{\footnotesize{}Department of Mathematics, Ludwig Maximilian University
of Munich, Germany}\\
{\footnotesize{}Emails: martin.christiansen@math.lmu.de, hainzl@math.lmu.de,
nam@math.lmu.de}}
\maketitle
\begin{abstract}
We prove a rigorous upper bound on the correlation energy of interacting
fermions in the mean-field regime for a wide class of interaction
potentials. Our result covers the Coulomb potential, and in this case
we obtain the analogue of the Gell-Mann--Brueckner formula $c_{1}\rho\log\left(\rho\right)+c_{2}\rho$
in the high density limit. We do this by refining the analysis of
our bosonization method to deal with singular potentials, and to capture
the exchange contribution which is absent in the purely bosonic picture.
\end{abstract}
\tableofcontents{}

\section{Introduction}

Although interacting Fermi gases have been studied extensively from
the beginning of quantum mechanics, their rigorous understanding remains
one of the major issues of condensed matter physics. From first principles,
a system of $N$ fermions in $\mathbb{R}^{3}$ can be described by
a Schr\"odinger equation in $\mathbb{R}^{3N}$, subject to the anti-symmetry
condition between the variables due to Pauli's exclusion principle.
However, this fundamental theory becomes very complex when $N\rightarrow\infty$,
leading to the need of various approximations. Justifying these approximations
is an important task of mathematical physics.

\medskip

One of the most basic approximations for fermions is the Hartree--Fock
(HF) theory. In HF theory, the particles are assumed to be independent,
namely the HF energy is computed by restricting the consideration
to Slater determinants. In spite of its simplicity, the HF theory
is used very successfully in computational physics and chemistry to
compute the ground state energy of atoms and molecules. The accuracy
of the HF energy (in comparison to the full quantum energy) for large
Coulomb systems was investigated in the 1990s by Fefferman and Seco
\cite{FerSec-90}, Bach \cite{Bach-92}, and Graf and Solovej \cite{GraSol-94}.

\medskip

On the other hand, for the \textit{electron gas} (e.g. jellium, a
homogeneous electron gas moving in a background of uniform positive
charge), the HF theory is essentially trivial in the high density
limit since the HF energy only contains an exponentially small correction
to the energy of the Fermi state, the ground state of the non-interacting
gas \cite{GonHaiLew-19}. Therefore, computing the \textit{correlation
energy}\footnote{This name comes from the fact that Slater determinants are the least
correlated wave functions under Pauli's exclusion principle.}, namely the correction to the HF energy, is a crucial task to understand
the effect of the interaction. It was already noticed by Wigner in
1934 \cite{Wigner-34} and confirmed by Heisenberg in 1947 \cite{Heisenberg-47}
that it would be very challenging to accomplish this task within perturbation
theory due to 
the long-range property of the Coulomb potential. Nevertheless,
a remarkable attempt in this direction was done by Macke in 1950 \cite{Macke-50}
when he used a partial resummation of the divergent series to predict
the leading order contribution $c_{1}\rho\log\left(\rho\right)$ of
the correlation energy (with density $\rho\rightarrow\infty$).

\medskip

A cornerstone in the correlation analysis of the electron gas is the
\textit{random phase approximation} (RPA) which was proposed by Bohm
and Pines in the 1950s \cite{BohPin-51,BohPin-52,BohPin-53,Pines-53}.
As an important consequence of the Bohm-Pines RPA theory, the electron
gas could be decoupled into collective plasmon excitations and quasi-electrons
that interacted via a screened Coulomb interaction. The latter fact
justified the independent particle approach commonly used for many-body
fermion systems. The justification of the RPA was a major question
in condensed matter and nuclear physics in the late 1950s and 1960s.
An important justification was given by Gell-Mann and Brueckner in
1957 \cite{GelBru-57} when they formally derived the RPA from a resummation
of Feynman diagrams where each term separately diverges but the sum
is convergent. More precisely, by considering the diagrams corresponding
to the interaction of pairs of fermions, one from inside and one from
outside the Fermi state, Gell-Mann and Brueckner were able to produce
the leading order contribution $c_{1}\rho\log\left(\rho\right)+c_{2}\rho$
of the correlation energy.

\medskip

Soon after the achievement of Gell-Mann and Brueckner, Sawada \cite{Sawada-57}
and Sawada--Brueckner--Fukuda--Brout \cite{SawBruFukBro-57} proposed
an alternative approach to the RPA where the pairs of electrons are
interpreted as bosons, leading to an effective Hamiltonian which is
quadratic in terms of the bosonic creation and annihilation operators.
Note that within the purely bosonic picture, quadratic Hamiltonians
can be diagonalized by Bogolubov transformations \cite{Bogolubov-47},
and hence their spectra can be computed explicitly. Therefore, the
Hamiltonian approach in \cite{Sawada-57,SawBruFukBro-57} is conceptually
more transparent than the resummation method in \cite{GelBru-57}.
Unfortunately the analysis in \cite{Sawada-57,SawBruFukBro-57} only
gives the contribution $c_{1}\rho\log\left(\rho\right)$ of the correlation
energy because the \textit{exchange contribution} of order $\rho$
is missed in the purely bosonic picture.

\medskip

Recently, the bosonization argument in \cite{Sawada-57,SawBruFukBro-57}
has been revisited and made rigorous in the mean-field regime with
smooth interaction potentials \cite{HaiPorRex-20,BNPSS-20,BNPSS-21,BNPSS-22,CHN-21,BPSS-21,CHN-22}.
In principle, if the interaction is sufficiently weak, then the non-bosonizable
terms of the interaction energy are negligible, and the quasi-bosonic
Hamiltonian can be analyzed with great precision. In particular, the
correlation energy has been successfully computed to the leading order
\cite{BNPSS-20,BNPSS-21,CHN-21,BPSS-21}. However, the boundedness
of interaction potentials is crucial for all of these works, and extending
the analysis to the electron gas remains a very interesting open question.

\medskip

In the present paper, we will give the first rigorous upper bound
to the correlation energy of the electron gas in the mean-field regime.
Our bound is consistent with the Gell-Mann--Brueckner formula $c_{1}\rho\log\left(\rho\right)+c_{2}\rho$
for jellium in the high density limit \cite{GelBru-57}. Although
our trial state argument is inspired by the bosonization method in
\cite{Sawada-57,SawBruFukBro-57}, we are able to capture correctly
the exchange contribution by carefully distinguishing the purely bosonic
picture and the quasi-bosonic one. On the mathematical side, we will
use the general method in our recent work \cite{CHN-21}, but several
new estimates are needed to deal with the singularity of the potential. The matching lower bound in the mean-field regime, as well as the
corresponding result in the thermodynamic limit, remain open, and
we hope to be able to come back to these issues in the future. 
\medskip

On the technical side, the key idea of \cite{CHN-21} is that while the bosonic property of fermionic pairs holds only in an average sense, this weak bosonic property is sufficient to extract correctly the correlation energy by implementing a quasi-bosonic Bogolubov transformation. The main contribution of the present paper is to show that this approach is also sufficient to extract the exchange correction to the purely bosonic computation. On the other hand, another bosonization method has been proposed in \cite{BNPSS-20}, where the bosonic property of fermionic pairs is strengthened by using suitable patches on the Fermi sphere for the quasi-bosonic creation and annihilation operators, making the comparison with the purely bosonic computation significantly easier. In fact, as explained in \cite{BPSS-21}, the approach in \cite{BNPSS-20} can be extended to give the leading order of the correlation energy upper bound for potentials satisfying  $\sum V_k^2|k| < \infty$. Although this condition only barely fails  for the Coulomb potential, there is a huge difference to the Coulomb case. While for $\sum V_k^2|k| < \infty$ the bosonic correlation contribution is of order $k_F$ and the exchange correlation is of lower order $o(k_F)$, for the Coulomb potential the exchange contribution raises to the order $k_F$, whereas the bosonic correlation behaves as $k_F \log(k_F)$, which makes the Coulomb case much more challenging (here $k_F$ is the radius of the Fermi ball). 
In particular, the method in \cite{BNPSS-20,BPSS-21} does not seem to capture the exchange contribution which is indeed important for the Coulomb potential.

\subsection{Main Result}

Let $\mathbb{T}^{3}=\left[0,2\pi\right]^{3}$
with periodic boundary conditions. Let $V:\mathbb{T}^{3}\rightarrow\mathbb{R}$ be defined by
\begin{equation}
V\left(x\right)=\frac{1}{\left(2\pi\right)^{3}}\sum_{k\in\mathbb{Z}_{\ast}^{3}}\hat{V}_{k}e^{ik\cdot x},\quad\mathbb{Z}_{\ast}^{3}=\mathbb{Z}^{3}\backslash\left\{ 0\right\} ,
\end{equation}
with Fourier coefficients satisfying
\begin{equation}
\hat{V}_{k}\geq0,\quad\hat{V}_{k}=\hat{V}_{-k},\quad\sum_{k\in\mathbb{Z}_{\ast}^{3}}\hat{V}_{k}^{2}<\infty.
\end{equation}
We implicitly assume that $\hat{V}_{0}=0$, or equivalently
that the ``background'' has been subtracted.

For $k_{F}>0$, let $N=\left|B_{F}\right|$ be the number of integer
points in the Fermi ball $B_{F}=\overline{B}\left(0,k_{F}\right)\cap\mathbb{Z}^{3}$
and consider the mean-field Hamiltonian
\begin{equation}
H_{N}=-\sum_{i=1}^{N}\Delta_{i}+k_{F}^{-1}\sum_{1\leq i<j\leq N}V\left(x_{i}-x_{j}\right)\label{eq:HamiltonianDefinition}
\end{equation}
on the fermionic space $\mathcal{H}_{N}=\bigwedge^{N}\mathfrak{h}$ with $\mathfrak{h}=L^{2}\left(\mathbb{T}^{3}\right)$\footnote{We consider spinless particles for simplicity. Including the spin
only requires slight modifications of the analysis.}. 
The leading order of the ground state energy of $H_{N}$ is given
by the Fermi state
\begin{equation}
\psi_{\mathrm{FS}}=\bigwedge_{p\in B_{F}}u_{p},\quad u_{p}\left(x\right)=\left(2\pi\right)^{-\frac{3}{2}}e^{ip\cdot x}.\label{eq:FermiStateDefinition}
\end{equation}
It is straightforward to find (see e.g. \cite[Eqs. (1.10) and (1.20)]{CHN-21})
\begin{equation} \label{eq:EPS}
E_{\mathrm{FS}}=\left\langle \psi_{\mathrm{FS}},H_{N}\psi_{\mathrm{FS}}\right\rangle = \sum_{p\in B_F} |p|^2 + \frac{1}{2(2\pi)^3} \sum_{k\in \mathbb{Z}^3_*} \hat V(k) \left( |L_k|-N\right)
\end{equation}
where for every $k\in\mathbb{Z}_{\ast}^{3}$, we denoted the \textit{lune} associated to $k$ by
\begin{equation}
L_{k}=\left(B_{F}+k\right)\backslash B_{F}=\left\{ p\in\mathbb{Z}^{3}\mid\left|p-k\right|\leq k_{F}<\left|p\right|\right\}. 
\end{equation}
Our main result concerns the corrections to the ground state energy. For every $k\in\mathbb{Z}_{\ast}^{3}$, define
\begin{equation}
\lambda_{k,p}=\frac{1}{2}\left(\left|p\right|^{2}-\left|p-k\right|^{2}\right),\quad \forall p\in L_{k}.
\end{equation}

We will prove the following:
\begin{thm}
\label{them:MainTheorem}As $k_{F}\rightarrow\infty$ it holds that
\[
\inf\sigma\left(H_{N}\right)\leq E_{\mathrm{FS}}+E_{\mathrm{corr},\mathrm{bos}}+E_{\mathrm{corr},\mathrm{ex}}+C\sqrt{\sum_{k\in\mathbb{Z}_{\ast}^{3}}\hat{V}_{k}^{2}\min\left\{ \left|k\right|,k_{F}\right\} }
\]
where
\[
E_{\mathrm{corr},\mathrm{bos}}=\frac{1}{\pi}\sum_{k\in\mathbb{Z}_{\ast}^{3}}\int_{0}^{\infty}F\left(\frac{\hat{V}_{k}k_{F}^{-1}}{\left(2\pi\right)^{3}}\sum_{p\in L_{k}}\frac{\lambda_{k,p}}{\lambda_{k,p}^{2}+t^{2}}\right)dt,\quad F\left(x\right)=\log\left(1+x\right)-x,
\]
is the bosonic contribution and
\[
E_{\mathrm{corr},\mathrm{ex}}=\frac{k_{F}^{-2}}{4\left(2\pi\right)^{6}}\sum_{k\in\mathbb{Z}_{\ast}^{3}}\sum_{p,q\in L_{k}}\frac{\hat{V}_{k}\hat{V}_{p+q-k}}{\lambda_{k,p}+\lambda_{k,q}}
\]
is the exchange contribution, for a constant $C>0$ depending only
on $\sum_{k\in\mathbb{Z}_{\ast}^{3}}\hat{V}_{k}^{2}$.
\end{thm}

Some remarks on our result:

\textbf{1.} Consider the Coulomb potential, $\hat{V}_{k}=g\left|k\right|^{-2}$
for a constant $g>0$. Following the analysis of \cite{GraSol-94}, we find
that
\begin{equation}
\inf\sigma\left(H_{N}\right)=E_{\mathrm{FS}}+o\left(k_{F}^{3}\right)
\end{equation}
where $E_{\mathrm{FS}}$ contains the kinetic energy of order $k_{F}^{5}$,
the direct interaction energy of order $k_{F}^{5}$ and the exchange
interaction energy of order $k_{F}^{3}$. Furthermore, it is straightforward
to adapt the proof in \cite{GonHaiLew-19} to see that the difference
between $E_{\mathrm{FS}}$ and the HF energy is exponentially small
as $k_{F}\rightarrow\infty$. Therefore our result really concerns
the correlation energy, which we bound from above by
\begin{equation} \label{eq:asymptotic-Ebos-Eex}
E_{\mathrm{corr},\mathrm{bos}}\sim - k_{F}\log\left(k_{F}\right)\quad\text{and}\quad E_{\mathrm{corr},\mathrm{ex}}\sim k_{F}
\end{equation}
plus the error term of order
\begin{equation} \label{eq:asymptotic-Eerror}
\sqrt{\sum_{k\in\mathbb{Z}_{\ast}^{3}}\hat{V}_{k}^{2}\min\left\{ \left|k\right|,k_{F}\right\} }\sim\sqrt{\log\left(k_{F}\right)}.
\end{equation}
In fact, it is easy to verify \eqref{eq:asymptotic-Eerror} using $\sum_{|k|\le k_F} \hat V_k^2 |k| \sim \log( k_F)$ and $\sum_{|k|\ge k_F} \hat V_k^2   \sim k_F^{-1}$. To see the leading order behavior $E_{\mathrm{corr},\mathrm{ex}}\sim k_F$ in  \eqref{eq:asymptotic-Ebos-Eex}, one may use that $\lambda_{k,p} \sim |k| \max\{|k|,k_F\} $ (in an average sense) and that $|L_k| \sim k_F^2 \min\left\{ |k|,k_{F}\right\}$. Moreover, from the expansion 
\begin{align}
\log(1+x)-x\approx -x^2/2 +o(x^3)_{x\to 0}
\end{align}
we have 
\begin{align} \label{eq:E-bos-expansion}
E_{\mathrm{corr},\mathrm{bos}} &\approx  -\frac{1}{4(2\pi)^6}\sum_{k\in \mathbb{Z}^3_*}  ( \hat V_k k_F^{-1})^2   \frac{2}{\pi}\int_0^\infty \left(\sum_{p\in L_{k}}\frac{\lambda_{k,p}}{\lambda_{k,p}^{2}+t^{2}}\right)^2 dt \nonumber\\
& = -\frac{1}{4(2\pi)^6}\sum_{k\in \mathbb{Z}^3_*} ( \hat V_k k_F^{-1})^2  \sum_{p,q\in L_k} \frac{1}{\lambda_{k,p} + \lambda_{k,q}},
\end{align}
and hence the asymptotic behavior $E_{\mathrm{corr},\mathrm{bos}}\sim - k_F \log(k_F)$ in  \eqref{eq:asymptotic-Ebos-Eex} follows similarly. 

\medskip

Note that the correlation energy $E_{\mathrm{corr},\mathrm{bos}}+E_{\mathrm{corr},\mathrm{ex}}$
in Theorem \ref{them:MainTheorem} is exactly the mean-field analogue
of the Gell-Mann--Brueckner formula $c_{1}\rho\log\left(\rho\right)+c_{2}\rho$
for jellium in the thermodynamic limit \cite{GelBru-57}. Indeed,
substituting $k_{F}^{-1}\hat{V}_{k}\rightarrow4\pi e^{2}\left|k\right|^{-2}$
and $\left(2\pi\right)^{3}\rightarrow$ the volume $\Omega$, $E_{\mathrm{corr},\mathrm{bos}}$
agrees with \cite[Eq. (34)]{SawBruFukBro-57} which is equivalent
with \cite[Eq. (19)]{GelBru-57} (accounting also for spin). In the thermodynamic limit, the right-hand side of \eqref{eq:E-bos-expansion} always diverges, no matter if we have the mean-field scaling or not, but the full expression on the left-hand side converges in either case. 

\medskip

Furthermore, we also obtain the exchange contribution $E_{\mathrm{corr},\mathrm{ex}}$,
which is the analogue of \cite[Eq. (9)]{GelBru-57}, which is completely
absent from the bosonic model of \cite{SawBruFukBro-57}. With the
same substitutions as above, the exchange contribution takes the form
\begin{align}
E_{\mathrm{corr},\mathrm{ex}} & =2\cdot\frac{1}{4\Omega^{2}}\sum_{k\in\mathbb{Z}_{\ast}^{3}}\sum_{p,q\in L_{k}}\frac{4\pi e^{2}}{\left|k\right|^{2}}\frac{4\pi e^{2}}{\left|p+q-k\right|^{2}}\frac{1}{\frac{1}{2}\left(\left|p\right|^{2}+\left|p-k\right|^{2}\right)+\frac{1}{2}\left(\left|q\right|^{2}+\left|q-k\right|^{2}\right)}\nonumber \\
 & =\frac{8\pi^{2}e^{4}}{\Omega^{2}}\sum_{k\in\mathbb{Z}_{\ast}^{3}}\sum_{p,q\in L_{k}}\frac{1}{\left|k\right|^{2}\left|p+q-k\right|^{2}k\cdot\left(p+q-k\right)}
\end{align}
which agrees with \cite[Eq. (9.14)]{Raimes-72} (noting that we take
$m=1/2$).

\medskip

\textbf{2.} If the potential satisfies $\sum_{k\in\mathbb{Z}_{\ast}^{3}}\hat{V}_{k}^{2}\left|k\right|<\infty$,
and so is less singular than the Coulomb potential, then the bosonic
contribution $E_{\mathrm{corr},\mathrm{bos}}$ is of order $k_{F}$,
while the exchange contribution is $o\left(k_{F}\right)$. In this
case, the upper bound
\begin{equation}
\inf\sigma\left(H_{N}\right)\leq E_{\mathrm{FS}}+E_{\mathrm{corr},\mathrm{bos}}+o\left(k_{F}\right)
\end{equation}
is already known; see \cite[Remark 1 after Theorem 1.3]{CHN-21} and
\cite[Appendix A]{BPSS-21}. Under the stronger condition $\sum\hat{V}_{k}\left|k\right|<\infty$
the matching lower bound was established in \cite{CHN-21,BPSS-21}
(see also \cite{BNPSS-20} and \cite{BNPSS-21} for previous results
on the upper and lower bounds, respectively, when $\hat{V}_{k}$ is
finitely supported). In comparison, the Coulomb potential is much
more challenging to analyze, since it leads to an additional logarithmic
factor in the bosonic contribution, and lifts the exchange contribution
to the order $k_{F}$. On the mathematical side, working with
the Coulomb potential thus requires a substantial refinement of the
bosonization method compared to the existing works.

\medskip

\textbf{3.} Although the case of the greatest physical interest is
the Coulomb potential, our result covers a far greater class of singular
potentials: Under the condition $\sum_{k\mathbb{Z}_{\ast}^{3}}\hat{V}_{k}^{2}<\infty$,
the error term $\sqrt{\sum_{k\in\mathbb{Z}_{\ast}^{3}}\hat{V}_{k}^{2}\min\left\{ \left|k\right|,k_{F}\right\} }$
is of order at most $O\left(\sqrt{k_{F}}\right)$, and so Theorem
\ref{them:MainTheorem} is always a meaningful result.

\subsection{Overview of the Proof}

We will construct a trial state by applying a quasi-bosonic Bogolubov
transformation to the Fermi state $\psi_{\mathrm{FS}}$. We will follow
the general formulation of the bosonization method in \cite{CHN-21}.
We quickly recall this here for the reader's convenience, after which
we explain the new components of the proof and the structure of the
rest of the paper.

\subsubsection*{Rewriting the Hamiltonian}

We will use the second quantization formalism in which we associate
to every plane wave state $u_{p}$ of equation (\ref{eq:FermiStateDefinition})
the creation and annihilation operators $c_{p}^{\ast}=a^{\ast}(u_p)$
and $c_{p}=a(u_p)$ on the fermionic Fock space. They
obey the canonical anti-commutation relations (CAR)
\begin{equation}
\left\{ c_{p},c_{q}\right\} =\left\{ c_{p}^{\ast},c_{q}^{\ast}\right\} =0,\quad\left\{ c_{p},c_{q}^{\ast}\right\} =\delta_{p,q},\quad  p,q\in\mathbb{Z}^{3}.
\end{equation}
The Hamiltonian $H_{N}$ of equation (\ref{eq:HamiltonianDefinition})
can then be written as
$
H_{N}=H_{\mathrm{kin}}+k_{F}^{-1}H_{\mathrm{int}}
$ 
where
\begin{equation}
H_{\mathrm{kin}}=\sum_{p\in\mathbb{Z}^{3}}\left|p\right|^{2}c_{p}^{\ast}c_{p},\quad H_{\mathrm{int}}=\frac{1}{2\left(2\pi\right)^{3}}\sum_{k\in\mathbb{Z}_{\ast}^{3}}\sum_{p,q\in\mathbb{Z}^{3}}\hat{V}_{k}c_{p+k}^{\ast}c_{q-k}^{\ast}c_{q}c_{p}.
\end{equation}
Note that the Fermi state $\psi_{\mathrm{FS}}$ obeys ($B_{F}^{c}$
denoting the complement of $B_{F}$ with respect to $\mathbb{Z}^{3}$)
\begin{equation}
c_{p}\psi_{\mathrm{FS}}=0=c_{q}^{\ast}\psi_{\mathrm{FS}},\quad p\in B_{F}^{c},\,q\in B_{F},
\end{equation}
and so it follows by the CAR that the kinetic energy of the Fermi
state is
\begin{equation}
\left\langle \psi_{\mathrm{FS}},H_{\mathrm{kin}}\psi_{\mathrm{FS}}\right\rangle =\sum_{p\in B_{F}}\left|p\right|^{2}.
\end{equation}
We define the \textit{localized kinetic operator} $H_{\mathrm{kin}}^{\prime}$
by
\begin{align}
H_{\mathrm{kin}}^{\prime} & =H_{\mathrm{kin}}-\left\langle \psi_{\mathrm{FS}},H_{\mathrm{kin}}\psi_{\mathrm{FS}}\right\rangle =\sum_{p\in B_{F}^{c}}\left|p\right|^{2}c_{p}^{\ast}c_{p}-\sum_{p\in B_{F}}\left|p\right|^{2}c_{p}c_{p}^{\ast}\label{eq:HKinPrimeDefinition}\\
 & =\sum_{p\in B_{F}^{c}}\left(\left|p\right|^{2}-k_{F}^{2}\right)c_{p}^{\ast}c_{p}+\sum_{p\in B_{F}}\left(k_{F}^{2}-\left|p\right|^{2}\right)c_{p}c_{p}^{\ast},\nonumber 
\end{align}
where we for the last identity used the ``particle-hole symmetry''
\begin{equation}
\mathcal{N}_{E}:=\sum_{p\in B_{F}^{c}}c_{p}^{\ast}c_{p}=\sum_{p\in B_{F}}c_{p}c_{p}^{\ast}\quad\text{on }\mathcal{H}_{N}.
\end{equation}
From the last identity of equation (\ref{eq:HKinPrimeDefinition})
it is clear that $H_{\mathrm{kin}}^{\prime}$ is non-negative.

We normal-order $H_{\mathrm{int}}$ with respect to $\psi_{\mathrm{FS}}$:
Using the CAR and the fact that $\sum_{p\in\mathbb{Z}^{3}}c_{p}^{\ast}c_{p}=\mathcal{N}=N$
on $\mathcal{H}_{N}$, it factorizes as
\begin{equation}
H_{\mathrm{int}}=\frac{1}{2\left(2\pi\right)^{3}}\sum_{k\in\mathbb{Z}_{\ast}^{3}}\hat{V}_{k}\left(\left(\sum_{p\in\mathbb{Z}^{3}}c_{p}^{\ast}c_{p+k}\right)^{\ast}\left(\sum_{q\in\mathbb{Z}^{3}}c_{q-k}^{\ast}c_{q}\right)-N\right).
\end{equation}
Decomposing for every $k\in\mathbb{Z}_{\ast}^{3}$
\begin{equation}
\sum_{p\in\mathbb{Z}^{3}}c_{p-k}^{\ast}c_{p}=B_{k}+B_{-k}^{\ast}+D_{k},\quad B_{k}=\sum_{p\in L_{k}}c_{p-k}^{\ast}c_{p},
\end{equation}
we can write
\begin{align}
H_{\mathrm{int}} & =\frac{1}{2\left(2\pi\right)^{3}}\sum_{k\in\mathbb{Z}_{\ast}^{3}}\hat{V}_{k}\left(\left(B_{k}+B_{-k}^{\ast}\right)^{\ast}\left(B_{k}+B_{-k}^{\ast}\right)-N\right)\\
 & +\frac{1}{2\left(2\pi\right)^{3}}\sum_{k\in\mathbb{Z}_{\ast}^{3}}\hat{V}_{k}\left(2\,\mathrm{Re}\left(\left(B_{k}+B_{-k}^{\ast}\right)^{\ast}D_{k}\right)+D_{k}^{\ast}D_{k}\right).\nonumber 
\end{align}
Using the CAR again it is easy to compute that
\begin{equation}
\left[B_{k},B_{k}^{\ast}\right]=\left|L_{k}\right|-\sum_{p\in L_{k}}\left(c_{p}^{\ast}c_{p}+c_{p-k}c_{p-k}^{\ast}\right)
\end{equation}
whence (using also that $\hat{V}_{k}=\hat{V}_{-k}$)
\begin{align}
H_{\mathrm{int}} & =-\frac{1}{2\left(2\pi\right)^{3}}\sum_{k\in\mathbb{Z}_{\ast}^{3}}\hat{V}_{k}\left(N-\left|L_{k}\right|\right)+\frac{1}{2\left(2\pi\right)^{3}}\sum_{k\in\mathbb{Z}_{\ast}^{3}}\hat{V}_{k}\left(2B_{k}^{\ast}B_{k}+B_{k}B_{-k}+B_{-k}^{\ast}B_{k}^{\ast}\right)\nonumber \\
 & +\frac{1}{2\left(2\pi\right)^{3}}\sum_{k\in\mathbb{Z}_{\ast}^{3}}\hat{V}_{k}\left(2\,\mathrm{Re}\left(\left(B_{k}+B_{-k}^{\ast}\right)^{\ast}D_{k}\right)+D_{k}^{\ast}D_{k}-\sum_{p\in L_{k}}\left(c_{p}^{\ast}c_{p}+c_{p-k}c_{p-k}^{\ast}\right)\right).
\end{align}
Note that the first sum is finite as $\left|L_{k}\right|=N$ for $\left|k\right|>2k_{F}$.
It is easily verified that $D_{k}\psi_{\mathrm{FS}}=D_{k}^{\ast}\psi_{\mathrm{FS}}=B_{k}\psi_{\mathrm{FS}}=0$,
so we deduce from this identity that
\begin{equation}
\left\langle \psi_{\mathrm{FS}},H_{\mathrm{int}}\psi_{\mathrm{FS}}\right\rangle =-\frac{1}{2\left(2\pi\right)^{3}}\sum_{k\in\mathbb{Z}_{\ast}^{3}}\hat{V}_{k}\left(N-\left|L_{k}\right|\right)
\end{equation}
and we summarize the calculations above in the following:
\begin{prop}
\label{prop:LocalizationofHN}It holds that
\[
H_{N}=E_{\mathrm{FS}}+H_{\mathrm{kin}}^{\prime}+\sum_{k\in\mathbb{Z}_{\ast}^{3}}\frac{\hat{V}_{k}k_{F}^{-1}}{2\left(2\pi\right)^{3}}\left(2B_{k}^{\ast}B_{k}+B_{k}B_{-k}+B_{-k}^{\ast}B_{k}^{\ast}\right)+\mathcal{C}+\mathcal{Q}
\]
where $E_{\mathrm{FS}}=\left\langle \psi_{\mathrm{FS}},H_{N}\psi_{\mathrm{FS}}\right\rangle $
and the \textup{cubic} and \textup{quartic} terms, $\mathcal{C}$
and $\mathcal{Q}$, are defined by
\begin{align*}
\mathcal{C} & =\frac{k_{F}^{-1}}{\left(2\pi\right)^{3}}\sum_{k\in\mathbb{Z}_{\ast}^{3}}\hat{V}_{k}\,\mathrm{Re}\left(\left(B_{k}+B_{-k}^{\ast}\right)^{\ast}D_{k}\right),\\
\mathcal{Q} & =\frac{k_{F}^{-1}}{2\left(2\pi\right)^{3}}\sum_{k\in\mathbb{Z}_{\ast}^{3}}\hat{V}_{k}\left(D_{k}^{\ast}D_{k}-\sum_{p\in L_{k}}\left(c_{p}^{\ast}c_{p}+c_{p-k}c_{p-k}^{\ast}\right)\right).
\end{align*}
\end{prop}

We will prove that the cubic and quartic terms are negligible, and
so the main contribution to the correlation energy comes from the
\textit{bosonizable terms}
\begin{equation} \label{eq:Heff}
H_{\mathrm{eff}}=H_{\mathrm{kin}}^{\prime}+\sum_{k\in\mathbb{Z}_{\ast}^{3}}\frac{\hat{V}_{k}k_{F}^{-1}}{2\left(2\pi\right)^{3}}\left(2B_{k}^{\ast}B_{k}+B_{k}B_{-k}+B_{-k}^{\ast}B_{k}^{\ast}\right).
\end{equation}
We will write these in terms of quasi-bosonic operators, which will
lead us to define a quasi-bosonic Bogolubov transformation that serves
to effectively diagonalize them.

\subsubsection*{The Quasi-Bosonic Quadratic Hamiltonian}

We define the \textit{excitation operators} $b_{k,p}^{\ast}$, $b_{k,p}$,
for $k\in\mathbb{Z}_{\ast}^{3}$ and $p\in L_{k}$, by
\begin{equation}
b_{k,p}=c_{p-k}^{\ast}c_{p},\quad b_{k,p}^{\ast}=c_{p}^{\ast}c_{p-k}.
\end{equation}
The name is due to the fact that $b_{k,p}^{\ast}$ acts by annihilating
a state with momentum $p-k\in B_{F}$ and creating a state with momentum
$p\in B_{F}^{c}$, i.e. it excites the state $p-k$ to the state $p$.

For the purpose of computations it is convenient to also introduce
a basis-independent notation for the quasi-bosonic operators. Consider
for $k\in\mathbb{Z}_{\ast}^{3}$ the auxilliary space $\ell^{2}(L_{k})$,
which we will consider only as a real vector space, with standard
orthonormal basis $\left(e_{p}\right)_{p\in L_{k}}$. For any $k\in\mathbb{Z}_{\ast}^{3}$
and $\varphi\in\ell^{2}(L_{k})$ we define the \textit{generalized
excitation operators} $b_{k}(\varphi)$ and $b_{k}^{\ast}(\varphi)$
by
\begin{equation}
b_{k}(\varphi)=\sum_{p\in L_{k}}\left\langle \varphi,e_{p}\right\rangle b_{k,p},\quad b_{k}^{\ast}(\varphi)=\sum_{p\in L_{k}}\left\langle e_{p},\varphi\right\rangle b_{k,p}^{\ast}.
\end{equation}
Note that the assignments $\varphi\mapsto b_{k}(\varphi),\,b_{k}^{\ast}(\varphi)$
are both linear (as we only consider $\ell^{2}(L_{k})$
as a real vector space). In this notation we simply have that $b_{k}\left(e_{p}\right)=b_{k,p}$. A short calculation using the CAR shows that these operators are quasi-bosonic
in the following sense:
\begin{lem}
\label{lemma:QuasiBosonicCommutationRelations}For any $k,l\in\mathbb{Z}_{\ast}^{3}$,
$\varphi\in\ell^{2}(L_{k})$ and $\psi\in\ell^{2}(L_l)$
it holds that
\begin{align*}
\left[b_{k}(\varphi),b_{l}\left(\psi\right)\right]  =\left[b_{k}^{\ast}(\varphi),b_{l}^{\ast}\left(\psi\right)\right]=0,\quad \left[b_{k}(\varphi),b_{l}^{\ast}\left(\psi\right)\right] =\delta_{k,l}\left\langle \varphi,\psi\right\rangle +\varepsilon_{k,l}\left(\varphi;\psi\right),
\end{align*}
where the \textup{exchange correction} $\varepsilon_{k,l}\left(\varphi;\psi\right)$
is given by
\[
\varepsilon_{k,l}\left(\varphi;\psi\right)=-\sum_{p\in L_{k}}\sum_{q\in L_{l}}\left\langle \varphi,e_{p}\right\rangle \left\langle e_{q},\psi\right\rangle \left(\delta_{p,q}c_{q-l}c_{p-k}^{\ast}+\delta_{p-k,q-l}c_{q}^{\ast}c_{p}\right).
\]
\end{lem}

Note that in the purely bosonic picture the exchange correction is
absent. In our quasi-bosonic case, these corrections are small but
non-zero; it will be important to keep careful track of them as it
is these that gives rise to the exchange contribution $E_{\mathrm{corr},\mathrm{ex}}$.

For any operators $A$, $B$ on $\ell^{2}(L_{k})$, we define
the \textit{associated quadratic operators} $Q_{1}^{k}(A)$,
$Q_{2}^{k}(B)$ on $\mathcal{H}_{N}$ by\footnote{Note that these definitions differ slightly from those of \cite{CHN-21}.
The main change is the definition of $Q_{1}^{k}(A)$; this
operator is what was denoted $\tilde{Q}_{1}^{k}(A)$ in
that paper.}
\begin{equation}
Q_{1}^{k}(A)=\sum_{p,q\in L_{k}}\left\langle e_{p},Ae_{q}\right\rangle b_{k,p}^{\ast}b_{k,q}=\sum_{p\in L_{k}}b_{k}^{\ast}(Ae_p)b_{k,p}
\end{equation}
and
\begin{equation}
Q_{2}^{k}(B)=\sum_{p,q\in L_{k}}\left\langle e_{p},Be_{q}\right\rangle \left(b_{k,p}b_{-k,-q}+b_{-k,-q}^{\ast}b_{k,p}^{\ast}\right)=\sum_{p\in L_{k}}\left(b_{k}(Be_p)b_{-k,-p}+b_{-k,-p}^{\ast}b_{k}^{\ast}(Be_p)\right).
\end{equation}
Defining the operator $P_{k}$ on $\ell^{2}(L_{k})$ by
\begin{equation}
P_{k}=\left|v_{k}\right\rangle \left\langle v_{k}\right|,\quad v_{k}=\sqrt{\frac{\hat{V}_{k}k_{F}^{-1}}{2\left(2\pi\right)^{3}}}\sum_{p\in L_{k}}e_{p}\in\ell^{2}(L_{k}),\quad\text{so that}\quad\left\langle e_{p},P_{k}e_{q}\right\rangle =\frac{\hat{V}_{k}k_{F}^{-1}}{2\left(2\pi\right)^{3}},
\end{equation}
we can express the interaction part of the bosonizable terms as
\begin{align}\label{eq:Non-KineticBosonizableTerms}
 & \quad\;\sum_{k\in\mathbb{Z}_{\ast}^{3}}\frac{\hat{V}_{k}k_{F}^{-1}}{2\left(2\pi\right)^{3}}\left(2B_{k}^{\ast}B_{k}+B_{k}B_{-k}+B_{-k}^{\ast}B_{k}^{\ast}\right)=\sum_{k\in\mathbb{Z}_{\ast}^{3}}\left(2\,Q_{1}^{k}(P_{k})+Q_{2}^{k}(P_{k})\right) \\
 & =\sum_{k\in\mathbb{Z}_{\ast}^{3}}\left(2\sum_{p,q\in L_{k}}\frac{\hat{V}_{k}k_{F}^{-1}}{2\left(2\pi\right)^{3}}b_{k,p}^{\ast}b_{k,q}+\sum_{p,q\in L_{k}}\frac{\hat{V}_{k}k_{F}^{-1}}{2\left(2\pi\right)^{3}}\left(b_{k,p}b_{-k,-q}+b_{-k,-q}^{\ast}b_{k,p}^{\ast}\right)\right) \nonumber
\end{align}
The localized kinetic operator $H_{\mathrm{kin}}^{\prime}$ cannot
be written exactly in a quadratic quasi-bosonic form, but due to the
commutation relation
\begin{equation}
\left[H_{\mathrm{kin}}^{\prime},b_{k,p}^{\ast}\right]=\left(\left|p\right|^{2}-\left|p-k\right|^{2}\right)b_{k,p}^{\ast}=2\lambda_{k,p}b_{k,p}^{\ast}\label{eq:HkinbkpCommutator}
\end{equation}
(see \cite[Eq. (1.76)]{CHN-21}) and the quasi-bosonicity of the $b_{k,p}^{\ast}$
operators, it is sensible to consider it analogous to a quadratic
operator of the form
\begin{equation}
\sum_{k\in\mathbb{Z}_{\ast}^{3}}\sum_{p\in L_{k}}2\lambda_{k,p}b_{k,p}^{\ast}b_{k,p}=\sum_{k\in\mathbb{Z}_{\ast}^{3}}2\,Q_{1}^{k}(h_{k})
\end{equation}
where the operators $h_{k}$ on $\ell^{2}(L_{k})$ are
simply defined by $h_{k}e_{p}=\lambda_{k,p}e_{p}$. In all we thus
consider the bosonizable terms as being analogous to a quasi-bosonic
quadratic operator as
\begin{equation}
H_{\mathrm{eff}}\approx\sum_{k\in\mathbb{Z}_{\ast}^{3}}\left(2\,Q_{1}^{k}(h_{k}+P_{k})+Q_{2}^{k}(P_{k})\right).\label{eq:QuadraticAnalogy}
\end{equation}

\subsubsection*{The Quasi-Bosonic Bogolubov Transformation}

If the quadratic Hamiltonian on the right-hand side of equation (\ref{eq:QuadraticAnalogy})
was exactly bosonic, it could be diagonalized by a Bogolubov transformation.
Motivated by this we define such a transformation in the quasi-bosonic
setting, while keeping careful track of the additional terms arising
from the exchange correction.

Let $K_{k}:\ell^{2}(L_{k})\rightarrow\ell^{2}(L_{k})$,
$k\in\mathbb{Z}_{\ast}^{3}$, be a collection of symmetric operators
satisfying
\begin{equation}
\left\langle e_{p},K_{k}e_{q}\right\rangle =\left\langle e_{-p},K_{-k}e_{-q}\right\rangle ,\quad k\in\mathbb{Z}_{\ast}^{3},\,p,q\in L_{k}.
\end{equation}
Then we define the \textit{associated quasi-bosonic Bogolubov kernel}
$\mathcal{K}$ on $\mathcal{H}_{N}$ by
\begin{align}
\mathcal{K} & =\frac{1}{2}\sum_{l\in\mathbb{Z}_{\ast}^{3}}\sum_{p,q\in L_{l}}\left\langle e_{p},K_{l}e_{q}\right\rangle \left(b_{l,p}b_{-l,-q}-b_{-l,-q}^{\ast}b_{l,p}^{\ast}\right)\label{eq:calKDefinition}\\
 & =\frac{1}{2}\sum_{l\in\mathbb{Z}_{\ast}^{3}}\sum_{q\in L_{l}}\left(b_{l}(K_{l}e_{q})b_{-l,-q}-b_{-l,-q}^{\ast}b_{l}^{\ast}(K_{l}e_{q})\right).\nonumber 
\end{align}
It is obvious from the second equation that $\mathcal{K}$ is skew-symmetric;
$\mathcal{K}$ thus generates a unitary transformation $e^{\mathcal{K}}:\mathcal{H}_{N}\rightarrow\mathcal{H}_{N}$
- the quasi-bosonic Bogolubov transformation.

We consider the case $\sum_{k\in\mathbb{Z}_{\ast}^{3}}\left\Vert K_{k}\right\Vert _{\mathrm{HS}}^{2}<\infty$,
in which case $\mathcal{K}$ is not only well-defined but even bounded
as an operator on $\mathcal{H}_{N}$, as we will prove in the next
section.

We choose the operators $\left(K_{k}\right)$ such that $e^{\mathcal{K}}$
would diagonalize the right-hand side of equation (\ref{eq:QuadraticAnalogy})
if it was exactly bosonic. As explained in \cite[Section 3]{CHN-21}
the diagonalizing kernel is
\begin{equation}
K_{k}=-\frac{1}{2}\log\left(h_{k}^{-\frac{1}{2}}\left(h_{k}^{\frac{1}{2}}\left(h_{k}+2P_{k}\right)h_{k}^{\frac{1}{2}}\right)^{\frac{1}{2}}h_{k}^{-\frac{1}{2}}\right).\label{eq:KkDefinition}
\end{equation}
Keeping careful track of the quasi-bosonic corrections, the action
of $e^{\mathcal{K}}$ on the bosonizable terms are as follows:
\begin{thm}
\label{thm:DiagonalizationoftheBosonizableTerms} Let $H_{\rm eff}$ be as in \eqref{eq:Heff}. Assume $\sum_{k\in\mathbb{Z}_{\ast}^{3}}\hat{V}_{k}^{2}<\infty$. 
Then $e^{\mathcal{K}}$ is well-defined and
\begin{align*}
 & \;\,e^{\mathcal{K}}H_{\mathrm{eff}}e^{-\mathcal{K}}=E_{\mathrm{corr},\mathrm{bos}}+H_{\mathrm{kin}}^{\prime}+2\sum_{k\in\mathbb{Z}_{\ast}^{3}}Q_{1}^{k}(e^{-K_{k}}h_{k}e^{-K_{k}}-h_{k})\\
 & +\sum_{k\in\mathbb{Z}_{\ast}^{3}}\int_{0}^{1}e^{\left(1-t\right)\mathcal{K}}\left(\varepsilon_{k}(\left\{ K_{k},B_{k}(t)\right\})+2\,\mathrm{Re}\left(\mathcal{E}_{k}^{1}(A_k(t))\right)+2\,\mathrm{Re}\left(\mathcal{E}_{k}^{2}(B_k(t))\right)\right)e^{-\left(1-t\right)\mathcal{K}}dt
\end{align*}
where for any symmetric operators $A_{k},B_{k}:\ell^{2}(L_{k})\rightarrow\ell^{2}(L_{k})$
we define
\begin{align*}
\varepsilon_{k}(A_{k}) & =-\sum_{p\in L_{k}}\left\langle e_{p},A_{k}e_{p}\right\rangle \left(c_{p}^{\ast}c_{p}+c_{p-k}c_{p-k}^{\ast}\right),\\
\mathcal{E}_{k}^{1}(A_k) & =\sum_{l\in\mathbb{Z}_{\ast}^{3}}\sum_{p\in L_{k}}\sum_{q\in L_{l}}b_{k}^{\ast}(A_k e_p)\left\{ \varepsilon_{k,l}(e_{p};e_{q}),b_{-l}^{\ast}(K_{-l}e_{-q})\right\}, \\
\mathcal{E}_{k}^{2}(B_{k}) & =\frac{1}{2}\sum_{l\in\mathbb{Z}_{\ast}^{3}}\sum_{p\in L_{k}}\sum_{q\in L_{l}}\left\{ b_{k}(B_k e_p),\left\{ \varepsilon_{-k,-l}(e_{-p};e_{-q}),b_{l}^{\ast}(K_{l}e_{q})\right\} \right\} ,
\end{align*}
and for $t\in\left[0,1\right]$ the operators $A_{k}(t),B_{k}(t):\ell^{2}(L_{k})\rightarrow\ell^{2}(L_{k})$
are given by
\begin{align*}
A_{k}(t) & =\frac{1}{2}\left(e^{tK_{k}}\left(h_{k}+2P_{k}\right)e^{tK_{k}}+e^{-tK_{k}}h_{k}e^{-tK_{k}}\right)-h_{k},\\
B_{k}(t) & =\frac{1}{2}\left(e^{tK_{k}}\left(h_{k}+2P_{k}\right)e^{tK_{k}}-e^{-tK_{k}}h_{k}e^{-tK_{k}}\right).
\end{align*}
\end{thm}

This result is essentially the same as \cite[Proposition 5.7]{CHN-21},
except that we now do not introduce a momentum cut-off and assume
only that $\sum_{k\in\mathbb{Z}_{\ast}^{3}}\hat{V}_{k}^{2}<\infty$.
For the readers convenience, we include in Appendix \ref{sec:DiagonalizationoftheBosonizableTerms}
the proof of the identity of Theorem \ref{thm:DiagonalizationoftheBosonizableTerms}
- that the condition $\sum_{k\in\mathbb{Z}_{\ast}^{3}}\hat{V}_{k}^{2}<\infty$
is sufficient to define $e^{\mathcal{K}}$ is proved in the next section.

\subsubsection*{Outline of the Paper}

Now we come to the main part of the paper. We will choose as our trial
state $\Psi=e^{-\mathcal{K}}\psi_{\mathrm{FS}}$. As mentioned the
cubic and quartic terms are negligible, so the energy of our trial
state energy is by Theorem \ref{thm:DiagonalizationoftheBosonizableTerms},
to leading order,
\begin{align}
 & \left\langle \Psi,H_{N}\Psi\right\rangle \approx E_{\mathrm{FS}}+E_{\mathrm{corr},\mathrm{bos}}\label{eq:TrialStateExpectation}\\
 & +\sum_{k\in\mathbb{Z}_{\ast}^{3}}\int_{0}^{1}\left\langle \psi_{\mathrm{FS}},e^{\left(1-t\right)\mathcal{K}}\left(\varepsilon_{k}(\left\{ K_{k},B_{k}(t)\right\})+2\,\mathrm{Re}\left(\mathcal{E}_{k}^{1}(A_k(t))\right)+2\,\mathrm{Re}\left(\mathcal{E}_{k}^{2}(B_k(t))\right)\right)e^{-\left(1-t\right)\mathcal{K}}\psi_{\mathrm{FS}}\right\rangle dt.\nonumber 
\end{align}
The main task will thus be to extract the exchange contribution $E_{\mathrm{corr},\mathrm{ex}}$
from this last term. The outline of the paper is as follows:

In Section \ref{sec:TheBogolubovKernel} we show that $e^{\mathcal{K}}$
is well-defined by proving that $\mathcal{K}$ is bounded under the
condition $\sum_{k\in\mathbb{Z}_{\ast}^{3}}\hat{V}_{k}^{2}<\infty$.
We do this by employing a type of higher-order fermionic estimate,
resulting in a bound of the form
\begin{equation}
\pm\mathcal{K}\leq C\sqrt{\sum_{k\in\mathbb{Z}_{\ast}^{3}}\hat{V}_{k}^{2}}\mathcal{N}_{E}
\end{equation}
which will also be crucial in allowing us to control $\mathcal{N}_{E}$
later.

In Section \ref{sec:AnalysisoftheOneBodyOperators} we establish various
bounds on the one-body operators $K_{k}$, $A_{k}(t)$
and $B_{k}(t)$. This is conceptually similar to the one-body
analysis in our previous paper \cite{CHN-21}, but we must refine
several estimates in order to establish control using only the assumption
that $\sum_{k\in\mathbb{Z}_{\ast}^{3}}\hat{V}_{k}^{2}<\infty$.

In Section \ref{sec:AnalysisoftheExchangeTerms} comes the main new
work: We engage in a detailed study of the exchange terms $\mathcal{\mathcal{E}}_{k}^{1}(A_k)$
and $\mathcal{\mathcal{E}}_{k}^{2}(B_k)$ so that we
can extract $E_{\mathrm{corr},\mathrm{ex}}$ from the last term of
equation (\ref{eq:TrialStateExpectation}), first in the form
\begin{equation}
\sum_{k\in\mathbb{Z}_{\ast}^{3}}\int_{0}^{1}\left\langle \psi_{\mathrm{FS}},2\,\mathrm{Re}\left(\mathcal{E}_{k}^{2}(B_k(t))\right)\psi_{\mathrm{FS}}\right\rangle dt,
\end{equation}
and then analyze this expression further to obtain the leading order
of this, which is precisely $E_{\mathrm{corr},\mathrm{ex}}$ as given
in Theorem \ref{them:MainTheorem}.

Finally in Section \ref{sec:EstimationoftheNonBosonizableTerms} we
control the non-bosonizable cubic and quartic terms, and bound the
number operator $\mathcal{N}_{E}$ and its powers by a Gronwall argument.
We end the paper by concluding Theorem \ref{them:MainTheorem}.

\medskip

\noindent
{\bf Acknowledgements.} We would like to thank the editor and the referees for helpful suggestions. MRC and PTN acknowledge the support from the Deutsche Forschungsgemeinschaft
(DFG project Nr. 426365943).

\section{\label{sec:TheBogolubovKernel}The Bogolubov Kernel}

We consider the kernel $\mathcal{K}$ defined by  (\ref{eq:calKDefinition}).
We prove the following:
\begin{prop} \label{prop:Kernel-cK-boundedness}
Let $K_{l}:\ell^{2}(L_l)\rightarrow\ell^{2}(L_l)$,
$l\in\mathbb{Z}_{\ast}^{3}$, be a collection of symmetric operators.
Then provided $\sum_{l\in\mathbb{Z}_{\ast}^{3}}\left\Vert K_{l}\right\Vert _{\mathrm{HS}}^{2}<\infty$,
the expression
\[
\mathcal{K}=\frac{1}{2}\sum_{l\in\mathbb{Z}_{\ast}^{3}}\sum_{p,q\in L_{l}}\left\langle e_{p},K_{l}e_{q}\right\rangle \left(b_{l,p}b_{-l,-q}-b_{-l,-q}^{\ast}b_{l,p}^{\ast}\right)
\]
defines a bounded operator $\mathcal{K}:\mathcal{H}_{N}\rightarrow\mathcal{H}_{N}$, and for any $\Psi,\Phi\in\mathcal{H}_{N}$ we have
$$
\left|\left\langle \Psi,\mathcal{K}\Phi\right\rangle \right|  \leq\sqrt{5}\sqrt{\sum_{l\in\mathbb{Z}_{\ast}^{3}}\left\Vert K_{l}\right\Vert _{\mathrm{HS}}^{2}}\sqrt{\left\langle \Psi,\left(\mathcal{N}_{E}+1\right)\Psi\right\rangle \left\langle \Phi,\left(\mathcal{N}_{E}+1\right)\Phi\right\rangle }.
$$
\end{prop}
Note that $\mathcal{N}_{E}=\sum_{p\in B_{F}^c} c_{p}^*c_{p}=\sum_{p\in B_{F}}c_{p}c_{p}^{\ast}\leq\left|B_{F}\right|=N$ on $\mathcal{H}_{N}$. Moreover, it was shown in \cite{CHN-21} (see also Theorem \ref{them:OneBodyEstimates}) that the kernels in (\ref{eq:KkDefinition}) satisfy $\left\Vert K_{k}\right\Vert _{\mathrm{HS}}\leq C\hat{V}_{k}$, and hence the boundedness of $\mathcal{K}$ follows from the assumption  $\sum_{k\in \mathbb{Z}_*^3} \hat V_k^2<\infty$. 
Let us write  
\begin{equation}
\mathcal{K}=\tilde{\mathcal{K}}-\tilde{\mathcal{K}}^{\ast}, 
\quad \tilde{\mathcal{K}}=\frac{1}{2}\sum_{l\in\mathbb{Z}_{\ast}^{3}}\sum_{p,q\in L_{l}}\left\langle e_{p},K_{l}e_{q}\right\rangle b_{l,p}b_{-l,-q},
\end{equation}
and focus on the boundedness of $\mathcal{K}$. Since 
\begin{align}
2\,\tilde{\mathcal{K}} =\sum_{l\in\mathbb{Z}_{\ast}^{3}}\sum_{p,q\in L_{l}}\left\langle e_{p},K_{l}e_{q}\right\rangle b_{l,p}c_{-q+l}^{\ast}c_{-q}
  =\sum_{q\in B_{F}^{c}}\left(\sum_{l\in\mathbb{Z}_{\ast}^{3}}\sum_{p\in L_{l}}1_{L_l}(q)\left\langle e_{p},K_{l}e_{q}\right\rangle b_{l,p}c_{-q+l}^{\ast}\right)c_{-q},
\end{align}
for  any $\Psi,\Phi\in\mathcal{H}_{N}$ we may estimate by the Cauchy--Schwarz inequality
\begin{align}\label{eq:calKtildeEstimate1}
|\langle \Psi,\tilde{\mathcal{K}}\Phi\rangle|  
& \leq\frac{1}{2}\sqrt{\sum_{q\in B_{F}^{c}}\left\Vert \sum_{l\in\mathbb{Z}_{\ast}^{3}}\sum_{p\in L_{l}}1_{L_l}(q)\left\langle K_{l}e_{q},e_{p}\right\rangle c_{-q+l}b_{l,p}^{\ast}\Psi\right\Vert ^{2}}\sqrt{\sum_{q\in B_{F}^{c}}\left\Vert c_{-q}\Phi\right\Vert ^{2}}\nonumber \\
 & = \frac{1}{2}\sqrt{\sum_{q\in B_{F}^{c}}\left\Vert \sum_{l\in\mathbb{Z}_{\ast}^{3}}\sum_{p\in L_{l}}1_{L_l}(q)\left\langle K_{l}e_{q},e_{p}\right\rangle c_{-q+l}b_{l,p}^{\ast}\Psi\right\Vert ^{2}}\sqrt{\left\langle \Phi,\mathcal{N}_{E}\Phi\right\rangle }. 
\end{align}
The operator appearing under the root can be written as
\begin{align}\label{eq:calKtildeEstimate2}
&  \quad\;\,\sum_{l\in\mathbb{Z}_{\ast}^{3}}\sum_{p\in L_{l}}1_{L_l}(q)\left\langle K_{l}e_{q},e_{p}\right\rangle c_{-q+l}b_{l,p}^{\ast}=\sum_{l\in\mathbb{Z}_{\ast}^{3}}\sum_{p\in L_{l}}1_{L_l}(q)\left\langle K_{l}e_{q},e_{p}\right\rangle c_{p}^{\ast}c_{p-l}c_{-q+l}, \nonumber\\
& =\sum_{p'\in B_{F}^{c}}\sum_{q',r'\in B_{F}}\left(\sum_{l\in\mathbb{Z}_{\ast}^{3}}\sum_{p\in L_{l}}\delta_{p',p}\delta_{q',p-l}\delta_{r',-q+l}1_{L_l}(q)\left\langle K_{l}e_{q},e_{p}\right\rangle \right)c_{p'}^{\ast}c_{q'}c_{r'}. 
\end{align}
Let us estimate the following general expression, with  some coefficients $A_{p,q,r}$,
\begin{equation}
\sum_{p\in B_{F}^{c}}\sum_{q,r\in B_{F}}A_{p,q,r}c_{p}^{\ast}c_{q}c_{r}. 
\end{equation}


\subsubsection*{A Higher Order Fermionic Estimate}

Note that the Cauchy--Schwarz inequality trivially implies that 
\begin{equation}
\left\Vert \sum A_{p}c_{p}\Psi\right\Vert \le \sum |A_p| \|c_p \Psi\| \le \sqrt{\sum |A_p|^2} \sqrt{\sum  \|c_p \Psi\|^2},
\end{equation}
but this is non-optimal for fermionic states. The ``standard fermionic estimate'' states that 
\begin{equation}
\left\Vert \sum A_{p}c_{p}\Psi\right\Vert ,\,\left\Vert \sum A_{p}c_{p}^{\ast}\Psi\right\Vert \leq\sqrt{\sum\left|A_{p}\right|^{2}}\left\Vert \Psi\right\Vert ,
\end{equation}
which can be proved by appealing to the CAR as follows: 
\begin{align} \label{eq:standard-fermionic-estimate}
\left(\sum A_{p}c_{p}\right)^{\ast}\left(\sum A_{q}c_{q}\right) 
 \le \left\{ \left(\sum A_{p}c_{p}\right)^{\ast},\left(\sum A_{q}c_{q}\right)\right\} 
=\sum\overline{A_{p}}A_{q}\left\{ c_{p}^{\ast},c_{q}\right\} 
=\sum\left|A_{p}\right|^{2}.
\end{align}


One can imagine generalizing this to quadratic expressions of the
form $\sum_{p,q}A_{p,q}c_{p}c_{q}$, but this fails since the CAR only yields a commutation relation for such expressions,
and not an anticommutation relation. 
 However, for cubic expressions, such
as $\sum_{p,q,r}A_{p,q,r}c_{p}^{\ast}c_{q}c_{r}$, the CAR does yield
an anticommutation relation, allowing the trick to be applied. The
anticommutator is of course not constant, but rather a combination
of quadratic, linear and constant expressions, but this still yields
a reduction in ``number operator order'', which will be crucial
for our estimation of $e^{\mathcal{K}}\mathcal{N}_{E}^{m}e^{-\mathcal{K}}$
later on. 
We will need the following basic
anticommutator:
\begin{lem} \label{lem:basic-anticommutator-cubic}
For any $p,p'\in B_{F}^{c}$ and $q,q',r,r'\in B_{F}$ it holds that
\begin{align*}
\left\{ \left(c_{p}^{\ast}c_{q}c_{r}\right)^{\ast},c_{p'}^{\ast}c_{q'}c_{r'}\right\}   =& \,\, \delta_{p,p'}c_{q'}c_{r'}c_{r}^{\ast}c_{q}^{\ast}+\delta_{q,q'}c_{p'}^{\ast}c_{r'}c_{r}^{\ast}c_{p}+\delta_{r,r'}c_{p'}^{\ast}c_{q'}c_{q}^{\ast}c_{p}\\
 & -\delta_{r,q'}c_{p'}^{\ast}c_{r'}c_{q}^{\ast}c_{p}-\delta_{r,q'}c_{p'}^{\ast}c_{r'}c_{q}^{\ast}c_{p}\\
 & -\delta_{q,q'}\delta_{r,r'}c_{p'}^{\ast}c_{p}-\delta_{p,p'}\delta_{r,r'}c_{q'}c_{q}^{\ast}-\delta_{p,p'}\delta_{q,q'}c_{r'}c_{r}^{\ast}\\
 & +\delta_{q,r'}\delta_{r,q'}c_{p'}^{\ast}c_{p}+\delta_{p,p'}\delta_{r,q'}c_{r'}c_{q}^{\ast}+\delta_{p,p'}\delta_{q,r'}c_{q'}c_{r}^{\ast}\\
 & +\delta_{p,p'}\delta_{q,q'}\delta_{r,r'}-\delta_{p,p'}\delta_{q,r'}\delta_{r,q'}. 
\end{align*}
\end{lem}

We can now conclude the desired bound:
\begin{prop} \label{prop:Apqrccc}
Let $A_{p,q,r}\in\mathbb{C}$ for $p\in B_{F}^{c}$ and $q,r\in B_{F}$
with $\sum_{p\in B_{F}^{c}}\sum_{q,r\in B_{F}}\left|A_{p,q,r}\right|^{2}<\infty$
be given. Then for any $\Psi\in\mathcal{H}_{N}$
\[
\left\Vert \sum_{p\in B_{F}^{c}}\sum_{q,r\in B_{F}}A_{p,q,r}c_{p}^{\ast}c_{q}c_{r}\Psi\right\Vert ^{2}\leq5\sum_{p\in B_{F}^{c}}\sum_{q,r\in B_{F}}\left|A_{p,q,r}\right|^{2}\left\langle \Psi,\left(\mathcal{N}_{E}+1\right)\Psi\right\rangle .
\]
\end{prop}

\textbf{Proof:} As in the proof of the standard fermionic estimate \eqref{eq:standard-fermionic-estimate}, 
we have
\begin{align*}
\left\Vert \sum_{p\in B_{F}^{c}}\sum_{q,r\in B_{F}}A_{p,q,r}c_{p}^{\ast}c_{q}c_{r}\Psi\right\Vert ^{2} 
 & \leq\sum_{p,p'\in B_{F}^{c}}\sum_{q,q',r,r'\in B_{F}}\overline{A_{p,q,r}}A_{p',q',r'}\left\langle \Psi,\left\{ \left(c_{p}^{\ast}c_{q}c_{r}\right)^{\ast},c_{p'}^{\ast}c_{q'}c_{r'}\right\} \Psi\right\rangle. 
\end{align*}
Hence, by the identity of Lemma \ref{lem:basic-anticommutator-cubic}, we bound the left-hand side by 
\begin{align}\label{eq:HigherFermionicEstimateTerms}
 & \sum_{p,p'\in B_{F}^{c}}\sum_{q,q',r,r'\in B_{F}}\overline{A_{p,q,r}}A_{p',q',r'}\left\langle \Psi,\left(\delta_{p,p'}c_{q'}c_{r'}c_{r}^{\ast}c_{q}^{\ast}+\delta_{q,q'}c_{p'}^{\ast}c_{r'}c_{r}^{\ast}c_{p}+\delta_{r,r'}c_{p'}^{\ast}c_{q'}c_{q}^{\ast}c_{p}\right)\Psi\right\rangle \nonumber \\
 &  -\sum_{p,p'\in B_{F}^{c}}\sum_{q,q',r,r'\in B_{F}}\overline{A_{p,q,r}}A_{p',q',r'}\left\langle \Psi,\left(\delta_{r,q'}c_{p'}^{\ast}c_{r'}c_{q}^{\ast}c_{p}+\delta_{r,q'}c_{p'}^{\ast}c_{r'}c_{q}^{\ast}c_{p}\right)\Psi\right\rangle \nonumber \\
 & -\sum_{p,p'\in B_{F}^{c}}\sum_{q,q',r,r'\in B_{F}}\overline{A_{p,q,r}}A_{p',q',r'}\left\langle \Psi,\left(\delta_{q,q'}\delta_{r,r'}c_{p'}^{\ast}c_{p}+\delta_{p,p'}\delta_{r,r'}c_{q'}c_{q}^{\ast}+\delta_{p,p'}\delta_{q,q'}c_{r'}c_{r}^{\ast}\right)\Psi\right\rangle \nonumber \\
 & +\sum_{p,p'\in B_{F}^{c}}\sum_{q,q',r,r'\in B_{F}}\overline{A_{p,q,r}}A_{p',q',r'}\left\langle \Psi,\left(\delta_{q,r'}\delta_{r,q'}c_{p'}^{\ast}c_{p}+\delta_{p,p'}\delta_{r,q'}c_{r'}c_{q}^{\ast}+\delta_{p,p'}\delta_{q,r'}c_{q'}c_{r}^{\ast}\right)\Psi\right\rangle \nonumber \\
 & +\sum_{p,p'\in B_{F}^{c}}\sum_{q,q',r,r'\in B_{F}}\overline{A_{p,q,r}}A_{p',q',r'}\left\langle \Psi,\left(\delta_{p,p'}\delta_{q,q'}\delta_{r,r'}-\delta_{p,p'}\delta_{q,r'}\delta_{r,q'}\right)\Psi\right\rangle .
\end{align}
We estimate the different types of expressions appearing above. Firstly,
by the standard fermionic estimate \eqref{eq:standard-fermionic-estimate}, 
\begin{align}
 & \sum_{p,p'\in B_{F}^{c}}\sum_{q,q',r,r'\in B_{F}}\overline{A_{p,q,r}}A_{p',q',r'}\left\langle \Psi,\left(\delta_{p,p'}c_{q'}c_{r'}c_{r}^{\ast}c_{q}^{\ast}\right)\Psi\right\rangle 
  =\sum_{p\in B_{F}^{c}}\left\Vert \sum_{q,r\in B_{F}}\overline{A_{p,q,r}}c_{r}^{\ast}c_{q}^{\ast}\Psi\right\Vert ^{2} \nonumber \\
 & \leq\sum_{p\in B_{F}^{c}}\left(\sum_{q\in B_{F}}\left\Vert \left(\sum_{r\in B_{F}}\overline{A_{p,q,r}}c_{r}^{\ast}\right)c_{q}^{\ast}\Psi\right\Vert \right)^{2}\leq\sum_{p\in B_{F}^{c}}\left(\sum_{q\in B_{F}}\sqrt{\sum_{r\in B_{F}}\left|A_{p,q,r}\right|^{2}}\left\Vert c_{q}^{\ast}\Psi\right\Vert \right)^{2}\nonumber \\
 & \leq\sum_{p\in B_{F}^{c}}\sum_{q,r\in B_{F}}\left|A_{p,q,r}\right|^{2}\left(\sum_{q\in B_{F}}\left\Vert c_{q}^{\ast}\Psi\right\Vert ^{2}\right)=\sum_{p\in B_{F}^{c}}\sum_{q,r\in B_{F}}\left|A_{p,q,r}\right|^{2}\left\langle \Psi,\mathcal{N}_{E}\Psi\right\rangle 
 \end{align}
and likewise for the other two terms on the first line of equation
(\ref{eq:HigherFermionicEstimateTerms}). For the terms on the second
line we similarly estimate
\begin{align}
 & \quad\,\left|\sum_{p,p'\in B_{F}^{c}}\sum_{q,q',r,r'\in B_{F}}\overline{A_{p,q,r}}A_{p',q',r'}\left\langle \Psi,\left(\delta_{r,q'}c_{p'}^{\ast}c_{r'}c_{q}^{\ast}c_{p}\right)\Psi\right\rangle \right|\nonumber \\
 & \leq\sum_{r\in B_{F}}\left\Vert \sum_{p'\in B_{F}^{c}}\sum_{r'\in B_{F}}A_{p',r,r'}c_{p'}c_{r'}^{\ast}\Psi\right\Vert \left\Vert \sum_{p\in B_{F}^{c}}\sum_{q\in B_{F}}\overline{A_{p,q,r}}c_{q}^{\ast}c_{p}\Psi\right\Vert \nonumber\\
 & \leq\sum_{p\in B_{F}^{c}}\sum_{r,r'\in B_{F}}\sqrt{\sum_{p'\in B_{F}^{c}}\left|A_{p',r,r'}\right|^{2}}\left\Vert c_{r'}^{\ast}\Psi\right\Vert \sqrt{\sum_{q\in B_{F}}\left|A_{p,q,r}\right|^{2}}\left\Vert c_{p}\Psi\right\Vert \nonumber \\
 & \leq\sum_{r\in B_{F}}\sqrt{\sum_{p\in B_{F}^{c}}\sum_{r'\in B_{F}}\left|A_{p,r,r'}\right|^{2}}\sqrt{\sum_{p\in B_{F}^{c}}\sum_{q\in B_{F}}\left|A_{p,q,r}\right|^{2}}\sqrt{\sum_{r'\in B_{F}}\left\Vert c_{r'}^{\ast}\Psi\right\Vert ^{2}}\sqrt{\sum_{p\in B_{F}^{c}}\left\Vert c_{p}\Psi\right\Vert ^{2}}\nonumber \\
 & \leq\sum_{p\in B_{F}^{c}}\sum_{q\in B_{F}}\left|A_{p,q,r}\right|^{2}\left\langle \Psi,\mathcal{N}_{E}\Psi\right\rangle . 
\end{align}
The terms on the third line of equation (\ref{eq:HigherFermionicEstimateTerms})
all factorize in a manifestly non-positive fashion, and so can be
dropped, while for the fourth line
\begin{align}
 & \quad\,\left|\sum_{p,p'\in B_{F}^{c}}\sum_{q,q',r,r'\in B_{F}}\overline{A_{p,q,r}}A_{p',q',r'}\left\langle \Psi,\left(\delta_{q,r'}\delta_{r,q'}c_{p'}^{\ast}c_{p}\right)\Psi\right\rangle \right|\nonumber\\
 & =\left|\sum_{q,r\in B_{F}}\left\langle \sum_{p'\in B_{F}^{c}}A_{p',r,q}c_{p'}\Psi,\sum_{p\in B_{F}^{c}}\overline{A_{p,q,r}}c_{p}\Psi\right\rangle \right|\leq\sum_{q,r\in B_{F}}\left\Vert \sum_{p'\in B_{F}^{c}}A_{p',r,q}c_{p'}\Psi\right\Vert \left\Vert \sum_{p\in B_{F}^{c}}\overline{A_{p,q,r}}c_{p}\Psi\right\Vert \nonumber \\
 & \leq\sum_{q,r\in B_{F}}\sqrt{\sum_{p'\in B_{F}^{c}}\left|A_{p',r,q}\right|^{2}}\sqrt{\sum_{p\in B_{F}^{c}}\left|A_{p,q,r}\right|^{2}}\left\Vert \Psi\right\Vert ^{2}\leq\sum_{p\in B_{F}^{c}}\sum_{q,r\in B_{F}}\left|A_{p,q,r}\right|^{2}\left\Vert \Psi\right\Vert ^{2}. 
\end{align}
Lastly, the terms on the fifth line are seen to simply be constant
and easily bounded by $\sum_{p\in B_{F}^{c}}\sum_{q,r\in B_{F}}\left|A_{p,q,r}\right|^{2}$,
whence the proposition follows.
$\hfill\square$

We can now conclude the following bound for $\tilde{\mathcal{K}}$, which in turn implies Proposition \ref{prop:Kernel-cK-boundedness}. 
\begin{prop}
\label{prop:calKTildeBound}For any $\Psi,\Phi\in\mathcal{H}_{N}$
it holds that
\[
|\langle \Psi,\tilde{\mathcal{K}}\Phi\rangle|\leq\frac{\sqrt{5}}{2}\sqrt{\sum_{l\in\mathbb{Z}_{\ast}^{3}}\left\Vert K_{l}\right\Vert _{\mathrm{HS}}^{2}}\sqrt{\left\langle \Psi,\left(\mathcal{N}_{E}+1\right)\Psi\right\rangle \left\langle \Phi,\mathcal{N}_{E}\Phi\right\rangle }.
\]
\end{prop}

\textbf{Proof:} By (\ref{eq:calKtildeEstimate1}) and
(\ref{eq:calKtildeEstimate2}), combined with the estimate of Proposition \ref{prop:Apqrccc}, we can bound
\begin{align}
|\langle \Psi,\tilde{\mathcal{K}}\Phi\rangle| & \leq\frac{\sqrt{5}}{2}\sqrt{\sum_{q\in B_{F}^{c}}\sum_{p'\in B_{F}^{c}}\sum_{q',r'\in B_{F}}\left|\sum_{l\in\mathbb{Z}_{\ast}^{3}}\sum_{p\in L_{l}}\delta_{p',p}\delta_{q',p-l}\delta_{r',-q+l}1_{L_l}(q)\left\langle K_{l}e_{q},e_{p}\right\rangle \right|^{2}}\\
 & \qquad\qquad\qquad\qquad\qquad\qquad\qquad\qquad\qquad\qquad\cdot\sqrt{\left\langle \Psi,\left(\mathcal{N}_{E}+1\right)\Psi\right\rangle \left\langle \Phi,\mathcal{N}_{E}\Phi\right\rangle }.\nonumber 
\end{align}
The sum inside the first square root is exactly equal to $\sum_{l\in\mathbb{Z}_{\ast}^{3}}\left\Vert K_{l}\right\Vert _{\mathrm{HS}}^{2}$. $\hfill\square$

\section{\label{sec:AnalysisoftheOneBodyOperators}Analysis of the One-Body
Operators}

In this section we analyze the operators $K_{k}$, $A_{k}(t)$
and $B_{k}(t)$ which appear in Theorem \ref{thm:DiagonalizationoftheBosonizableTerms},
obtaining the following:
\begin{thm}
\label{them:OneBodyEstimates}For any $k\in\mathbb{Z}_{\ast}^{3}$
it holds that
\[
\left\Vert K_{k}\right\Vert _{\mathrm{HS}}\leq C\hat{V}_{k}\min\,\{1,k_{F}^{2}\left|k\right|^{-2}\}.
\]
Moreover, for all $p,q\in L_{k}$ and $t\in\left[0,1\right]$, 
\begin{align*}
\left|\left\langle e_{p},K_{k}e_{q}\right\rangle \right| & \leq C\frac{\hat{V}_{k}k_{F}^{-1}}{\lambda_{k,p}+\lambda_{k,q}},\\
\left|\left\langle e_{p},\left(-K_{k}\right)e_{q}\right\rangle -\frac{\hat{V}_{k}k_{F}^{-1}}{2\left(2\pi\right)^{3}}\frac{1}{\lambda_{k,p}+\lambda_{k,q}}\right| & \leq C\frac{\hat{V}_{k}^{2}k_{F}^{-1}}{\lambda_{k,p}+\lambda_{k,q}},\\
\left|\left\langle e_{p},A_{k}(t)e_{q}\right\rangle \right|,\left|\left\langle e_{p},B_{k}(t)e_{q}\right\rangle \right| & \leq C\left(1+\hat{V}_{k}^{2}\right)\hat{V}_{k}k_{F}^{-1},\\
\left|\left\langle e_{p},\left\{ K_{k},B_{k}(t)\right\} e_{q}\right\rangle \right| & \leq C\left(1+\hat{V}_{k}^{2}\right)\hat{V}_{k}^{2}k_{F}^{-1},\\
\left|\left\langle e_{p},\left(\int_{0}^{1}B_{k}(t)dt\right)e_{q}\right\rangle -\frac{\hat{V}_{k}k_{F}^{-1}}{4\left(2\pi\right)^{3}}\right| & \leq C\left(1+\hat{V}_{k}\right)\hat{V}_{k}^{2}k_{F}^{-1},
\end{align*}
for a constant $C>0$ independent of all relevant quantities.
\end{thm}

The analysis of this section is similar to that of \cite[Section 7]{CHN-21},
but compared to that section, the estimates of this section are considerably
more precise: We quantify the error of the upper bound on $\left\langle e_{p},\left(-K_{k}\right)e_{q}\right\rangle$,
obtain elementwise estimates for $A_{k}(t)$ and $B_{k}(t)$
(rather than only estimates for the norm $\left\Vert \cdot\right\Vert _{\infty,2}$ as in \cite{CHN-21}), and determine the leading term of the operator
$\int_{0}^{1}B_{k}(t)dt$ 
which will be needed to extract the exchange contribution in the next
section.

\subsection{Matrix Element Estimates for $K$-Quantities}

To ease the notation we will abstract the problem slightly: Instead
of $\ell^{2}(L_{k})$ we consider a general $n$-dimensional
Hilbert space $\left(V,\left\langle \cdot,\cdot\right\rangle \right)$,
let $h:V\rightarrow V$ be a positive self-adjoint operator on $V$
with eigenbasis $\left(x_{i}\right)_{i=1}^{n}$ and eigenvalues $\left(\lambda_{i}\right)_{i=1}^{n}$,
and let $v\in V$ be any vector such that $\left\langle x_{i},v\right\rangle \geq0$
for all $1\leq i\leq n$, and let $P_{w}\left(\cdot\right)=\left\langle w,\cdot\right\rangle w$ be the projection onto $w\in V$. 
Theorem \ref{them:OneBodyEstimates} will
then be obtained at the end by insertion of the particular operators
$h_{k}$ and $P_{k}$.

We define $K:V\rightarrow V$ by
\begin{equation}
K=-\frac{1}{2}\log\left(h^{-\frac{1}{2}}\left(h^{\frac{1}{2}}\left(h+2P_{v}\right)h^{\frac{1}{2}}\right)^{\frac{1}{2}}h^{-\frac{1}{2}}\right)=-\frac{1}{2}\log\left(h^{-\frac{1}{2}}\left(h^{2}+2P_{h^{\frac{1}{2}}v}\right)^{\frac{1}{2}}h^{-\frac{1}{2}}\right).
\end{equation}
As $\big(h^{2}+2P_{h^{\frac{1}{2}}v}\big)^{\frac{1}{2}} \ge h$ we see that $K\le 0$. In \cite[Section 7.2]{CHN-21} we proved the following result. 
\begin{prop}
\label{prop:e-2Ke2KElementEstimates}For all $1\leq i,j\leq n$ it
holds that
\[
\frac{2}{1+2\left\langle v,h^{-1}v\right\rangle }\frac{\left\langle x_{i},v\right\rangle \left\langle v,x_{j}\right\rangle }{\lambda_{i}+\lambda_{j}}\leq\left\langle x_{i},\left(e^{-2K}-1\right)x_{j}\right\rangle ,\left\langle x_{i},\left(1-e^{2K}\right)x_{j}\right\rangle \leq2\frac{\left\langle x_{i},v\right\rangle \left\langle v,x_{j}\right\rangle }{\lambda_{i}+\lambda_{j}}.
\]
\end{prop}

Below it will be more convenient to consider the hyperbolic functions
$\sinh\left(-2K\right)$ and $\cosh\left(-2K\right)$ rather than
$e^{-2K}$ and $e^{2K}$. The previous proposition implies the following
for these operators:
\begin{cor}
\label{coro:HyperbolicBounds}For any $1\leq i,j\leq n$ it holds
that
\begin{align*}
\left\langle x_{i},\sinh\left(-2K\right)x_{j}\right\rangle  & \leq2\frac{\left\langle x_{i},v\right\rangle \left\langle v,x_{j}\right\rangle }{\lambda_{i}+\lambda_{j}},\\
\left\langle x_{i},\left(\cosh\left(-2K\right)-1\right)x_{j}\right\rangle  & \leq\frac{2\left\langle v,h^{-1}v\right\rangle }{1+2\left\langle v,h^{-1}v\right\rangle }\frac{\left\langle x_{i},v\right\rangle \left\langle v,x_{j}\right\rangle }{\lambda_{i}+\lambda_{j}}.
\end{align*}
\end{cor}

\textbf{Proof:} These bounds follow from Proposition \ref{prop:e-2Ke2KElementEstimates} and the identities 
\begin{align}
\sinh\left(-2K\right)&=\frac{1}{2}\left(\left(e^{-2K}-1\right)+\left(1-e^{2K}\right)\right),\\
\cosh\left(-2K\right)-1&=\frac{1}{2}\left(\left(e^{-2K}-1\right)-\left(1-e^{2K}\right)\right).\nonumber \tag*{$\square$}
\end{align}

Now we extend our elementwise estimates to more general operators.
These estimates are similar to those of Proposition 7.10 of \cite{CHN-21},
but more precise. First we consider $K$ itself:
\begin{prop}
\label{prop:KElementEstimates}For any $1\leq i,j\leq n$ it holds
that
\[
\frac{1}{1+2\left\langle v,h^{-1}v\right\rangle }\frac{\left\langle x_{i},v\right\rangle \left\langle v,x_{j}\right\rangle }{\lambda_{i}+\lambda_{j}}\leq\left\langle x_{i},\left(-K\right)x_{j}\right\rangle \leq\frac{\left\langle x_{i},v\right\rangle \left\langle v,x_{j}\right\rangle }{\lambda_{i}+\lambda_{j}}.
\]
\end{prop}

\textbf{Proof:} 
From the identity 
\begin{equation}
-x=\frac{1}{2}\sum_{m=1}^{\infty}\frac{1}{m}\left(1-e^{2x}\right)^{m},\quad x\leq0,
\end{equation}
which follows by the Mercator series, we thus have that $-K=\frac{1}{2}\sum_{m=1}^{\infty}\frac{1}{m}\left(1-e^{2K}\right)^{m}$.
Noting that Proposition \ref{prop:e-2Ke2KElementEstimates} in particular
implies that $\left\langle x_{i},\left(1-e^{2K}\right)x_{j}\right\rangle \geq0$
for all $1\leq i,j\leq n$, whence also $\left\langle x_{i},\left(1-e^{2K}\right)^{m}x_{j}\right\rangle \geq0$
for any $m\in\mathbb{N}$, we may estimate
\begin{align}
\left\langle x_{i},\left(-K\right)x_{j}\right\rangle  & =\frac{1}{2}\sum_{m=1}^{\infty}\frac{1}{m}\left\langle x_{i},\left(1-e^{2K}\right)^{m}x_{j}\right\rangle \geq\frac{1}{2}\left\langle x_{i},\left(1-e^{2K}\right)x_{j}\right\rangle \\
 & \geq\frac{1}{1+2\left\langle v,h^{-1}v\right\rangle }\frac{\left\langle x_{i},v\right\rangle \left\langle v,x_{j}\right\rangle }{\lambda_{i}+\lambda_{j}}\nonumber 
\end{align}
which is the lower bound. This similarly implies that $\left\langle x_{i},\left(-K\right)^{m}x_{j}\right\rangle \geq0$
for all $1\leq i,j\leq n$, $m\in\mathbb{N}$, so the upper bound
now also follows from Proposition \ref{prop:e-2Ke2KElementEstimates}
by noting that
\begin{equation}
\frac{\left\langle x_{i},v\right\rangle \left\langle v,x_{j}\right\rangle }{\lambda_{i}+\lambda_{j}}\geq\frac{1}{2}\left\langle x_{i},\left(e^{-2K}-1\right)x_{j}\right\rangle =\frac{1}{2}\sum_{m=1}^{\infty}\frac{1}{m!}\left\langle x_{i},\left(-2K\right)^{m}x_{j}\right\rangle \geq\left\langle x_{i},\left(-K\right)x_{j}\right\rangle .
\end{equation}
The proof of Proposition \ref{prop:KElementEstimates} is complete. 
$\hfill\square$

\medskip

The fact that $\left\langle x_{i},\left(-K\right)^{m}x_{j}\right\rangle \geq0$
for all $1\leq i,j\leq n$, $m\in\mathbb{N}$, has the important consequence
that for any such $i$ and $j$, the functions
\begin{equation}
t\mapsto\left\langle x_{i},\sinh\left(-tK\right)x_{j}\right\rangle ,\,\left\langle x_{i},\left(\sinh\left(-tK\right)+tK\right)x_{j}\right\rangle ,\,\left\langle x_{i},\left(\cosh\left(-tK\right)-1\right)x_{j}\right\rangle 
\end{equation}
are non-negative and convex for $t\in\left[0,\infty\right)$, as follows
by considering the Taylor expansions of the operators involved. This
allows us to extend the bounds of Corollary \ref{coro:HyperbolicBounds}
to arbitrary $t\in\left[0,1\right]$:
\begin{prop}
\label{prop:tDependentElementEstimates}For all $1\leq i,j\leq n$
and $t\in\left[0,1\right]$ it holds that
\begin{align*}
\frac{1}{1+2\left\langle v,h^{-1}v\right\rangle }\frac{\left\langle x_{i},v\right\rangle \left\langle v,x_{j}\right\rangle }{\lambda_{i}+\lambda_{j}}t\leq\left\langle x_{i},\sinh\left(-tK\right)x_{j}\right\rangle  & \leq\frac{\left\langle x_{i},v\right\rangle \left\langle v,x_{j}\right\rangle }{\lambda_{i}+\lambda_{j}}t,\\
0\leq\left\langle x_{i},\left(\cosh\left(-tK\right)-1\right)x_{j}\right\rangle  & \leq\frac{\left\langle v,h^{-1}v\right\rangle }{1+2\left\langle v,h^{-1}v\right\rangle }\frac{\left\langle x_{i},v\right\rangle \left\langle v,x_{j}\right\rangle }{\lambda_{i}+\lambda_{j}},\\
\left|\left\langle x_{i},\left(e^{tK}-1\right)x_{j}\right\rangle \right| & \leq\frac{\left\langle x_{i},v\right\rangle \left\langle v,x_{j}\right\rangle }{\lambda_{i}+\lambda_{j}}.
\end{align*}
\end{prop}

\textbf{Proof:} By the noted convexity we immediately conclude the
upper bounds
\begin{align}
\left\langle x_{i},\sinh\left(-tK\right)x_{j}\right\rangle  & \leq\frac{t}{2}\left\langle x_{i},\sinh\left(-2K\right)x_{j}\right\rangle \leq\frac{\left\langle x_{i},v\right\rangle \left\langle v,x_{j}\right\rangle }{\lambda_{i}+\lambda_{j}}t\\
\left\langle x_{i},\left(\cosh\left(-tK\right)-1\right)x_{j}\right\rangle  & \leq\frac{t}{2}\left\langle x_{i},\left(\cosh\left(-2K\right)-1\right)x_{j}\right\rangle \leq\frac{\left\langle v,h^{-1}v\right\rangle }{1+2\left\langle v,h^{-1}v\right\rangle }\frac{\left\langle x_{i},v\right\rangle \left\langle v,x_{j}\right\rangle }{\lambda_{i}+\lambda_{j}}\nonumber 
\end{align}
and by non-negativity of $\left\langle x_{i},\left(\sinh\left(-tK\right)+tK\right)x_{j}\right\rangle $
and Proposition \ref{prop:KElementEstimates}, the lower bound
\begin{equation}
\left\langle x_{i},\sinh\left(-tK\right)x_{j}\right\rangle \geq\left\langle x_{i},\left(-tK\right)x_{j}\right\rangle \geq\frac{1}{1+2\left\langle v,h^{-1}v\right\rangle }\frac{\left\langle x_{i},v\right\rangle \left\langle v,x_{j}\right\rangle }{\lambda_{i}+\lambda_{j}}t.
\end{equation}
Lastly we can apply the non-negativity of the hyperbolic operators
to conclude the bound for $e^{tK}-1$ as
\begin{align}
&\left|\left\langle x_{i},\left(e^{tK}-1\right)x_{j}\right\rangle \right|  =\left|\left\langle x_{i},\left(\left(\cosh\left(-tK\right)-1\right)-\sinh\left(-tK\right)\right)x_{j}\right\rangle \right|\\
 & \leq\max\left\{ \left\langle x_{i},\left(\cosh\left(-tK\right)-1\right)x_{j}\right\rangle ,\left\langle x_{i},\sinh\left(-tK\right)x_{j}\right\rangle \right\} \leq\frac{\left\langle x_{i},v\right\rangle \left\langle v,x_{j}\right\rangle }{\lambda_{i}+\lambda_{j}}. \nonumber \tag*{$\square$}
\end{align}

\subsection{Matrix Element Estimates for $A(t)$ and $B(t)$}

We now consider operators $A(t),B(t):V\rightarrow V$
defined by
\begin{align}
A(t) &=\frac{1}{2}\left(e^{tK}\left(h+2P_{v}\right)e^{tK}+e^{-tK}he^{-tK}\right)-h,\\
B(t) & =\frac{1}{2}\left(e^{tK}\left(h+2P_{v}\right)e^{tK}-e^{-tK}he^{-tK}\right),\nonumber 
\end{align}
for $t\in\left[0,1\right]$. We decompose these as
\begin{align}
A(t) & =A_{h}(t)+e^{tK}P_{v}e^{tK},\quad B(t) =\left(1-t\right)P_{v}+B_{h}(t)+e^{tK}P_{v}e^{tK}-P_{v}
\end{align}
with
\begin{align}
C_{K}(t)&=\cosh\left(-tK\right)-1, \quad S_{K}(t)=\sinh\left(-tK\right), \nonumber\\
A_{h}(t) & =\cosh\left(-tK\right)h\cosh\left(-tK\right)+\sinh\left(-tK\right)h\sinh\left(-tK\right)-h\nonumber \\
 & =\left\{ h,C_{K}(t)\right\} +S_{K}(t)h\,S_{K}(t)+C_{K}(t)h\,C_{K}(t),\nonumber \\
B_{h}(t) & =-\sinh\left(-tK\right)h\cosh\left(-tK\right)-\cosh\left(-tK\right)h\sinh\left(-tK\right)+tP_{v}\\
 & =tP_{v}-\left\{ h,S_{K}(t)\right\} -S_{K}(t)h\,C_{K}(t)-C_{K}(t)h\,S_{K}(t).\nonumber 
\end{align}
We begin by estimating the $e^{tK}P_{v}e^{tK}$ terms:
\begin{prop}
\label{prop:etKPetK-PEstimate}For all $1\leq i,j\leq n$ and $t\in\left[0,1\right]$
it holds that
\[
\left|\left\langle x_{i},\left(e^{tK}P_{v}e^{tK}-P_{v}\right)x_{j}\right\rangle \right|\leq\left(2+\left\langle v,h^{-1}v\right\rangle \right)\left\langle v,h^{-1}v\right\rangle \left\langle x_{i},v\right\rangle \left\langle v,x_{j}\right\rangle .
\]
\end{prop}

\textbf{Proof:} Writing
\begin{equation}
e^{tK}P_{v}e^{tK}-P_{v}=\left\{ P_{v},e^{tK}-1\right\} +\left(e^{tK}-1\right)P_{v}\left(e^{tK}-1\right)
\end{equation}
we see that
\begin{align}
\left\langle x_{i},\left(e^{tK}P_{v}e^{tK}-P_{v}\right)x_{j}\right\rangle  & =\left\langle x_{i},v\right\rangle \left\langle \left(e^{tK}-1\right)v,x_{j}\right\rangle +\left\langle x_{i},\left(e^{tK}-1\right)v\right\rangle \left\langle v,x_{j}\right\rangle \\
 & +\left\langle x_{i},\left(e^{tK}-1\right)v\right\rangle \left\langle \left(e^{tK}-1\right)v,x_{j}\right\rangle .\nonumber 
\end{align}
Now, by Proposition \ref{prop:tDependentElementEstimates} we can
for any $1\leq i\leq n$ estimate
\begin{align}
\left|\left\langle x_{i},\left(e^{tK}-1\right)v\right\rangle \right| & =\left|\sum_{j=1}^{n}\left\langle x_{i},\left(e^{tK}-1\right)x_{j}\right\rangle \left\langle x_{j},v\right\rangle \right|\leq\sum_{j=1}^{n}\frac{\left\langle x_{i},v\right\rangle \left\langle v,x_{j}\right\rangle }{\lambda_{i}+\lambda_{j}}\left\langle x_{j},v\right\rangle \\
 & \leq\left\langle x_{i},v\right\rangle \sum_{j=1}^{n}\frac{\left|\left\langle x_{j},v\right\rangle \right|^{2}}{\lambda_{j}}=\left\langle x_{i},v\right\rangle \left\langle v,h^{-1}v\right\rangle \nonumber 
\end{align}
whence the claim follows.
$\hfill\square$

\medskip

Note that for $\left\langle x_{i},e^{tK}P_{v}e^{tK}x_{j}\right\rangle $
this in particular implies the bound
\begin{equation}
\left|\left\langle x_{i},e^{tK}P_{v}e^{tK}x_{j}\right\rangle \right|\leq\left(1+\left\langle v,h^{-1}v\right\rangle \right)^{2}\left\langle x_{i},v\right\rangle \left\langle v,x_{j}\right\rangle .\label{eq:etKPetKEstimate}
\end{equation}
We now consider $A_{h}(t)$ and $B_{h}(t)$:
\begin{prop}
\label{prop:AhBhEstimate}For all $1\leq i,j\leq n$ and $t\in\left[0,1\right]$
it holds that
\[
\left|\left\langle x_{i},A_{h}(t)x_{j}\right\rangle \right|,\left|\left\langle x_{i},B_{h}(t)x_{j}\right\rangle \right|\leq4\left\langle v,h^{-1}v\right\rangle \left\langle x_{i},v\right\rangle \left\langle v,x_{j}\right\rangle .
\]
\end{prop}

\textbf{Proof:} The estimates of Proposition \ref{prop:tDependentElementEstimates}
imply that
\begin{align}
&\left|\left\langle x_{i},\left\{ h,C_{K}(t)\right\} x_{j}\right\rangle \right|  =\left(\lambda_{i}+\lambda_{j}\right)\left|\left\langle x_{i},C_{K}(t)x_{j}\right\rangle \right| \\
&\leq\left(\lambda_{i}+\lambda_{j}\right)\frac{\left\langle v,h^{-1}v\right\rangle }{1+2\left\langle v,h^{-1}v\right\rangle }\frac{\left\langle x_{i},v\right\rangle \left\langle v,x_{j}\right\rangle }{\lambda_{i}+\lambda_{j}} \leq\left\langle v,h^{-1}v\right\rangle \left\langle x_{i},v\right\rangle \left\langle v,x_{j}\right\rangle,\nonumber 
 \end{align}
 and
 \begin{align}
&\left|\left\langle x_{i},S_{K}(t)h\,S_{K}(t)x_{j}\right\rangle \right|  = \left|\sum_{k=1}^{n}\lambda_{k}\left\langle x_{i},S_{K}(t)x_{k}\right\rangle \left\langle x_{k},S_{K}(t)x_{j}\right\rangle \right| \\
& \leq \sum_{k=1}^{n}\lambda_{k}\frac{\left\langle x_{i},v\right\rangle \left\langle v,x_{k}\right\rangle }{\lambda_{i}+\lambda_{k}}\frac{\left\langle x_{k},v\right\rangle \left\langle v,x_{j}\right\rangle }{\lambda_{k}+\lambda_{j}} \leq\left\langle x_{i},v\right\rangle \left\langle v,x_{j}\right\rangle \sum_{k=1}^{n}\frac{\left|\left\langle x_{k},v\right\rangle \right|^{2}}{\lambda_{k}}=\left\langle v,h^{-1}v\right\rangle \left\langle x_{i},v\right\rangle \left\langle v,x_{j}\right\rangle \nonumber.
\end{align}
The latter estimate only relied on the inequality
\begin{equation}
\left|\left\langle x_{i},S_{K}(t)x_{j}\right\rangle \right|\leq\frac{\left\langle x_{i},v\right\rangle \left\langle v,x_{j}\right\rangle }{\lambda_{i}+\lambda_{j}},
\end{equation}
which is also true for $C_{K}(t)$, so the terms $C_{K}(t)h\,C_{K}(t)$, $C_{K}(t)h\,S_{K}(t)$ and $S_{K}(t)h\,C_{K}(t)$ also obey this estimate. It thus only remains to bound $tP_{v}-\left\{ h,S_{K}(t)\right\} $.
From Proposition \ref{prop:tDependentElementEstimates} we see that
\begin{equation}
\frac{\left\langle x_{i},v\right\rangle \left\langle v,x_{j}\right\rangle }{1+2\left\langle v,h^{-1}v\right\rangle }t\leq\left\langle x_{i},\left\{ h,S_{K}(t)\right\} x_{j}\right\rangle \leq\left\langle x_{i},v\right\rangle \left\langle v,x_{j}\right\rangle t
\end{equation}
whence
\begin{align}
 & \quad\;\,\left|\left\langle x_{i},\left(tP_{v}-\left\{ h,S_{K}(t)\right\} \right)x_{j}\right\rangle \right|=\left\langle x_{i},P_{v}x_{j}\right\rangle t-\left\langle x_{i},\left\{ h,S_{K}(t)\right\} x_{j}\right\rangle \\
 & \leq\left(1-\frac{1}{1+2\left\langle v,h^{-1}v\right\rangle }\right)\left\langle x_{i},v\right\rangle \left\langle v,x_{j}\right\rangle t
  \leq2\left\langle v,h^{-1}v\right\rangle \left\langle x_{i},v\right\rangle \left\langle v,x_{j}\right\rangle . \nonumber \tag*{$\square$}
\end{align}

Combining equation (\ref{eq:etKPetKEstimate}) and Proposition \ref{prop:AhBhEstimate}
we conclude the following:
\begin{prop}
\label{prop:AtBtEstimates}For all $1\leq i,j\leq n$ and $t\in\left[0,1\right]$
it holds that
\[
\left|\left\langle x_{i},A(t)x_{j}\right\rangle \right|,\,\left|\left\langle x_{i},B(t)x_{j}\right\rangle \right|\leq3\left(1+\left\langle v,h^{-1}v\right\rangle \right)^{2}\left\langle x_{i},v\right\rangle \left\langle v,x_{j}\right\rangle .
\]
\end{prop}

\subsubsection*{Analysis of $\left\{ K,B(t)\right\} $ and $\int_{0}^{1}B(t)dt$}

We end by estimating $\left\{ K,B(t)\right\} $ and $\int_{0}^{1}B(t)dt$,
the latter of which will be needed for the analysis of the exchange
contribution in the next section.

\begin{prop}
\label{prop:KBtEstimate}For all $1\leq i,j\leq n$ and $t\in\left[0,1\right]$
it holds that
\[
\left|\left\langle x_{i},\left\{ K,B(t)\right\} x_{j}\right\rangle \right|\leq6\left(1+\left\langle v,h^{-1}v\right\rangle \right)^{2}\left\langle v,h^{-1}v\right\rangle \left\langle x_{i},v\right\rangle \left\langle v,x_{j}\right\rangle .
\]
\end{prop}

\textbf{Proof:} Using the Propositions \ref{prop:KElementEstimates}
and \ref{prop:AtBtEstimates} we see that
\begin{align}
&\left|\left\langle x_{i},KB(t)x_{j}\right\rangle \right| =\left|\sum_{k=1}^{n}\left\langle x_{i},Kx_{k}\right\rangle \left\langle x_{k},B(t)x_{j}\right\rangle \right| \nonumber\\
& \leq3\left(1+\left\langle v,h^{-1}v\right\rangle \right)^{2}\sum_{k=1}^{n}\frac{\left\langle x_{i},v\right\rangle \left\langle v,x_{k}\right\rangle }{\lambda_{i}+\lambda_{k}}\left\langle x_{k},v\right\rangle \left\langle v,x_{j}\right\rangle \nonumber \\
 & \leq3\left(1+\left\langle v,h^{-1}v\right\rangle \right)^{2}\sum_{k=1}^{n}\frac{\left|\left\langle x_{k},v\right\rangle \right|^{2}}{\lambda_{k}}\left\langle x_{i},v\right\rangle \left\langle v,x_{j}\right\rangle \\
 & =3\left(1+\left\langle v,h^{-1}v\right\rangle \right)^{2}\left\langle v,h^{-1}v\right\rangle \left\langle x_{i},v\right\rangle \left\langle v,x_{j}\right\rangle .\nonumber 
\end{align}
This estimate is also valid for $\left|\left\langle x_{i},B(t)Kx_{j}\right\rangle \right|$
whence the claim follows.
$\hfill\square$

\begin{prop}
For all $1\leq i,j\leq n$ it holds that
\[
\left|\left\langle x_{i},\left(\int_{0}^{1}B(t)dt\right)x_{j}\right\rangle -\frac{1}{2}\left\langle x_{i},v\right\rangle \left\langle v,x_{j}\right\rangle \right|\leq\left(6+\left\langle v,h^{-1}v\right\rangle \right)\left\langle v,h^{-1}v\right\rangle \left\langle x_{i},v\right\rangle \left\langle v,x_{j}\right\rangle .
\]
\end{prop}

\textbf{Proof:} Noting that $\frac{1}{2}\left\langle x_{i},v\right\rangle \left\langle v,x_{j}\right\rangle =\frac{1}{2}\left\langle x_{i},P_{v}x_{j}\right\rangle $
and that
\begin{align}
\int_{0}^{1}B(t)dt-\frac{1}{2}P_{v} & =\int_{0}^{1}\left(\left(1-t\right)P_{v}+B_{h}(t)+e^{tK}P_{v}e^{tK}-P_{v}\right)dt-\frac{1}{2}P_{v}\\
 & =\int_{0}^{1}\left(B_{h}(t)+e^{tK}P_{v}e^{tK}-P_{v}\right)dt\nonumber 
\end{align}
we can estimate using the Propositions \ref{prop:etKPetK-PEstimate}
and \ref{prop:AhBhEstimate} that
\begin{align}
\left|\left\langle x_{i},\left(\int_{0}^{1}B(t)dt-\frac{1}{2}P_{v}\right)x_{j}\right\rangle \right| & \leq\int_{0}^{1} \left( \left|\left\langle x_{i},B_{h}(t)x_{j}\right\rangle \right| + \left|\left\langle x_{i},\left(e^{tK}P_{v}e^{tK}-P_{v}\right)x_{j}\right\rangle \right| \right) dt\\
 & \leq\left(6+\left\langle v,h^{-1}v\right\rangle \right)\left\langle v,h^{-1}v\right\rangle \left\langle x_{i},v\right\rangle \left\langle v,x_{j}\right\rangle . \nonumber \tag*{$\square$}
\end{align}

\subsubsection*{Insertion of the Particular Operators $h_{k}$ and $P_{k}$}

Recall that the particular operators we must consider are $h_{k},P_{k}:\ell^{2}(L_{k})\rightarrow\ell^{2}(L_{k})$
defined by
\begin{equation}
\begin{array}{ccccccc}
h_{k}e_{p} & = & \lambda_{k,p}e_{p}, &  & \lambda_{k,p} & = & \frac{1}{2}\left(\left|p\right|^{2}-\left|p-k\right|^{2}\right),\\
P_{k}(\cdot) & = & \left\langle v_{k},\cdot\right\rangle v_{k}, &  & v_{k} & = & \sqrt{\frac{\hat{V}_{k}k_{F}^{-1}}{2\left(2\pi\right)^{3}}}\sum_{p\in L_{k}}e_{p}.
\end{array}
\end{equation}
For these we have that
\begin{equation}
\left\langle v_{k},h_{k}^{-1}v_{k}\right\rangle =\frac{\hat{V}_{k}k_{F}^{-1}}{2\left(2\pi\right)^{3}}\sum_{p\in L_{k}}\frac{1}{\lambda_{k,p}}.
\end{equation}
In \cite{CHN-21} the following estimates for sums of the form $\sum_{p\in L_{k}}\lambda_{k,p}^{\beta}$
were proved:
\begin{prop}
\label{prop:RiemannSumEstimates}For any $k\in\mathbb{Z}_{\ast}^{3}$
and $\beta\in\left[-1,0\right]$ it holds that
\[
\sum_{p\in L_{k}}\lambda_{k,p}^{\beta}\leq C_{\beta}\begin{cases}
k_{F}^{2+\beta}\left|k\right|^{1+\beta} & \left|k\right|\leq2k_{F}\\
k_{F}^{3}\left|k\right|^{2\beta} & \left|k\right|>2k_{F}
\end{cases}
\]
for a constant $C_{\beta}>0$ independent of $k$ and $k_{F}$.
\end{prop}

In particular, it holds that 
\begin{align}\sum_{p\in L_{k}}\lambda_{k,p}^{-1}\leq Ck_{F}\min\,\{1,k_{F}^{2}\left|k\right|^{-2}\},
\end{align}
so $\left\langle v_{k},h_{k}^{-1}v_{k}\right\rangle \leq C\hat{V}_{k}.$ Additionally, 
\begin{align}\left\langle e_{p},v_{k}\right\rangle \left\langle v_{k},e_{q}\right\rangle =\frac{\hat{V}_{k}k_{F}^{-1}}{2\left(2\pi\right)^{3}}.\end{align}
Inserting these quantities into the statements of the Propositions
\ref{prop:KElementEstimates}, \ref{prop:AtBtEstimates} and \ref{prop:KBtEstimate}
yields Theorem \ref{them:OneBodyEstimates}, noting also that by Proposition
\ref{prop:KElementEstimates}
\begin{align}
\left\Vert K_{k}\right\Vert _{\mathrm{HS}} & =\sqrt{\sum_{p,q\in L_{k}}\left|\left\langle e_{p},K_{k}e_{q}\right\rangle \right|^{2}}\leq\frac{\hat{V}_{k}k_{F}^{-1}}{2\left(2\pi\right)^{3}}\sqrt{\sum_{p,q\in L_{k}}\frac{1}{\left(\lambda_{k,p}+\lambda_{k,q}\right)^{2}}}\leq\frac{\hat{V}_{k}k_{F}^{-1}}{2\left(2\pi\right)^{3}}\sum_{p\in L_{k}}\frac{1}{\lambda_{k,p}}\\
 & \leq C\hat{V}_{k}\min\,\{1,k_{F}^{2}\left|k\right|^{-2}\}.\nonumber 
\end{align}

\section{\label{sec:AnalysisoftheExchangeTerms}Analysis of the Exchange Terms}

In this section we analyze the \textit{exchange terms}, by which we
mean the quantities of the expression
\begin{equation}
\sum_{k\in\mathbb{Z}_{\ast}^{3}}\int_{0}^{1}e^{\left(1-t\right)\mathcal{K}}\left(\varepsilon_{k}(\left\{ K_{k},B_{k}(t)\right\})+2\,\mathrm{Re}\left(\mathcal{E}_{k}^{1}(A_k(t))\right)+2\,\mathrm{Re}\left(\mathcal{E}_{k}^{2}(B_k(t))\right)\right)e^{-\left(1-t\right)\mathcal{K}}dt
\end{equation}
which appears in Theorem \ref{thm:DiagonalizationoftheBosonizableTerms}. The name is apt as these enter our calculations due to the presence
of the exchange correction $\varepsilon_{k,l}\left(p;q\right)$ of
the quasi-bosonic commutation relations (see Lemma \ref{lemma:QuasiBosonicCommutationRelations}). To be precise, we will consider in this section the operators
$\varepsilon_{k}(\left\{ K_{k},B_{k}(t)\right\})$,
$\mathcal{E}_{k}^{1}(A_k(t))$ and $\mathcal{E}_{k}^{2}(B_k(t))$, and the effect of the integration will be handled in the next section.
The main result of this section is the following estimates for them:
\begin{thm}
\label{them:ExchangeTermsEstimates}For any $\Psi\in\mathcal{H}_{N}$
and $t\in\left[0,1\right]$ it holds that
\begin{align*}
\left|\sum_{k\in\mathbb{Z}_{\ast}^{3}}\left\langle \Psi,\varepsilon_{k}(\left\{ K_{k},B_{k}(t)\right\})\Psi\right\rangle \right| & \leq Ck_{F}^{-1}\left\langle \Psi,\mathcal{N}_{E}\Psi\right\rangle, \\
\sum_{k\in\mathbb{Z}_{\ast}^{3}}\left|\left\langle \Psi,\mathcal{E}_{k}^{1}(A_k(t))\Psi\right\rangle \right| & \leq C\sqrt{\sum_{k\in\mathbb{Z}_{\ast}^{3}}\hat{V}_{k}^{2}\min\left\{ \left|k\right|,k_{F}\right\} }\left\langle \Psi,\left(\mathcal{N}_{E}^{3}+1\right)\Psi\right\rangle, \\
\sum_{k\in\mathbb{Z}_{\ast}^{3}}\left|\left\langle \Psi,\left(\mathcal{E}_{k}^{2}(B_k(t))-\left\langle \psi_{\mathrm{FS}},\mathcal{E}_{k}^{2}(B_k(t))\psi_{\mathrm{FS}}\right\rangle \right)\Psi\right\rangle \right| & \leq C\sqrt{\sum_{k\in\mathbb{Z}_{\ast}^{3}}\hat{V}_{k}^{2}\min\left\{ \left|k\right|,k_{F}\right\} }\left\langle \Psi,\mathcal{N}_{E}^{3}\Psi\right\rangle 
\end{align*}
for a constant $C>0$ depending only on $\sum_{k\in\mathbb{Z}_{\ast}^{3}}\hat{V}_{k}^{2}$.
\end{thm}
The constant terms in the final estimate of the theorem give the \textit{exchange contribution}
\begin{equation}
\sum_{k\in\mathbb{Z}_{\ast}^{3}}\int_{0}^{1}\left\langle \psi_{\mathrm{FS}},2\,\mathrm{Re}\left(\mathcal{E}_{k}^{2}(B_k(t))\right)\psi_{\mathrm{FS}}\right\rangle dt.
\end{equation}
It is not generally negligible for singular potentials $V$, and the leading behavior is given by
by\begin{prop}
\label{prop:LeadingExchangeContribution}It holds that
\[
\left|\sum_{k\in\mathbb{Z}_{\ast}^{3}}\int_{0}^{1}\left\langle \psi_{\mathrm{FS}},2\,\mathrm{Re}\left(\mathcal{E}_{k}^{2}(B_k(t))\right)\psi_{\mathrm{FS}}\right\rangle dt-E_{\mathrm{corr},\mathrm{ex}}\right|\leq C\sqrt{\sum_{k\in\mathbb{Z}_{\ast}^{3}}\hat{V}_{k}^{2}\min\left\{ \left|k\right|,k_{F}\right\} }
\]
for a constant $C>0$ depending only on $\sum_{k\in\mathbb{Z}_{\ast}^{3}}\hat{V}_{k}^{2}$,
where
\[
E_{\mathrm{corr},\mathrm{ex}}=\frac{k_{F}^{-2}}{4\left(2\pi\right)^{6}}\sum_{k\in\mathbb{Z}_{\ast}^{3}}\sum_{p,q\in L_{k}}\frac{\hat{V}_{k}\hat{V}_{p+q-k}}{\lambda_{k,p}+\lambda_{k,q}}.
\]
\end{prop}

\subsubsection*{Analysis of $\varepsilon_{k}$ Terms}

Let us first consider terms of the form $\sum_{k\in\mathbb{Z}_{\ast}^{3}}\varepsilon_{k}(A_{k})$,
where we recall that 
\begin{equation}
\varepsilon_{k}(A_{k})=-\sum_{p\in L_{k}}\left\langle e_{p},A_{k}e_{p}\right\rangle \left(c_{p}^{\ast}c_{p}+c_{p-k}c_{p-k}^{\ast}\right).
\end{equation}
When summing over $k\in\mathbb{Z}_{\ast}^{3}$, we can split the sum
into two parts and interchange the summations as follows:
\begin{align}
&-\sum_{k\in\mathbb{Z}_{\ast}^{3}}\varepsilon_{k}(A_{k})  =\sum_{k\in\mathbb{Z}_{\ast}^{3}}\sum_{p\in L_{k}}\left\langle e_{p},A_{k}e_{p}\right\rangle c_{p}^{\ast}c_{p}+\sum_{k\in\mathbb{Z}_{\ast}^{3}}\sum_{q\in\left(L_{k}-k\right)}\left\langle e_{q+k},A_{k}e_{q+k}\right\rangle c_{q}c_{q}^{\ast}\\
 & =\sum_{p\in B_{F}^{c}}\left(\sum_{k\in\mathbb{Z}_{\ast}^{3}}1_{L_k}(p)\left\langle e_{p},A_{k}e_{p}\right\rangle \right)c_{p}^{\ast}c_{p}+\sum_{q\in B_{F}}\left(\sum_{k\in\mathbb{Z}_{\ast}^{3}}1_{L_k}(q+k)\left\langle e_{q+k},A_{k}e_{q+k}\right\rangle \right)c_{q}c_{q}^{\ast}.\nonumber 
\end{align}
Recalling that $\mathcal{N}_{E}=\sum_{p\in B_{F}^{c}}c_{p}^{\ast}c_{p}=\sum_{q\in B_{F}}c_{q}c_{q}^{\ast}$ on $\mathcal{H}_{N}$, we can then immediately conclude that
\begin{align}
\pm\sum_{k\in\mathbb{Z}_{\ast}^{3}}\varepsilon_{k}(A_{k}) & \leq\left(\sup_{p\in B_{F}^{c}}\sum_{k\in\mathbb{Z}_{\ast}^{3}}1_{L_k}(p)\left|\left\langle e_{p},A_{k}e_{p}\right\rangle \right|+\sup_{q\in B_{F}}\sum_{k\in\mathbb{Z}_{\ast}^{3}}1_{L_k}(q+k)\left|\left\langle e_{q+k},A_{k}e_{q+k}\right\rangle \right|\right)\mathcal{N}_{E}\nonumber \\
 & \leq2\left(\sum_{k\in\mathbb{Z}_{\ast}^{3}}\sup_{p\in L_{k}}\left|\left\langle e_{p},A_{k}e_{p}\right\rangle \right|\right)\mathcal{N}_{E}.
\end{align}
By the estimates of the previous section we thus obtain the first
estimate of Theorem \ref{them:ExchangeTermsEstimates}:
\begin{prop}
For any $\Psi\in\mathcal{H}_{N}$ and $t\in\left[0,1\right]$ it holds
that
\[
\left|\sum_{k\in\mathbb{Z}_{\ast}^{3}}\left\langle \Psi,\varepsilon_{k}(\left\{ K_{k},B_{k}(t)\right\})\Psi\right\rangle \right|\leq Ck_{F}^{-1}\left\langle \Psi,\mathcal{N}_{E}\Psi\right\rangle 
\]
for a constant $C>0$ depending only on $\sum_{k\in\mathbb{Z}_{\ast}^{3}}\hat{V}_{k}^{2}$.
\end{prop}

\textbf{Proof:} By Theorem \ref{them:OneBodyEstimates} we have that
\begin{equation}
\left|\left\langle e_{p},\left\{ K_{k},B_{k}(t)\right\} e_{q}\right\rangle \right|\leq C\left(1+\hat{V}_{k}^{2}\right)\hat{V}_{k}^{2}k_{F}^{-1},\quad k\in\mathbb{Z}_{\ast}^{3},\,p,q\in L_{k},
\end{equation}
for a constant $C>0$ independent of all quantities, so
\begin{align}
 & \quad\;\left|\sum_{k\in\mathbb{Z}_{\ast}^{3}}\left\langle \Psi,\varepsilon_{k}(\left\{ K_{k},B_{k}(t)\right\})\Psi\right\rangle \right|\leq2\left(\sum_{k\in\mathbb{Z}_{\ast}^{3}}\sup_{p\in L_{k}}\left|\left\langle e_{p},\left\{ K_{k},B_{k}(t)\right\} e_{p}\right\rangle \right|\right)\left\langle \Psi,\mathcal{N}_{E}\Psi\right\rangle \\
 & \leq Ck_{F}^{-1}\sum_{k\in\mathbb{Z}_{\ast}^{3}}\left(1+\hat{V}_{k}^{2}\right)\hat{V}_{k}^{2}\left\langle \Psi,\mathcal{N}_{E}\Psi\right\rangle \leq Ck_{F}^{-1}\left(1+\Vert\hat{V}\Vert_{\infty}^{2}\right)\sum_{k\in\mathbb{Z}_{\ast}^{3}}\hat{V}_{k}^{2}\left\langle \Psi,\mathcal{N}_{E}\Psi\right\rangle .\nonumber 
\end{align}
As $\Vert\hat{V}\Vert_{\infty}^{2}\leq\Vert\hat{V}\Vert_{2}^{2}=\sum_{k\in\mathbb{Z}_{\ast}^{3}}\hat{V}_{k}^{2}$
the claim follows.
$\hfill\square$

\subsection{Analysis of $\mathcal{E}_{k}^{1}$ Terms}

We consider terms of the form
\begin{equation}
\mathcal{E}_{k}^{1}(A_k)=\sum_{l\in\mathbb{Z}_{\ast}^{3}}\sum_{p\in L_{k}}\sum_{q\in L_{l}}b_{k}^{\ast}(A_k e_p)\left\{ \varepsilon_{k,l}(e_{p};e_{q}),b_{-l}^{\ast}(K_{-l}e_{-q})\right\} .
\end{equation}
Recalling that 
\begin{equation}
\varepsilon_{k,l}(e_{p};e_{q})=-\left(\delta_{p,q}c_{q-l}c_{p-k}^{\ast}+\delta_{p-k,q-l}c_{q}^{\ast}c_{p}\right)
\end{equation}
we see that $\mathcal{E}_{k}^{1}(A_k)$ splits into two
sums as
\begin{align}
-\mathcal{E}_{k}^{1}(A_k) & =\sum_{l\in\mathbb{Z}_{\ast}^{3}}\sum_{p\in L_{k}}\sum_{q\in L_{l}}b_{k}^{\ast}(A_k e_p)\left\{ \delta_{p,q}c_{q-l}c_{p-k}^{\ast},b_{-l}^{\ast}(K_{-l}e_{-q})\right\} \nonumber \\
 &\quad +\sum_{l\in\mathbb{Z}_{\ast}^{3}}\sum_{p\in\left(L_{k}-k\right)}\sum_{q\in\left(L_{l}-l\right)}b_{k}^{\ast}(A_{k}e_{p+k})\left\{ \delta_{p,q}c_{q+l}^{\ast}c_{p+k},b_{-l}^{\ast}(K_{-l}e_{-q-l})\right\}\nonumber  \\
 & =\sum_{l\in\mathbb{Z}_{\ast}^{3}}\sum_{p\in L_{k}\cap L_{l}}b_{k}^{\ast}(A_k e_p)\left\{ c_{p-l}c_{p-k}^{\ast},b_{-l}^{\ast}(K_{-l}e_{-p})\right\} \nonumber \\
 & \quad+\sum_{l\in\mathbb{Z}_{\ast}^{3}}\sum_{p\in\left(L_{k}-k\right)\cap\left(L_{l}-l\right)}b_{k}^{\ast}(A_{k}e_{p+k})\left\{ c_{p+l}^{\ast}c_{p+k},b_{-l}^{\ast}\left(K_{-l}e_{-p-l}\right)\right\} .
\end{align}
The two sums on the right-hand side have the same ``schematic form'':
They can  be written as
\begin{equation}
\sum_{l\in\mathbb{Z}_{\ast}^{3}}\sum_{p\in S_{k}\cap S_{l}}b_{k}^{\ast}\left(A_{k}e_{p_{1}}\right)\left\{ \tilde{c}_{p_{2}}^{\ast}\tilde{c}_{p_{3}},b_{-l}^{\ast}\left(K_{-l}e_{p_{4}}\right)\right\} ,\quad\tilde{c}_{p}=\begin{cases}
c_{p} & p\in B_{F}^{c}\\
c_{p}^{\ast} & p\in B_{F}
\end{cases},\label{eq:calE1FirstSchematicForm}
\end{equation}
where the index set is either the lune $S_{k}=L_{k}$ or the corresponding
hole states $S_{k}=L_{k}-k,$ and depending on this index set the
variables $p_{1},p_{2},p_{3},p_{4}$ are given by
\begin{equation}
\left(p_{1},p_{2},p_{3},p_{4}\right)=\begin{cases}
\left(p,p-l,p-k,-p\right) & S_{k}=L_{k}\\
\left(p+k,p+l,p+k,-p-l\right) & S_{k}=L_{k}-k
\end{cases}.
\end{equation}
Note that in either case $p_{1}$, $p_{3}$ only depend on $p$ and
$k$, while $p_{2}$, $p_{4}$ depend only on $p$ and $l$. Additionally,
$p_{1}$ is always an element of $L_{k}$ and $p_{4}$ is always an
element of $L_{-l}$.

Since $b_{k,p}=c_{p-k}^{\ast}c_{p}=\tilde{c}_{p-k}\tilde{c}_{p}$
it is easily seen that $\left[b,\tilde{c}\right]=0$, so in normal-ordering
(with respect to $\psi_{\mathrm{FS}}$) the summand of equation (\ref{eq:calE1FirstSchematicForm})
we find
\begin{align}
 & \quad\;b_{k}^{\ast}\left(A_{k}e_{p_{1}}\right)\left\{ \tilde{c}_{p_{2}}^{\ast}\tilde{c}_{p_{3}},b_{-l}^{\ast}\left(K_{-l}e_{p_{4}}\right)\right\} \nonumber \\
 & =b_{k}^{\ast}\left(A_{k}e_{p_{1}}\right)\tilde{c}_{p_{2}}^{\ast}\tilde{c}_{p_{3}}b_{-l}^{\ast}\left(K_{-l}e_{p_{4}}\right)+b_{k}^{\ast}\left(A_{k}e_{p_{1}}\right)b_{-l}^{\ast}\left(K_{-l}e_{p_{4}}\right)\tilde{c}_{p_{2}}^{\ast}\tilde{c}_{p_{3}}\\
 & =2\,\tilde{c}_{p_{2}}^{\ast}b_{k}^{\ast}\left(A_{k}e_{p_{1}}\right)b_{-l}^{\ast}\left(K_{-l}e_{p_{4}}\right)\tilde{c}_{p_{3}}+\tilde{c}_{p_{2}}^{\ast}b_{k}^{\ast}\left(A_{k}e_{p_{1}}\right)\left[\tilde{c}_{p_{3}},b_{-l}^{\ast}\left(K_{-l}e_{p_{4}}\right)\right].\nonumber 
\end{align}
To bound a sum of the form $\sum_{k\in\mathbb{Z}_{\ast}^{3}}\mathcal{E}_{1}^{k}(A_k)$
it thus suffices to estimate the two schematic forms
\begin{align}
 & \sum_{k,l\in\mathbb{Z}_{\ast}^{3}}\sum_{p\in S_{k}\cap S_{l}}\tilde{c}_{p_{2}}^{\ast}b_{k}^{\ast}\left(A_{k}e_{p_{1}}\right)b_{-l}^{\ast}\left(K_{-l}e_{p_{4}}\right)\tilde{c}_{p_{3}},\label{eq:calE1SchematicForms}\\
 & \sum_{k,l\in\mathbb{Z}_{\ast}^{3}}\sum_{p\in S_{k}\cap S_{l}}\tilde{c}_{p_{2}}^{\ast}b_{k}^{\ast}\left(A_{k}e_{p_{1}}\right)\left[b_{-l}\left(K_{-l}e_{p_{4}}\right),\tilde{c}_{p_{3}}^{\ast}\right]^{\ast}.\nonumber 
\end{align}

\subsubsection*{Preliminary Estimates}

We prepare for the estimation of these schematic forms by deriving
some auxilliary bounds for the operators involved. Recall that for any $k\in\mathbb{Z}_{\ast}^{3}$ and $\varphi\in\ell^{2}(L_{k})$,
\begin{equation}
b_{k}(\varphi)=\sum_{p\in L_{k}}\left\langle \varphi,e_{p}\right\rangle b_{k,p}=\sum_{p\in L_{k}}\left\langle \varphi,e_{p}\right\rangle c_{p-k}^{\ast}c_{p}.
\end{equation}
Denote $\mathcal{N}_{k}=\sum_{p\in L_{k}}b_{k,p}^{\ast}b_{k,p}$. We can bound both $b_{k}(\varphi)$ and $b_{k}^{\ast}(\varphi)$
as follows:
\begin{prop}
\label{prop:bbastEstimates}For any $k\in\mathbb{Z}_{\ast}^{3}$,
$\varphi\in\ell^{2}(L_{k})$ and $\Psi\in\mathcal{H}_{N}$
it holds that
\[
\left\Vert b_{k}(\varphi)\Psi\right\Vert \leq\left\Vert \varphi\right\Vert \Vert\mathcal{N}_{k}^{\frac{1}{2}}\Psi\Vert,\quad\left\Vert b_{k}^{\ast}(\varphi)\Psi\right\Vert \leq\left\Vert \varphi\right\Vert \Vert\left(\mathcal{N}_{k}+1\right)^{\frac{1}{2}}\Psi\Vert. 
\]
\end{prop}

\textbf{Proof:} By the triangle and Cauchy-Schwarz inequalities we
immediately obtain
\begin{equation} \label{eq:bvarphi}
\left\Vert b_{k}(\varphi)\Psi\right\Vert \leq\sum_{p\in L_{k}}\left|\left\langle \varphi,e_{p}\right\rangle \right|\left\Vert b_{k,p}\Psi\right\Vert \leq\left\Vert \varphi\right\Vert \sqrt{\sum_{p\in L_{k}}\left\Vert b_{k,p}\Psi\right\Vert ^{2}}=\left\Vert \varphi\right\Vert \Vert\mathcal{N}_{k}^{\frac{1}{2}}\Psi\Vert
\end{equation}
and the bound for $\left\Vert b_{k}^{\ast}(\varphi)\Psi\right\Vert $
now follows from \eqref{eq:bvarphi} and the fact that 
\begin{align}  \label{eq:varepsilon_{k,k}<=0}
\varepsilon_{k,k}\left(\varphi;\varphi\right)=\left[b_{k}(\varphi),b_{k}^{\ast}(\varphi)\right]-\|\varphi\|^2=-\sum_{p\in L_{k}}\left|\left\langle e_{p},\varphi\right\rangle \right|^{2}\left(c_{p-k}c_{p-k}^{\ast}+c_{p}^{\ast}c_{p}\right)\leq0.  
\end{align}
$\hfill\square$

It is straightforward to see that $\mathcal{N}_{k}\le \mathcal{N}_{E}$. Moreover, by rearranging
the summations, 
\begin{align}
\sum_{k\in\mathbb{Z}_{\ast}^{3}}\mathcal{N}_{k}  =\sum_{k\in\mathbb{Z}_{\ast}^{3}}\sum_{p\in L_{k}}c_{p}^{\ast}c_{p-k}c_{p-k}^{\ast}c_{p} 
  =\sum_{p\in B_{F}^{c}}c_{p}^{\ast}c_{p}\sum_{k\in\left(B_{F}+p\right)}c_{p-k}c_{p-k}^{\ast} 
 =\mathcal{N}_{E}^{2}
\end{align}
on $\mathcal{H}_{N}$. We also note that for any $\Psi\in\mathcal{H}_{N}$
and $p\in\mathbb{Z}^{3}$
\begin{align}
\Vert\mathcal{N}_{k}^{\frac{1}{2}}\tilde{c}_{p}\Psi\Vert & \leq\Vert\tilde{c}_{p}\mathcal{N}_{k}^{\frac{1}{2}}\Psi\Vert\leq\Vert\tilde{c}_{p}\mathcal{N}_{E}^{\frac{1}{2}}\Psi\Vert\\
\Vert\left(\mathcal{N}_{k}+1\right)^{\frac{1}{2}}\tilde{c}_{p}\Psi\Vert & \leq\Vert\tilde{c}_{p}\left(\mathcal{N}_{k}+1\right)^{\frac{1}{2}}\Psi\Vert\leq\Vert\tilde{c}_{p}\left(\mathcal{N}_{E}+1\right)^{\frac{1}{2}}\Psi\Vert,\nonumber 
\end{align}
as follows by the inequality (considering $p\in B_{F}^{c}$ for definiteness)
\begin{align}
\tilde{c}_{p}^{\ast}\mathcal{N}_{k}\tilde{c}_{p} & =\sum_{q\in L_{k}}c_{p}^{\ast}c_{q}^{\ast}c_{q-k}c_{q-k}^{\ast}c_{q}c_{p}=\sum_{q\in L_{k}}c_{q}^{\ast}c_{q-k}c_{q-k}^{\ast}\left(c_{q}c_{p}^{\ast}-\delta_{p,q}\right)c_{p}\\
 & =\mathcal{N}_{k}c_{p}^{\ast}c_{p}-1_{L_k}(p)c_{p}^{\ast}c_{p-k}c_{p-k}^{\ast}c_{p}\leq\mathcal{N}_{k}c_{p}^{\ast}c_{p}\nonumber 
\end{align}
and the fact that $\left[\tilde{c}_{p}^{\ast}c_{p},\mathcal{N}_{k}\right]=0=\left[\tilde{c}_{p}^{\ast}c_{p},\mathcal{N}_{E}\right]$.
Similarly
\begin{equation}
\Vert\mathcal{N}_{E}^{\frac{1}{2}}\tilde{c}_{p}\Psi\Vert\leq\Vert\tilde{c}_{p}\mathcal{N}_{E}^{\frac{1}{2}}\Psi\Vert,\quad\Vert\left(\mathcal{N}_{E}+1\right)^{\frac{1}{2}}\tilde{c}_{p}\Psi\Vert\leq\Vert\tilde{c}_{p}\left(\mathcal{N}_{E}+1\right)^{\frac{1}{2}}\Psi\Vert.
\end{equation}
To analyze the commutator term $\left[b_{-l}\left(K_{-l}e_{p_{4}}\right),\tilde{c}_{p_{3}}^{\ast}\right]$
we calculate a general identity: For any $l\in\mathbb{Z}_{\ast}^{3}$,
$\psi\in\ell^{2}(L_l)$ and $p\in\mathbb{Z}^{3}$
\begin{align}
\left[b_{l}\left(\psi\right),\tilde{c}_{p}^{\ast}\right] 
 =\begin{cases}
-1_{L_l}(p+l)\left\langle \psi,e_{p+l}\right\rangle \tilde{c}_{p+l} & p\in B_{F}\\
1_{L_l}(p)\left\langle \psi,e_{p}\right\rangle \tilde{c}_{p-l} & p\in B_{F}^{c}
\end{cases},\label{eq:bcastCommutator}
\end{align}
so for our particular commutator we obtain
\begin{equation}
\left[b_{-l}\left(K_{-l}e_{p_{4}}\right),\tilde{c}_{p_{3}}^{\ast}\right]=\begin{cases}
-1_{L_{-l}}(p_{3}-l)\left\langle K_{-l}e_{p_{4}},e_{p_{3}-l}\right\rangle \tilde{c}_{p_{3}-l} & S_{k}=L_{k}\\
1_{L_{-l}}(p_{3})\left\langle K_{-l}e_{p_{4}},e_{p_{3}}\right\rangle \tilde{c}_{p_{3}+l} & S_{k}=L_{k}-k
\end{cases}.\label{eq:calE1CommutatorTermIdentity}
\end{equation}
It will be crucial to our estimates that the prefactors obey the following:
\begin{prop}
\label{prop:calE1KSplitEstimate}For any $k,l\in\mathbb{Z}_{\ast}^{3}$
and $p\in S_{k}\cap S_{l}$ it holds that
\begin{align*}
\left|1_{L_{-l}}(p_{3}-l)\left\langle K_{-l}e_{p_{4}},e_{p_{3}-l}\right\rangle \right| &\leq C\hat{V}_{-l}k_{F}^{-1}\frac{1_{L_{-k}}(p_{2}-k)1_{L_{-l}}(p_{3}-l)}{\sqrt{\lambda_{k,p_{1}}+\lambda_{-k,p_{2}-k}}\sqrt{\lambda_{-l,p_{3}-l}+\lambda_{-l,p_{4}}}},\quad S_{k}=L_{k},\\
\left|1_{L_{-l}}(p_{3})\left\langle K_{-l}e_{p_{4}},e_{p_{3}}\right\rangle \right| &\leq C\hat{V}_{-l}k_{F}^{-1}\frac{1_{L_{-k}}(p_{2})1_{L_{-l}}(p_{3})}{\sqrt{\lambda_{k,p_{1}}+\lambda_{-k,p_{2}}}\sqrt{\lambda_{-l,p_{3}}+\lambda_{-l,p_{4}}}},\quad S_{k}=L_{k}-k.
\end{align*}
\end{prop}

\textbf{Proof:} Recall that $p_{1},p_{2},p_{3},p_{4}$ are given by
\begin{equation}
\left(p_{1},p_{2},p_{3},p_{4}\right)=\begin{cases}
\left(p,p-l,p-k,-p\right) & S_{k}=L_{k}\\
\left(p+k,p+l,p+k,-p-l\right) & S_{k}=L_{k}-k
\end{cases}.
\end{equation}
From this we see that for any $p\in S_{k}\cap S_{l}$
\begin{align}
\begin{cases}
1_{L_{-l}}(p_{3}-l) & S_{k}=L_{k}\\
1_{L_{-l}}(p_{3}) & S_{k}=L_{k}-k
\end{cases} 
=\begin{cases}
1_{L_{-k}}(p_{2}-k) & S_{k}=L_{k}\\
1_{L_{-k}}(p_{2}) & S_{k}=L_{k}-k
\end{cases} 
\end{align}
where the assumption that $p\in S_{k}\cap S_{l}$ enters to ensure
that $1_{B_{F}}(p-k)=1=1_{B_{F}}(p-l)$ or $1_{B_{F}^{c}}(p+k)=1=1_{B_{F}^{c}}(p+l)$,
respectively. Importantly this also implies that, when combined with
such an indicator function, we also have the identity
\begin{align}
 & \quad\,\begin{cases}
\lambda_{-l,p_{3}-l}+\lambda_{-l,p_{4}} & S_{k}=L_{k}\\
\lambda_{-l,p_{3}}+\lambda_{-l,p_{4}} & S_{k}=L_{k}-k
\end{cases}
 =\begin{cases}
\lambda_{k,p_{1}}+\lambda_{-k,p_{2}-k} & S_{k}=L_{k}\\
\lambda_{k,p_{1}}+\lambda_{-k,p_{2}} & S_{k}=L_{k}-k
\end{cases}. 
\end{align}
The claim now follows by applying these identities to the estimates
\begin{align}
\left|1_{L_{-l}}(p_{3}-l)\left\langle K_{-l}e_{p_{4}},e_{p_{3}-l}\right\rangle \right| & \leq C\frac{1_{L_{-l}}(p_{3}-l)\hat{V}_{-l}k_{F}^{-1}}{\lambda_{-l,p_{3}-l}+\lambda_{-l,p_{4}}},\qquad\;\;S_{k}=L_{k},\\
\left|1_{L_{-l}}(p_{3})\left\langle K_{-l}e_{p_{4}},e_{p_{3}}\right\rangle \right| & \leq C\frac{1_{L_{-l}}(p_{3})\hat{V}_{-l}k_{F}^{-1}}{\lambda_{-l,p_{3}}+\lambda_{-l,p_{4}}},\qquad\qquad S_{k}=L_{k}-k,\nonumber 
\end{align}
which are given by Theorem \ref{them:OneBodyEstimates}.
$\hfill\square$

\medskip

Below we will only use the simpler bound
\begin{equation}
\begin{cases}
\left|1_{L_{-l}}(p_{3}-l)\left\langle K_{-l}e_{p_{4}},e_{p_{3}-l}\right\rangle \right| & S_{k}=L_{k}\\
\left|1_{L_{-l}}(p_{3})\left\langle K_{-l}e_{p_{4}},e_{p_{3}}\right\rangle \right| & S_{k}=L_{k}-k
\end{cases}\leq C\frac{\hat{V}_{-l}k_{F}^{-1}}{\sqrt{\lambda_{k,p_{1}}\lambda_{-l,p_{4}}}}\label{eq:calE1SimpleSplitInequality}
\end{equation}
but for the $\mathcal{E}_{k}^{2}$ terms the more general ones will
be needed.

\subsubsection*{Estimation of $\sum_{k\in\mathbb{Z}_{\ast}^{3}}\mathcal{E}_{k}^{1}(A_k(t))$}

Now the main estimate of this subsection:
\begin{prop}
\label{prop:calE1Estimates}For any collection of symmetric operators
$(A_k)$ and $\Psi\in\mathcal{H}_{N}$ it holds that
\begin{align*}
\sum_{k,l\in\mathbb{Z}_{\ast}^{3}}\sum_{p\in S_{k}\cap S_{l}}\left|\left\langle \Psi,\tilde{c}_{p_{2}}^{\ast}b_{k}^{\ast}\left(A_{k}e_{p_{1}}\right)b_{-l}^{\ast}\left(K_{-l}e_{p_{4}}\right)\tilde{c}_{p_{3}}\Psi\right\rangle \right| & \leq C\sqrt{\sum_{k\in\mathbb{Z}_{\ast}^{3}}\max_{p\in L_{k}}\left\Vert A_{k}e_{p}\right\Vert ^{2}}\Vert\left(\mathcal{N}_{E}+1\right)^{\frac{3}{2}}\Psi\Vert^{2}\\
\sum_{k,l\in\mathbb{Z}_{\ast}^{3}}\sum_{p\in S_{k}\cap S_{l}}\left|\left\langle \Psi,\tilde{c}_{p_{2}}^{\ast}b_{k}^{\ast}\left(A_{k}e_{p_{1}}\right)\left[b_{-l}\left(K_{-l}e_{p_{4}}\right),\tilde{c}_{p_{3}}^{\ast}\right]^{\ast}\tilde{c}_{p_{3}}\Psi\right\rangle \right| & \leq Ck_{F}^{-\frac{1}{2}}\sqrt{\sum_{k\in\mathbb{Z}_{\ast}^{3}}\Vert A_{k}h_{k}^{-\frac{1}{2}}\Vert_{\mathrm{HS}}^{2}}\left\Vert \left(\mathcal{N}_{E}+1\right)\Psi\right\Vert ^{2}.
\end{align*}
\end{prop}

\textbf{Proof:} Using the triangle and Cauchy-Schwarz inequalities
and Proposition \ref{prop:bbastEstimates} we estimate
\begin{align}
 & \quad\,\sum_{k,l\in\mathbb{Z}_{\ast}^{3}}\sum_{p\in S_{k}\cap S_{l}}\left|\left\langle \Psi,\tilde{c}_{p_{2}}^{\ast}b_{k}^{\ast}\left(A_{k}e_{p_{1}}\right)b_{-l}^{\ast}\left(K_{-l}e_{p_{4}}\right)\tilde{c}_{p_{3}}\Psi\right\rangle \right|\nonumber \\
 & \leq\sum_{k,l\in\mathbb{Z}_{\ast}^{3}}\sum_{p\in S_{k}\cap S_{l}}\left\Vert b_{k}\left(A_{k}e_{p_{1}}\right)\tilde{c}_{p_{2}}\Psi\right\Vert \left\Vert b_{-l}^{\ast}\left(K_{-l}e_{p_{4}}\right)\tilde{c}_{p_{3}}\Psi\right\Vert \nonumber \\
 & \leq\sum_{k\in\mathbb{Z}_{\ast}^{3}}\sum_{p\in S_{k}}\sum_{l\in\mathbb{Z}_{\ast}^{3}}1_{S_{l}}(p)\left\Vert A_{k}e_{p_{1}}\right\Vert \left\Vert K_{-l}e_{p_{4}}\right\Vert \Vert\mathcal{N}_{k}^{\frac{1}{2}}\tilde{c}_{p_{2}}\Psi\Vert\Vert\left(\mathcal{N}_{-l}+1\right)^{\frac{1}{2}}\tilde{c}_{p_{3}}\Psi\Vert\\
 & \leq\sum_{k\in\mathbb{Z}_{\ast}^{3}}\left(\max_{p\in L_{k}}\left\Vert A_{k}e_{p}\right\Vert \right)\sum_{p\in S_{k}}\Vert\tilde{c}_{p_{3}}\left(\mathcal{N}_{E}+1\right)^{\frac{1}{2}}\Psi\Vert\sqrt{\sum_{l\in\mathbb{Z}_{\ast}^{3}}1_{S_{l}}(p)\left\Vert K_{-l}e_{p_{4}}\right\Vert ^{2}}\sqrt{\sum_{l\in\mathbb{Z}_{\ast}^{3}}1_{S_{l}}(p)\Vert\tilde{c}_{p_{2}}\mathcal{N}_{k}^{\frac{1}{2}}\Psi\Vert^{2}}\nonumber \\
 & \leq\sum_{k\in\mathbb{Z}_{\ast}^{3}}\left(\max_{p\in L_{k}}\left\Vert A_{k}e_{p}\right\Vert \right)\Vert\mathcal{N}_{E}^{\frac{1}{2}}\mathcal{N}_{k}^{\frac{1}{2}}\Psi\Vert\sqrt{\sum_{p\in S_{k}} \Vert \tilde{c}_{p_{3}}\left(\mathcal{N}_{E}+1\right)^{\frac{1}{2}}\Psi\Vert ^{2}}\sqrt{\sum_{p\in S_{k}}\sum_{l\in\mathbb{Z}_{\ast}^{3}}1_{S_{l}}(p)\left\Vert K_{-l}e_{p_{4}}\right\Vert ^{2}}\nonumber \\
 & \leq\sqrt{\sum_{k\in\mathbb{Z}_{\ast}^{3}}\max_{p\in L_{k}}\left\Vert A_{k}e_{p}\right\Vert ^{2}}\sqrt{\sum_{l\in\mathbb{Z}_{\ast}^{3}}\left\Vert K_{l}\right\Vert _{\mathrm{HS}}^{2}}\left\Vert \left(\mathcal{N}_{E}+1\right)\Psi\right\Vert \sqrt{\sum_{k\in\mathbb{Z}_{\ast}^{3}}\Vert\mathcal{N}_{E}^{\frac{1}{2}}\mathcal{N}_{k}^{\frac{1}{2}}\Psi\Vert^{2}}\nonumber \\
 & =\sqrt{\sum_{k\in\mathbb{Z}_{\ast}^{3}}\max_{p\in L_{k}}\left\Vert A_{k}e_{p}\right\Vert ^{2}}\sqrt{\sum_{l\in\mathbb{Z}_{\ast}^{3}}\left\Vert K_{l}\right\Vert _{\mathrm{HS}}^{2}}\left\Vert \left(\mathcal{N}_{E}+1\right)\Psi\right\Vert \Vert\mathcal{N}_{E}^{\frac{3}{2}}\Psi\Vert\nonumber 
\end{align}
and the first bound now follows by recalling that $\left\Vert K_{l}\right\Vert _{\mathrm{HS}}^{2}\leq C\hat{V}_{l}$.
For the second we have by the equations (\ref{eq:calE1CommutatorTermIdentity})
and (\ref{eq:calE1SimpleSplitInequality}) that
\begin{align}
 & \quad\,\sum_{k,l\in\mathbb{Z}_{\ast}^{3}}\sum_{p\in S_{k}\cap S_{l}}\left|\left\langle \Psi,\tilde{c}_{p_{2}}^{\ast}b_{k}^{\ast}\left(A_{k}e_{p_{1}}\right)\left[b_{-l}\left(K_{-l}e_{p_{4}}\right),\tilde{c}_{p_{3}}^{\ast}\right]^{\ast}\Psi\right\rangle \right|\nonumber \\
 & \leq\sum_{k,l\in\mathbb{Z}_{\ast}^{3}}\sum_{p\in S_{k}\cap S_{l}}\left\Vert \left[b_{-l}\left(K_{-l}e_{p_{4}}\right),\tilde{c}_{p_{3}}^{\ast}\right]\tilde{c}_{p_{2}}\Psi\right\Vert \left\Vert b_{k}^{\ast}\left(A_{k}e_{p_{1}}\right)\Psi\right\Vert \nonumber \\
 & \leq C\sum_{l\in\mathbb{Z}_{\ast}^{3}}\sum_{p\in S_{l}}\sum_{k\in\mathbb{Z}_{\ast}^{3}}1_{S_{k}}(p)\left\Vert A_{k}e_{p_{1}}\right\Vert \frac{\hat{V}_{-l}k_{F}^{-1}}{\sqrt{\lambda_{k,p_{1}}\lambda_{-l,p_{4}}}}\left\Vert \tilde{c}_{p_{3}\mp l}\tilde{c}_{p_{2}}\Psi\right\Vert \Vert\left(\mathcal{N}_{k}+1\right)^{\frac{1}{2}}\Psi\Vert\\
 & \leq Ck_{F}^{-1}\Vert\left(\mathcal{N}_{E}+1\right)^{\frac{1}{2}}\Psi\Vert\sum_{p}\sum_{l\in\mathbb{Z}_{\ast}^{3}}\frac{1_{S_{l}}(p)\hat{V}_{-l}}{\sqrt{\lambda_{-l,p_{4}}}}\sqrt{\sum_{k\in\mathbb{Z}_{\ast}^{3}}1_{S_{k}}(p)\Vert A_{k}h_{k}^{-\frac{1}{2}}e_{p_{1}}\Vert^{2}}\sqrt{\sum_{k\in\mathbb{Z}_{\ast}^{3}}1_{S_{k}}(p)\left\Vert \tilde{c}_{p_{3}\mp l}\tilde{c}_{p_{2}}\Psi\right\Vert ^{2}}\nonumber \\
 & \leq Ck_{F}^{-1}\Vert\left(\mathcal{N}_{E}+1\right)^{\frac{1}{2}}\Psi\Vert\sum_{p}\sqrt{\sum_{k\in\mathbb{Z}_{\ast}^{3}}1_{S_{k}}(p)\Vert A_{k}h_{k}^{-\frac{1}{2}}e_{p_{1}}\Vert^{2}}\sqrt{\sum_{l\in\mathbb{Z}_{\ast}^{3}}1_{S_{l}}(p)\frac{\hat{V}_{-l}^{2}}{\lambda_{-l,p_{4}}}}\sqrt{\sum_{l\in\mathbb{Z}_{\ast}^{3}}1_{S_{l}}(p)\Vert\tilde{c}_{p_{2}}\mathcal{N}_{E}^{\frac{1}{2}}\Psi\Vert^{2}}\nonumber \\
 & \leq Ck_{F}^{-1}\Vert\left(\mathcal{N}_{E}+1\right)^{\frac{1}{2}}\Psi\Vert\left\Vert \mathcal{N}_{E}\Psi\right\Vert \sqrt{\sum_{k\in\mathbb{Z}_{\ast}^{3}}\sum_{p\in S_{k}}\Vert A_{k}h_{k}^{-\frac{1}{2}}e_{p_{1}}\Vert^{2}}\sqrt{\sum_{l\in\mathbb{Z}_{\ast}^{3}}\hat{V}_{-l}^{2}\sum_{p\in S_{l}}\frac{1}{\lambda_{-l,p_{4}}}}\nonumber \\
 & \leq Ck_{F}^{-1}\sqrt{\sum_{k\in\mathbb{Z}_{\ast}^{3}}\Vert A_{k}h_{k}^{-\frac{1}{2}}\Vert_{\mathrm{HS}}^{2}}\sqrt{\sum_{l\in\mathbb{Z}_{\ast}^{3}}\hat{V}_{l}^{2}\sum_{p\in L_{l}}\frac{1}{\lambda_{l,p}}}\Vert\left(\mathcal{N}_{E}+1\right)^{\frac{1}{2}}\Psi\Vert\left\Vert \mathcal{N}_{E}\Psi\right\Vert \nonumber 
\end{align}
where we used $\left\Vert A_{k}e_{p_{1}}\right\Vert \lambda_{k,p_{1}}^{-\frac{1}{2}}=\Vert A_{k}h_{k}^{-\frac{1}{2}}e_{p_{1}}\Vert$.
The claim  follows by $\sum_{p\in L_{l}}\lambda_{l,p}^{-1}\leq Ck_{F}$.
$\hfill\square$

\medskip

The bound on $\sum_{k\in\mathbb{Z}_{\ast}^{3}}\mathcal{E}_{k}^{1}(A_k(t))$
of Theorem \ref{them:ExchangeTermsEstimates} now follows by our matrix
element estimates:
\begin{prop}
\label{prop:ParticularcalE1Estimates}For any $\Psi\in\mathcal{H}_{N}$
and $t\in\left[0,1\right]$ it holds that
\[
\sum_{k\in\mathbb{Z}_{\ast}^{3}}\left|\left\langle \Psi,\mathcal{E}_{k}^{1}(A_k(t))\Psi\right\rangle \right|\leq C\sqrt{\sum_{k\in\mathbb{Z}_{\ast}^{3}}\hat{V}_{k}^{2}\min\left\{ \left|k\right|,k_{F}\right\} }\left\langle \Psi,\left(\mathcal{N}_{E}^{3}+1\right)\Psi\right\rangle 
\]
for a constant $C>0$ depending only on $\sum_{k\in\mathbb{Z}_{\ast}^{3}}\hat{V}_{k}^{2}$.
\end{prop}

\textbf{Proof:} By Theorem \ref{them:OneBodyEstimates} we have
\begin{equation}
\left|\left\langle e_{p},A_{k}(t)e_{q}\right\rangle \right|\leq C\left(1+\hat{V}_{k}^{2}\right)\hat{V}_{k}k_{F}^{-1},\quad k\in\mathbb{Z}_{\ast}^{3},\,p,q\in L_{k}.
\end{equation}
Combining with $\left|L_{k}\right|\leq C\min\left\{ k_{F}^{2}\left|k\right|,k_{F}^{3}\right\} $ 
since $\sum_{q\in L_{k}}\lambda_{k,q}^{-1}\leq Ck_{F}$, we get 
\begin{align}
\sum_{k\in\mathbb{Z}_{\ast}^{3}}\max_{p\in L_{k}}\left\Vert A_{k}(t)e_{p}\right\Vert ^{2} & \leq \frac{C}{k_{F}^{2}}\sum_{k\in\mathbb{Z}_{\ast}^{3}}\left(1+\hat{V}_{k}^{2}\right)^{2}\hat{V}_{k}^{2}\left|L_{k}\right| 
  =C\left(1+\Vert\hat{V}\Vert_{\infty}^{4}\right)\sum_{k\in\mathbb{Z}_{\ast}^{3}}\hat{V}_{k}^{2}\min\left\{ \left|k\right|,k_{F}\right\} \nonumber\\
 \sum_{k\in\mathbb{Z}_{\ast}^{3}}\Vert A_{k}(t)h_{k}^{-\frac{1}{2}}\Vert_{\mathrm{HS}}^{2} & =\sum_{k\in\mathbb{Z}_{\ast}^{3}}\sum_{p,q\in L_{k}}\left|\left\langle e_{p},A_{k}(t)h_{k}^{-\frac{1}{2}}e_{q}\right\rangle \right|^{2}\leq Ck_{F}^{-2}\sum_{k\in\mathbb{Z}_{\ast}^{3}}\left(1+\hat{V}_{k}^{2}\right)^{2}\hat{V}_{k}^{2}\left|L_{k}\right|\sum_{q\in L_{k}}\frac{1}{\lambda_{k,q}}\nonumber \\
 & \leq Ck_{F}\left(1+\Vert\hat{V}\Vert_{\infty}^{4}\right)\sum_{k\in\mathbb{Z}_{\ast}^{3}}\hat{V}_{k}^{2}\min\left\{ \left|k\right|,k_{F}\right\}. 
\end{align}
Inserting
these estimates into Proposition \ref{prop:calE1Estimates} yields
the claim.
$\hfill\square$

\subsection{Analysis of $\mathcal{E}_{k}^{2}$ Terms}

Now we come to the terms
\begin{equation}
\mathcal{E}_{k}^{2}(B_{k})=\frac{1}{2}\sum_{l\in\mathbb{Z}_{\ast}^{3}}\sum_{p\in L_{k}}\sum_{q\in L_{l}}\left\{ b_{k}(B_k e_p),\left\{ \varepsilon_{-k,-l}(e_{-p};e_{-q}),b_{l}^{\ast}(K_{l}e_{q})\right\} \right\} .
\end{equation}
We will analyze these similarly to the $\mathcal{E}_{k}^{1}(A_k)$
terms. Noting that
\begin{equation}
\varepsilon_{-k,-l}(e_{-p};e_{-q})=-\left(\delta_{p,q}c_{-q+l}c_{-p+k}^{\ast}+\delta_{p-k,q-l}c_{-q}^{\ast}c_{-p}\right)
\end{equation}
we find that $\mathcal{E}_{k}^{2}(B_{k})$ splits into
two sums as
\begin{align}
-2\,\mathcal{E}_{k}^{2}(B_{k}) & =\sum_{l\in\mathbb{Z}_{\ast}^{3}}\sum_{p\in L_{k}}\sum_{q\in L_{l}}\left\{ b_{k}(B_k e_p),\left\{ \delta_{p,q}c_{-q+l}c_{-p+k}^{\ast},b_{l}^{\ast}(K_{l}e_{q})\right\} \right\} \nonumber \\
 & \quad +\sum_{l\in\mathbb{Z}_{\ast}^{3}}\sum_{p\in\left(L_{k}-k\right)}\sum_{q\in\left(L_{l}-l\right)}\left\{ b_{k}\left(B_{k}e_{p+k}\right),\left\{ \delta_{p,q}c_{-q-l}^{\ast}c_{-p-k},b_{l}^{\ast}\left(K_{l}e_{q+l}\right)\right\} \right\} \\
 & =\sum_{l\in\mathbb{Z}_{\ast}^{3}}\sum_{p\in L_{k}\cap L_{l}}\left\{ b_{k}(B_k e_p),\left\{ c_{-p+l}c_{-p+k}^{\ast},b_{l}^{\ast}\left(K_{l}e_{p}\right)\right\} \right\} \nonumber \\
 & \quad +\sum_{l\in\mathbb{Z}_{\ast}^{3}}\sum_{p\in\left(L_{k}-k\right)\cap\left(L_{l}-l\right)}\left\{ b_{k}\left(B_{k}e_{p+k}\right),\left\{ c_{-p-l}^{\ast}c_{-p-k},b_{l}^{\ast}\left(K_{l}e_{p+l}\right)\right\} \right\} \nonumber 
\end{align}
and again these share a common schematic form, namely
\begin{equation}
\sum_{l\in\mathbb{Z}_{\ast}^{3}}\sum_{p\in S_{k}\cap S_{l}}\left\{ b_{k}\left(B_{k}e_{p_{1}}\right),\left\{ \tilde{c}_{p_{2}}^{\ast}\tilde{c}_{p_{3}},b_{l}^{\ast}\left(K_{l}e_{p_{4}}\right)\right\} \right\} 
\end{equation}
where the momenta are now
\begin{equation}
\left(p_{1},p_{2},p_{3},p_{4}\right)=\begin{cases}
\left(p,-p+l,-p+k,p\right) & S_{k}=L_{k}\\
\left(p+k,-p-l,-p-k,p+l\right) & S_{k}=L_{k}-k
\end{cases}.
\end{equation}
Again $p_{1}$, $p_{3}$ only depend on $p$ and $k$ while $p_{2}$,
$p_{4}$ only depend on $p$ and $l$.

We normal order the summand: As
\begin{align}
 & \quad\;b_{k}\left(B_{k}e_{p_{1}}\right)\left\{ \tilde{c}_{p_{2}}^{\ast}\tilde{c}_{p_{3}},b_{l}^{\ast}\left(K_{l}e_{p_{4}}\right)\right\} \nonumber \\
 & =\tilde{c}_{p_{2}}^{\ast}b_{k}\left(B_{k}e_{p_{1}}\right)\left\{ \tilde{c}_{p_{3}},b_{l}^{\ast}\left(K_{l}e_{p_{4}}\right)\right\} +\left[b_{k}\left(B_{k}e_{p_{1}}\right),\tilde{c}_{p_{2}}^{\ast}\right]\left\{ \tilde{c}_{p_{3}},b_{l}^{\ast}\left(K_{l}e_{p_{4}}\right)\right\} \nonumber \\
 & =2\,\tilde{c}_{p_{2}}^{\ast}b_{k}\left(B_{k}e_{p_{1}}\right)b_{l}^{\ast}\left(K_{l}e_{p_{4}}\right)\tilde{c}_{p_{3}}+\tilde{c}_{p_{2}}^{\ast}b_{k}\left(B_{k}e_{p_{1}}\right)\left[b_{l}\left(K_{l}e_{p_{4}}\right),\tilde{c}_{p_{3}}^{\ast}\right]^{\ast}\nonumber \\
 &\quad +2\left[b_{k}\left(B_{k}e_{p_{1}}\right),\tilde{c}_{p_{2}}^{\ast}\right]b_{l}^{\ast}\left(K_{l}e_{p_{4}}\right)\tilde{c}_{p_{3}}+\left[b_{k}\left(B_{k}e_{p_{1}}\right),\tilde{c}_{p_{2}}^{\ast}\right]\left[b_{l}\left(K_{l}e_{p_{4}}\right),\tilde{c}_{p_{3}}^{\ast}\right]^{\ast}\\
 & =2\,\tilde{c}_{p_{2}}^{\ast}b_{l}^{\ast}\left(K_{l}e_{p_{4}}\right)b_{k}\left(B_{k}e_{p_{1}}\right)\tilde{c}_{p_{3}}+2\,\tilde{c}_{p_{2}}^{\ast}\left[b_{k}\left(B_{k}e_{p_{1}}\right),b_{l}^{\ast}\left(K_{l}e_{p_{4}}\right)\right]\tilde{c}_{p_{3}}\nonumber \\
 &\quad +\tilde{c}_{p_{2}}^{\ast}\left[b_{l}\left(K_{l}e_{p_{4}}\right),\tilde{c}_{p_{3}}^{\ast}\right]^{\ast}b_{k}\left(B_{k}e_{p_{1}}\right)+\tilde{c}_{p_{2}}^{\ast}\left[b_{k}\left(B_{k}e_{p_{1}}\right),\left[b_{l}\left(K_{l}e_{p_{4}}\right),\tilde{c}_{p_{3}}^{\ast}\right]^{\ast}\right]\nonumber \\
 &\quad +2\,b_{l}^{\ast}\left(K_{l}e_{p_{4}}\right)\left[b_{k}\left(B_{k}e_{p_{1}}\right),\tilde{c}_{p_{2}}^{\ast}\right]\tilde{c}_{p_{3}}+2\left[b_{l}\left(K_{l}e_{p_{4}}\right),\left[b_{k}\left(B_{k}e_{p_{1}}\right),\tilde{c}_{p_{2}}^{\ast}\right]^{\ast}\right]^{\ast}\tilde{c}_{p_{3}}\nonumber \\
 &\quad -\left[b_{l}\left(K_{l}e_{p_{4}}\right),\tilde{c}_{p_{3}}^{\ast}\right]^{\ast}\left[b_{k}\left(B_{k}e_{p_{1}}\right),\tilde{c}_{p_{2}}^{\ast}\right]+\left\{ \left[b_{k}\left(B_{k}e_{p_{1}}\right),\tilde{c}_{p_{2}}^{\ast}\right],\left[b_{l}\left(K_{l}e_{p_{4}}\right),\tilde{c}_{p_{3}}^{\ast}\right]^{\ast}\right\} \nonumber 
\end{align}
and simply
\begin{align}
 & \quad\;\left\{ \tilde{c}_{p_{2}}^{\ast}\tilde{c}_{p_{3}},b_{l}^{\ast}\left(K_{l}e_{p_{4}}\right)\right\} b_{k}\left(B_{k}e_{p_{1}}\right)=\tilde{c}_{p_{2}}^{\ast}\left\{ \tilde{c}_{p_{3}},b_{l}^{\ast}\left(K_{l}e_{p_{4}}\right)\right\} b_{k}\left(B_{k}e_{p_{1}}\right)\\
 & =2\,\tilde{c}_{p_{2}}^{\ast}b_{l}^{\ast}\left(K_{l}e_{p_{4}}\right)b_{k}\left(B_{k}e_{p_{1}}\right)\tilde{c}_{p_{3}}+\tilde{c}_{p_{2}}^{\ast}\left[b_{l}\left(K_{l}e_{p_{4}}\right),\tilde{c}_{p_{3}}^{\ast}\right]^{\ast}b_{k}\left(B_{k}e_{p_{1}}\right)\nonumber 
\end{align}
the summand decomposes into 8 schematic forms as
\begin{align}
 & \quad\left\{ b_{k}\left(B_{k}e_{p_{1}}\right),\left\{ \tilde{c}_{p_{2}}^{\ast}\tilde{c}_{p_{3}},b_{l}^{\ast}\left(K_{l}e_{p_{4}}\right)\right\} \right\} \nonumber \\
 & =4\,\tilde{c}_{p_{2}}^{\ast}b_{l}^{\ast}\left(K_{l}e_{p_{4}}\right)b_{k}\left(B_{k}e_{p_{1}}\right)\tilde{c}_{p_{3}}+2\,\tilde{c}_{p_{2}}^{\ast}\left[b_{k}\left(B_{k}e_{p_{1}}\right),b_{l}^{\ast}\left(K_{l}e_{p_{4}}\right)\right]\tilde{c}_{p_{3}}\nonumber \\
 & +2\,\tilde{c}_{p_{2}}^{\ast}\left[b_{l}\left(K_{l}e_{p_{4}}\right),\tilde{c}_{p_{3}}^{\ast}\right]^{\ast}b_{k}\left(B_{k}e_{p_{1}}\right)+2\,b_{l}^{\ast}\left(K_{l}e_{p_{4}}\right)\left[b_{k}\left(B_{k}e_{p_{1}}\right),\tilde{c}_{p_{2}}^{\ast}\right]\tilde{c}_{p_{3}}\\
 & +\tilde{c}_{p_{2}}^{\ast}\left[b_{k}\left(B_{k}e_{p_{1}}\right),\left[b_{l}\left(K_{l}e_{p_{4}}\right),\tilde{c}_{p_{3}}^{\ast}\right]^{\ast}\right]+2\left[b_{l}\left(K_{l}e_{p_{4}}\right),\left[b_{k}\left(B_{k}e_{p_{1}}\right),\tilde{c}_{p_{2}}^{\ast}\right]^{\ast}\right]^{\ast}\tilde{c}_{p_{3}}\nonumber \\
 & -\left[b_{l}\left(K_{l}e_{p_{4}}\right),\tilde{c}_{p_{3}}^{\ast}\right]^{\ast}\left[b_{k}\left(B_{k}e_{p_{1}}\right),\tilde{c}_{p_{2}}^{\ast}\right]+\left\{ \left[b_{k}\left(B_{k}e_{p_{1}}\right),\tilde{c}_{p_{2}}^{\ast}\right],\left[b_{l}\left(K_{l}e_{p_{4}}\right),\tilde{c}_{p_{3}}^{\ast}\right]^{\ast}\right\} .\nonumber 
\end{align}
Of these it should be noted that only the last one is proportional
to a constant (i.e. does not contain any creation or annihilation
operators). As the rest annihilate $\psi_{\mathrm{FS}}$, it follows
that (when summed) the constant term yields precisely $\left\langle \psi_{\mathrm{FS}},\mathcal{E}_{k}^{2}(B_{k})\psi_{\mathrm{FS}}\right\rangle $,
whence bounding the other terms amounts to estimating the operator
\begin{equation}
\mathcal{E}_{k}^{2}(B_{k})-\left\langle \psi_{\mathrm{FS}},\mathcal{E}_{k}^{2}(B_{k})\psi_{\mathrm{FS}}\right\rangle 
\end{equation}
as in the statement of Theorem \ref{them:ExchangeTermsEstimates}.

\subsubsection*{Estimation of the Top Terms}

We begin by bounding the ``top'' terms
\[
\sum_{k,l\in\mathbb{Z}_{\ast}^{3}}\sum_{p\in S_{k}\cap S_{l}}\tilde{c}_{p_{2}}^{\ast}b_{l}^{\ast}\left(K_{l}e_{p_{4}}\right)b_{k}\left(B_{k}e_{p_{1}}\right)\tilde{c}_{p_{3}}\quad\text{and}\quad\sum_{k,l\in\mathbb{Z}_{\ast}^{3}}\sum_{p\in S_{k}\cap S_{l}}\tilde{c}_{p_{2}}^{\ast}\left[b_{k}\left(B_{k}e_{p_{1}}\right),b_{l}^{\ast}\left(K_{l}e_{p_{4}}\right)\right]\tilde{c}_{p_{3}}.
\]
By the quasi-bosonic commutation relations, the commutator term reduces
to
\begin{align}
 & \quad\sum_{k,l\in\mathbb{Z}_{\ast}^{3}}\sum_{p\in S_{k}\cap S_{l}}\tilde{c}_{p_{2}}^{\ast}\left[b_{k}\left(B_{k}e_{p_{1}}\right),b_{l}^{\ast}\left(K_{l}e_{p_{4}}\right)\right]\tilde{c}_{p_{3}}\\
 & =\sum_{k\in\mathbb{Z}_{\ast}^{3}}\sum_{p\in S_{k}}\left\langle B_{k}e_{p_{1}},K_{k}e_{p_{1}}\right\rangle \tilde{c}_{p_{3}}^{\ast}\tilde{c}_{p_{3}}+\sum_{k,l\in\mathbb{Z}_{\ast}^{3}}\sum_{p\in S_{k}\cap S_{l}}\tilde{c}_{p_{2}}^{\ast}\varepsilon_{k,l}\left(B_{k}e_{p_{1}};K_{l}e_{p_{4}}\right)\tilde{c}_{p_{3}}\nonumber 
\end{align}
where we used that $p_{1}=p_{4}$ and $p_{2}=p_{3}$ when $k=l$.
Now, the exchange correction of the second sum splits as
\begin{align}
-\varepsilon_{k,l}\left(B_{k}e_{p_{1}};K_{l}e_{p_{4}}\right) & =\sum_{q\in L_{k}}\sum_{q'\in L_{l}}\left\langle B_{k}e_{p_{1}},e_{q}\right\rangle \left\langle e_{q'},K_{l}e_{p_{4}}\right\rangle \left(\delta_{q,q'}c_{q'-l}c_{q-k}^{\ast}+\delta_{q-k,q'-l}c_{q'}^{\ast}c_{q}\right)\nonumber \\
 & =\sum_{q\in L_{k}\cap L_{l}}\left\langle B_{k}e_{p_{1}},e_{q}\right\rangle \left\langle e_{q},K_{l}e_{p_{4}}\right\rangle \tilde{c}_{q-l}^{\ast}\tilde{c}_{q-k}\\
 & +\sum_{q\in\left(L_{k}-k\right)\cap\left(L_{l}-l\right)}\left\langle B_{k}e_{p_{1}},e_{q+k}\right\rangle \left\langle e_{q+l},K_{l}e_{p_{4}}\right\rangle \tilde{c}_{q+l}^{\ast}\tilde{c}_{q+k}\nonumber 
\end{align}
which are both of the schematic form $\sum_{q\in S_{k}^{\prime}\cap S_{l}^{\prime}}\left\langle B_{k}e_{p_{1}},e_{q_{1}}\right\rangle \left\langle e_{q_{4}},K_{l}e_{p_{4}}\right\rangle \tilde{c}_{q_{2}}^{\ast}\tilde{c}_{q_{3}}$.

To estimate $\sum_{k,l\in\mathbb{Z}_{\ast}^{3}}\sum_{p\in S_{k}\cap S_{l}}\tilde{c}_{p_{2}}^{\ast}\varepsilon_{k,l}\left(B_{k}e_{p_{1}};K_{l}e_{p_{4}}\right)\tilde{c}_{p_{3}}$
it thus suffices to consider
\begin{equation}
\sum_{k,l\in\mathbb{Z}_{\ast}^{3}}\sum_{p\in S_{k}\cap S_{l}}\sum_{q\in S_{k}^{\prime}\cap S_{l}^{\prime}}\left\langle B_{k}e_{p_{1}},e_{q_{1}}\right\rangle \left\langle e_{q_{4}},K_{l}e_{p_{4}}\right\rangle \tilde{c}_{p_{2}}^{\ast}\tilde{c}_{q_{2}}^{\ast}\tilde{c}_{q_{3}}\tilde{c}_{p_{3}}.
\end{equation}
The estimates for the top terms are as follows:
\begin{prop}
\label{prop:calE2TopEstimates}For any collection of symmetric operators
$(B_k)$ and $\Psi\in\mathcal{H}_{N}$ it holds that
\begin{align*}
\sum_{k,l\in\mathbb{Z}_{\ast}^{3}}\sum_{p\in S_{k}\cap S_{l}}\left|\left\langle \Psi,\tilde{c}_{p_{2}}^{\ast}b_{l}^{\ast}\left(K_{l}e_{p_{4}}\right)b_{k}\left(B_{k}e_{p_{1}}\right)\tilde{c}_{p_{3}}\Psi\right\rangle \right| & \leq C\sqrt{\sum_{k\in\mathbb{Z}_{\ast}^{3}}\max_{p\in L_{k}}\left\Vert B_{k}e_{p}\right\Vert ^{2}}\Vert\mathcal{N}_{E}^{\frac{3}{2}}\Psi\Vert^{2}\\
\sum_{k,l\in\mathbb{Z}_{\ast}^{3}}\sum_{p\in S_{k}\cap S_{l}}\left|\left\langle \Psi,\tilde{c}_{p_{2}}^{\ast}\left[b_{k}\left(B_{k}e_{p_{1}}\right),b_{l}^{\ast}\left(K_{l}e_{p_{4}}\right)\right]\tilde{c}_{p_{3}}\Psi\right\rangle \right| & \leq C\sqrt{\sum_{k\in\mathbb{Z}_{\ast}^{3}}\sum_{p\in L_{k}}\max_{q\in L_{k}}\left|\left\langle e_{p},B_{k}e_{q}\right\rangle \right|^{2}}\left\Vert \mathcal{N}_{E}\Psi\right\Vert ^{2}
\end{align*}
for a constant $C>0$ depending only on $\sum_{k\in\mathbb{Z}_{\ast}^{3}}\hat{V}_{k}^{2}$.
\end{prop}

\textbf{Proof:} The first term we can estimate as in Proposition \ref{prop:calE1Estimates}
by
\begin{align}
 & \quad\,\sum_{k,l\in\mathbb{Z}_{\ast}^{3}}\sum_{p\in S_{k}\cap S_{l}}\left|\left\langle \Psi,\tilde{c}_{p_{2}}^{\ast}b_{l}^{\ast}\left(K_{l}e_{p_{4}}\right)b_{k}\left(B_{k}e_{p_{1}}\right)\tilde{c}_{p_{3}}\Psi\right\rangle \right|\nonumber \\
 & \leq\sum_{k,l\in\mathbb{Z}_{\ast}^{3}}\sum_{p\in S_{k}\cap S_{l}}\left\Vert b_{l}\left(K_{l}e_{p_{4}}\right)\tilde{c}_{p_{2}}\Psi\right\Vert \left\Vert b_{k}\left(B_{k}e_{p_{1}}\right)\tilde{c}_{p_{3}}\Psi\right\Vert \nonumber \\
 & \leq\sum_{k\in\mathbb{Z}_{\ast}^{3}}\sum_{p\in S_{k}}\sum_{l\in\mathbb{Z}_{\ast}^{3}}1_{S_{l}}(p)\left\Vert B_{k}e_{p_{1}}\right\Vert \left\Vert K_{l}e_{p_{4}}\right\Vert \Vert\mathcal{N}_{l}^{\frac{1}{2}}\tilde{c}_{p_{2}}\Psi\Vert\Vert\mathcal{N}_{k}^{\frac{1}{2}}\tilde{c}_{p_{3}}\Psi\Vert\\
 & \leq\sum_{k\in\mathbb{Z}_{\ast}^{3}}\left(\max_{p\in L_{k}}\left\Vert B_{k}e_{p}\right\Vert \right)\sum_{p\in S_{k}}\Vert\tilde{c}_{p_{3}}\mathcal{N}_{k}^{\frac{1}{2}}\Psi\Vert\sqrt{\sum_{l\in\mathbb{Z}_{\ast}^{3}}1_{S_{l}}(p)\left\Vert K_{l}e_{p_{4}}\right\Vert ^{2}}\sqrt{\sum_{l\in\mathbb{Z}_{\ast}^{3}}1_{S_{l}}(p)\Vert\tilde{c}_{p_{2}}\mathcal{N}_{E}^{\frac{1}{2}}\Psi\Vert^{2}}\nonumber \\
 & \leq\left\Vert \mathcal{N}_{E}\Psi\right\Vert \sum_{k\in\mathbb{Z}_{\ast}^{3}}\left(\max_{p\in L_{k}}\left\Vert B_{k}e_{p}\right\Vert \right)\sqrt{\sum_{p\in S_{k}}\Vert\tilde{c}_{p_{3}}\mathcal{N}_{k}^{\frac{1}{2}}\Psi\Vert^{2}}\sqrt{\sum_{p\in S_{k}}\sum_{l\in\mathbb{Z}_{\ast}^{3}}1_{S_{l}}(p)\left\Vert K_{l}e_{p_{4}}\right\Vert ^{2}}\nonumber \\
 & \leq\sqrt{\sum_{l\in\mathbb{Z}_{\ast}^{3}}\left\Vert K_{l}\right\Vert _{\mathrm{HS}}^{2}}\left\Vert \mathcal{N}_{E}\Psi\right\Vert \sum_{k\in\mathbb{Z}_{\ast}^{3}}\left(\max_{p\in L_{k}}\left\Vert B_{k}e_{p}\right\Vert \right)\Vert\mathcal{N}_{E}^{\frac{1}{2}}\mathcal{N}_{k}^{\frac{1}{2}}\Psi\Vert\nonumber
\end{align}
and obviously $\Vert\mathcal{N}_{E}^{\frac{1}{2}}\mathcal{N}_{k}^{\frac{1}{2}}\Psi\Vert  \le \Vert \mathcal{N}_{E}\Psi \Vert  \Vert \mathcal{N}_{E}^{\frac{3}{2}}\Psi \Vert$.  For the commutator term we have 
\begin{align}
\sum_{k\in\mathbb{Z}_{\ast}^{3}}\sum_{p\in S_{k}}\left|\left\langle B_{k}e_{p_{1}},K_{k}e_{p_{1}}\right\rangle \left\langle \Psi,\tilde{c}_{p_{3}}^{\ast}\tilde{c}_{p_{3}}\Psi\right\rangle \right| & \leq\sum_{k\in\mathbb{Z}_{\ast}^{3}}\max_{p\in L_{k}}\left|\left\langle B_{k}e_{p},K_{k}e_{p}\right\rangle \right|\sum_{p\in S_{k}}\left\langle \Psi,\tilde{c}_{p_{3}}^{\ast}\tilde{c}_{p_{3}}\Psi\right\rangle \\
 & \leq\sum_{k\in\mathbb{Z}_{\ast}^{3}}\max_{p\in L_{k}}\left|\left\langle e_{p},B_{k}K_{k}e_{p}\right\rangle \right|\left\langle \Psi,\mathcal{N}_{E}\Psi\right\rangle. \nonumber 
\end{align}
By the matrix element estimate for $K_{k}$ of Theorem \ref{them:OneBodyEstimates}
we have for any $p\in L_{k}$ that
\begin{align}
\left|\left\langle B_{k}e_{p},K_{k}e_{p}\right\rangle \right| & \leq\sum_{q\in L_{k}}\left|\left\langle B_{k}e_{p},e_{q}\right\rangle \right|\left|\left\langle e_{q},K_{k}e_{p}\right\rangle \right|\leq C\sum_{q\in L_{k}}\left|\left\langle e_{p},B_{k}e_{q}\right\rangle \right|\frac{\hat{V}_{k}k_{F}^{-1}}{\lambda_{k,q}+\lambda_{k,p}}\\
 & \leq C\hat{V}_{k}k_{F}^{-1}\left(\max_{q\in L_{k}}\left|\left\langle e_{p},B_{k}e_{q}\right\rangle \right|\right)\sum_{q\in L_{k}}\frac{1}{\lambda_{k,q}}\leq C\hat{V}_{k}\max_{q\in L_{k}}\left|\left\langle e_{p},B_{k}e_{q}\right\rangle \right|\nonumber 
\end{align}
since $\sum_{q\in L_{k}}\lambda_{k,q}^{-1}\leq Ck_{F}$. Consequently
\begin{align}
\sum_{k\in\mathbb{Z}_{\ast}^{3}}\sum_{p\in S_{k}}\left|\left\langle B_{k}e_{p_{1}},K_{k}e_{p_{1}}\right\rangle \left\langle \Psi,\tilde{c}_{p_{3}}^{\ast}\tilde{c}_{p_{3}}\Psi\right\rangle \right| & \leq C\sum_{k\in\mathbb{Z}_{\ast}^{3}}\hat{V}_{k}\left(\max_{p,q\in L_{k}}\left|\left\langle e_{p},B_{k}e_{q}\right\rangle \right|\right)\left\langle \Psi,\mathcal{N}_{E}\Psi\right\rangle \\
 & \leq C\sqrt{\sum_{k\in\mathbb{Z}_{\ast}^{3}}\hat{V}_{k}^{2}}\sqrt{\sum_{k\in\mathbb{Z}_{\ast}^{3}}\max_{p,q\in L_{k}}\left|\left\langle e_{p},B_{k}e_{q}\right\rangle \right|^{2}}\left\langle \Psi,\mathcal{N}_{E}\Psi\right\rangle \nonumber 
\end{align}
and clearly 
\begin{align}\max_{p,q\in L_{k}}\left|\left\langle e_{p},B_{k}e_{q}\right\rangle \right|^{2}\leq\sum_{p\in L_{k}}\max_{q\in L_{k}}\left|\left\langle e_{p},B_{k}e_{q}\right\rangle \right|^{2}.
\end{align}
Finally
\begin{align}
 & \quad\,\sum_{k,l\in\mathbb{Z}_{\ast}^{3}}\sum_{p\in S_{k}\cap S_{l}}\sum_{q\in S_{k}^{\prime}\cap S_{l}^{\prime}}\left|\left\langle B_{k}e_{p_{1}},e_{q_{1}}\right\rangle \left\langle e_{q_{4}},K_{l}e_{p_{4}}\right\rangle \left\langle \Psi,\tilde{c}_{p_{2}}^{\ast}\tilde{c}_{q_{2}}^{\ast}\tilde{c}_{q_{3}}\tilde{c}_{p_{3}}\Psi\right\rangle \right|\nonumber \\
 & \leq\sum_{k,l\in\mathbb{Z}_{\ast}^{3}}\sum_{p\in S_{k}\cap S_{l}}\sum_{q\in S_{k}^{\prime}\cap S_{l}^{\prime}}\left|\left\langle B_{k}e_{p_{1}},e_{q_{1}}\right\rangle \right|\left|\left\langle e_{q_{4}},K_{l}e_{p_{4}}\right\rangle \right|\left\Vert \tilde{c}_{q_{2}}\tilde{c}_{p_{2}}\Psi\right\Vert \left\Vert \tilde{c}_{q_{3}}\tilde{c}_{p_{3}}\Psi\right\Vert \\
 & \leq\sqrt{\sum_{k,l\in\mathbb{Z}_{\ast}^{3}}\sum_{p\in S_{k}\cap S_{l}}\sum_{q\in S_{k}^{\prime}\cap S_{l}^{\prime}}\left|\left\langle B_{k}e_{p_{1}},e_{q_{1}}\right\rangle \right|^{2}\left\Vert \tilde{c}_{q_{2}}\tilde{c}_{p_{2}}\Psi\right\Vert ^{2}}\sqrt{\sum_{k,l\in\mathbb{Z}_{\ast}^{3}}\sum_{p\in S_{k}\cap S_{l}}\sum_{q\in S_{k}^{\prime}\cap S_{l}^{\prime}}\left|\left\langle e_{q_{4}},K_{l}e_{p_{4}}\right\rangle \right|^{2}\left\Vert \tilde{c}_{q_{3}}\tilde{c}_{p_{3}}\Psi\right\Vert ^{2}}\nonumber \\
 & \leq\sqrt{\sum_{k\in\mathbb{Z}_{\ast}^{3}}\sum_{p\in S_{k}}\max_{q\in L_{k}}\left|\left\langle e_{p_{1}},B_{k}e_{q}\right\rangle \right|^{2}\sum_{l\in\mathbb{Z}_{\ast}^{3}}1_{S_{l}}(p)\Vert\tilde{c}_{p_{2}}\mathcal{N}_{E}^{\frac{1}{2}}\Psi\Vert^{2}}\sqrt{\sum_{l\in\mathbb{Z}_{\ast}^{3}}\sum_{p\in S_{l}}\left\Vert K_{l}e_{p_{4}}\right\Vert ^{2}\sum_{k\in\mathbb{Z}_{\ast}^{3}}1_{S_{k}}(p)\left\Vert \tilde{c}_{p_{3}}\Psi\right\Vert ^{2}}\nonumber \\
 & \leq\sqrt{\sum_{k\in\mathbb{Z}_{\ast}^{3}}\sum_{p\in L_{k}}\max_{q\in L_{k}}\left|\left\langle e_{p},B_{k}e_{q}\right\rangle \right|^{2}}\sqrt{\sum_{l\in\mathbb{Z}_{\ast}^{3}}\left\Vert K_{l}\right\Vert _{\mathrm{HS}}^{2}}\Vert\mathcal{N}_{E}^{\frac{1}{2}}\Psi\Vert\left\Vert \mathcal{N}_{E}\Psi\right\Vert \nonumber 
\end{align}
whence the claim follows as $\left\Vert K_{l}\right\Vert _{\mathrm{HS}}\leq C\hat{V}_{l}$.
$\hfill\square$

\subsubsection*{Estimation of the Single Commutator Terms}

For the single commutator terms
\[
\sum_{k,l\in\mathbb{Z}_{\ast}^{3}}\sum_{p\in S_{k}\cap S_{l}}\tilde{c}_{p_{2}}^{\ast}\left[b_{l}\left(K_{l}e_{p_{4}}\right),\tilde{c}_{p_{3}}^{\ast}\right]^{\ast}b_{k}\left(B_{k}e_{p_{1}}\right)\quad\text{and}\quad\sum_{k,l\in\mathbb{Z}_{\ast}^{3}}\sum_{p\in S_{k}\cap S_{l}}b_{l}^{\ast}\left(K_{l}e_{p_{4}}\right)\left[b_{k}\left(B_{k}e_{p_{1}}\right),\tilde{c}_{p_{2}}^{\ast}\right]\tilde{c}_{p_{3}}
\]
we note that by equation (\ref{eq:bcastCommutator}), the commutator
$\left[b_{l}\left(K_{l}e_{p_{4}}\right),\tilde{c}_{p_{3}}^{\ast}\right]$
is given by
\begin{equation}
\left[b_{l}\left(K_{l}e_{p_{4}}\right),\tilde{c}_{p_{3}}^{\ast}\right]=\begin{cases}
-1_{L_{l}}(p_{3}+l)\left\langle K_{l}e_{p_{4}},e_{p_{3}+l}\right\rangle \tilde{c}_{p_{3}+l} & S_{k}=L_{k}\\
1_{L_{l}}(p_{3})\left\langle K_{l}e_{p_{4}},e_{p_{3}}\right\rangle \tilde{c}_{p_{3}-l} & S_{k}=L_{k}-k
\end{cases}.\label{eq:Klep4cp3Commutator}
\end{equation}
The prefactors again obey an estimate as in Proposition \ref{prop:calE1KSplitEstimate}:
\begin{prop}
\label{prop:calE2KSplitEstimate}For any $k,l\in\mathbb{Z}_{\ast}^{3}$
and $p\in S_{k}\cap S_{l}$ it holds that
\begin{align*}
\left|1_{L_{l}}(p_{3}+l)\left\langle K_{l}e_{p_{4}},e_{p_{3}+l}\right\rangle \right| &\leq C\hat{V}_{l}k_{F}^{-1}\frac{1_{L_{k}}(p_{2}+k)1_{L_{l}}(p_{3}+l)}{\sqrt{\lambda_{k,p_{1}}+\lambda_{k,p_{2}+k}}\sqrt{\lambda_{l,p_{3}+l}+\lambda_{l,p_{4}}}},\quad S_{k}=L_{k},\\
\left|1_{L_{l}}(p_{3})\left\langle K_{l}e_{p_{4}},e_{p_{3}}\right\rangle \right| &\leq C\hat{V}_{l}k_{F}^{-1}\frac{1_{L_{k}}(p_{2})1_{L_{l}}(p_{3})}{\sqrt{\lambda_{k,p_{1}}+\lambda_{k,p_{2}}}\sqrt{\lambda_{l,p_{3}}+\lambda_{l,p_{4}}}},\quad S_{k}=L_{k}-k.
\end {align*}
\end{prop}

The proof is essentially the same as that of Proposition \ref{prop:calE1KSplitEstimate}
(indeed, this proposition can be obtained directly from the former
by appropriate substition, but some care must be used since the $p_{i}$'s
differ in their definition).

For the single commutator terms we again only need the simpler bound
\begin{equation}
\begin{cases}
\left|1_{L_{l}}(p_{3}+l)\left\langle K_{l}e_{p_{4}},e_{p_{3}+l}\right\rangle \right| & S_{k}=L_{k}\\
\left|1_{L_{l}}(p_{3})\left\langle K_{l}e_{p_{4}},e_{p_{3}}\right\rangle \right| & S_{k}=L_{k}-k
\end{cases}\leq C\frac{\hat{V}_{l}k_{F}^{-1}}{\sqrt{\lambda_{k,p_{1}}\lambda_{l,p_{4}}}}
\end{equation}
but the full one will be needed for the double commutator terms below.
Now the estimate:
\begin{prop}
\label{prop:calE2SingleCommutatorEstimates}For any collection of
symmetric operators $(B_k)$ and $\Psi\in\mathcal{H}_{N}$
it holds that
\begin{align*}
\sum_{k,l\in\mathbb{Z}_{\ast}^{3}}\sum_{p\in S_{k}\cap S_{l}}\left|\left\langle \Psi,\tilde{c}_{p_{2}}^{\ast}\left[b_{l}\left(K_{l}e_{p_{4}}\right),\tilde{c}_{p_{3}}^{\ast}\right]^{\ast}b_{k}\left(B_{k}e_{p_{1}}\right)\Psi\right\rangle \right| & \leq Ck_{F}^{-\frac{1}{2}}\sqrt{\sum_{k\in\mathbb{Z}_{\ast}^{3}}\Vert B_{k}h_{k}^{-\frac{1}{2}}\Vert_{\mathrm{HS}}^{2}}\left\Vert \mathcal{N}_{E}\Psi\right\Vert ^{2},\\
\sum_{k,l\in\mathbb{Z}_{\ast}^{3}}\sum_{p\in S_{k}\cap S_{l}}\left|\left\langle \Psi,b_{l}^{\ast}\left(K_{l}e_{p_{4}}\right)\left[b_{k}\left(B_{k}e_{p_{1}}\right),\tilde{c}_{p_{2}}^{\ast}\right]\tilde{c}_{p_{3}}\Psi\right\rangle \right| & \leq C\sqrt{\sum_{k\in\mathbb{Z}_{\ast}^{3}}\sum_{p\in L_{k}}\max_{q\in L_{k}}\left|\left\langle e_{p},B_{k}e_{q}\right\rangle \right|^{2}}\left\Vert \mathcal{N}_{E}\Psi\right\Vert ^{2}
\end{align*}
for a constant $C>0$ depending only on $\sum_{k\in\mathbb{Z}_{\ast}^{3}}\hat{V}_{k}^{2}$.
\end{prop}

\textbf{Proof:} As in the second estimate of Proposition \ref{prop:calE1Estimates}
we have
\begin{align}
 & \quad\,\sum_{k,l\in\mathbb{Z}_{\ast}^{3}}\sum_{p\in S_{k}\cap S_{l}}\left|\left\langle \Psi,\tilde{c}_{p_{2}}^{\ast}\left[b_{l}\left(K_{l}e_{p_{4}}\right),\tilde{c}_{p_{3}}^{\ast}\right]^{\ast}b_{k}\left(B_{k}e_{p_{1}}\right)\Psi\right\rangle \right|\nonumber \\
 & \leq\sum_{k,l\in\mathbb{Z}_{\ast}^{3}}\sum_{p\in S_{k}\cap S_{l}}\left\Vert \left[b_{l}\left(K_{l}e_{p_{4}}\right),\tilde{c}_{p_{3}}^{\ast}\right]\tilde{c}_{p_{2}}\Psi\right\Vert \left\Vert b_{k}\left(B_{k}e_{p_{1}}\right)\Psi\right\Vert \nonumber \\
 & \leq C\sum_{l\in\mathbb{Z}_{\ast}^{3}}\sum_{p\in S_{l}}\sum_{k\in\mathbb{Z}_{\ast}^{3}}1_{S_{k}}(p)\left\Vert B_{k}e_{p_{1}}\right\Vert \frac{\hat{V}_{l}k_{F}^{-1}}{\sqrt{\lambda_{k,p_{1}}\lambda_{l,p_{4}}}}\left\Vert \tilde{c}_{p_{3}\pm l}\tilde{c}_{p_{2}}\Psi\right\Vert \Vert\mathcal{N}_{k}^{\frac{1}{2}}\Psi\Vert\\
 & \leq Ck_{F}^{-1}\Vert\mathcal{N}_{E}^{\frac{1}{2}}\Psi\Vert\sum_{p}\sum_{l\in\mathbb{Z}_{\ast}^{3}}\frac{1_{S_{l}}(p)\hat{V}_{l}}{\sqrt{\lambda_{l,p_{4}}}}\sqrt{\sum_{k\in\mathbb{Z}_{\ast}^{3}}1_{S_{k}}(p)\Vert B_{k}h_{k}^{-\frac{1}{2}}e_{p_{1}}\Vert^{2}}\sqrt{\sum_{k\in\mathbb{Z}_{\ast}^{3}}1_{S_{k}}(p)\left\Vert \tilde{c}_{p_{3}\pm l}\tilde{c}_{p_{2}}\Psi\right\Vert ^{2}}\nonumber \\
 & \leq Ck_{F}^{-1}\Vert\mathcal{N}_{E}^{\frac{1}{2}}\Psi\Vert\sum_{p}\sqrt{\sum_{k\in\mathbb{Z}_{\ast}^{3}}1_{S_{k}}(p)\Vert B_{k}h_{k}^{-\frac{1}{2}}e_{p_{1}}\Vert^{2}}\sqrt{\sum_{l\in\mathbb{Z}_{\ast}^{3}}1_{S_{l}}(p)\frac{\hat{V}_{l}^{2}}{\lambda_{l,p_{4}}}}\sqrt{\sum_{l\in\mathbb{Z}_{\ast}^{3}}1_{S_{l}}(p)\Vert\tilde{c}_{p_{2}}\mathcal{N}_{E}^{\frac{1}{2}}\Psi\Vert^{2}}\nonumber \\
 & \leq Ck_{F}^{-1}\Vert\mathcal{N}_{E}^{\frac{1}{2}}\Psi\Vert\left\Vert \mathcal{N}_{E}\Psi\right\Vert \sqrt{\sum_{k\in\mathbb{Z}_{\ast}^{3}}\sum_{p\in S_{k}}\Vert B_{k}h_{k}^{-\frac{1}{2}}e_{p_{1}}\Vert^{2}}\sqrt{\sum_{l\in\mathbb{Z}_{\ast}^{3}}\hat{V}_{l}^{2}\sum_{p\in S_{l}}\frac{1}{\lambda_{l,p_{4}}}}\nonumber \\
 & \leq Ck_{F}^{-\frac{1}{2}}\sqrt{\sum_{k\in\mathbb{Z}_{\ast}^{3}}\Vert B_{k}h_{k}^{-\frac{1}{2}}\Vert_{\mathrm{HS}}^{2}}\sqrt{\sum_{l\in\mathbb{Z}_{\ast}^{3}}\hat{V}_{l}^{2}}\Vert\mathcal{N}_{E}^{\frac{1}{2}}\Psi\Vert\left\Vert \mathcal{N}_{E}\Psi\right\Vert .\nonumber 
\end{align}
By equation (\ref{eq:bcastCommutator}) it holds that
\begin{equation}
\left[b_{k}\left(B_{k}e_{p_{1}}\right),\tilde{c}_{p_{2}}^{\ast}\right]=\begin{cases}
-1_{L_{k}}(p_{2}+k)\left\langle B_{k}e_{p_{1}},e_{p_{2}+k}\right\rangle \tilde{c}_{p_{2}+k} & p\in B_{F}\\
1_{L_{k}}(p_{2})\left\langle B_{k}e_{p_{1}},e_{p_{2}}\right\rangle \tilde{c}_{p_{2}-k} & p\in B_{F}^{c}
\end{cases}\label{eq:Bkep1cp2Commutator}
\end{equation}
so the second term can be bounded as
\begin{align}
 & \quad\,\sum_{k,l\in\mathbb{Z}_{\ast}^{3}}\sum_{p\in S_{k}\cap S_{l}}\left|\left\langle \Psi,b_{l}^{\ast}\left(K_{l}e_{p_{4}}\right)\left[b_{k}\left(B_{k}e_{p_{1}}\right),\tilde{c}_{p_{2}}^{\ast}\right]\tilde{c}_{p_{3}}\Psi\right\rangle \right|\nonumber \\
 & \leq\sum_{k,l\in\mathbb{Z}_{\ast}^{3}}\sum_{p\in S_{k}\cap S_{l}}\left\Vert b_{l}\left(K_{l}e_{p_{4}}\right)\Psi\right\Vert \left\Vert \left[b_{k}\left(B_{k}e_{p_{1}}\right),\tilde{c}_{p_{2}}^{\ast}\right]\tilde{c}_{p_{3}}\Psi\right\Vert \nonumber \\
 & \leq\sum_{k\in\mathbb{Z}_{\ast}^{3}}\sum_{p\in S_{k}}\sum_{l\in\mathbb{Z}_{\ast}^{3}}1_{S_{l}}(p)\left(\max_{q\in L_{k}}\left|\left\langle e_{p_{1}},B_{k}e_{q}\right\rangle \right|\right)\left\Vert K_{l}e_{p_{4}}\right\Vert \Vert\mathcal{N}_{l}^{\frac{1}{2}}\Psi\Vert\left\Vert \tilde{c}_{p_{2}\pm k}\tilde{c}_{p_{3}}\Psi\right\Vert \nonumber \\
 & \leq\Vert\mathcal{N}_{E}^{\frac{1}{2}}\Psi\Vert\sum_{p}\sum_{k\in\mathbb{Z}_{\ast}^{3}}1_{S_{k}}(p)\left(\max_{q\in L_{k}}\left|\left\langle e_{p_{1}},B_{k}e_{q}\right\rangle \right|\right)\sqrt{\sum_{l\in\mathbb{Z}_{\ast}^{3}}1_{S_{l}}(p)\left\Vert K_{l}e_{p_{4}}\right\Vert ^{2}}\sqrt{\sum_{l\in\mathbb{Z}_{\ast}^{3}}1_{S_{l}}(p)\left\Vert \tilde{c}_{p_{2}\pm k}\tilde{c}_{p_{3}}\Psi\right\Vert ^{2}}\nonumber \\
 & \leq\Vert\mathcal{N}_{E}^{\frac{1}{2}}\Psi\Vert\sum_{p}\sqrt{\sum_{l\in\mathbb{Z}_{\ast}^{3}}1_{S_{l}}(p)\left\Vert K_{l}e_{p_{4}}\right\Vert ^{2}}\sqrt{\sum_{k\in\mathbb{Z}_{\ast}^{3}}1_{S_{k}}(p)\left(\max_{q\in L_{k}}\left|\left\langle e_{p_{1}},B_{k}e_{q}\right\rangle \right|^{2}\right)}\sqrt{\sum_{k\in\mathbb{Z}_{\ast}^{3}}1_{S_{k}}(p)\Vert\tilde{c}_{p_{3}}\mathcal{N}_{E}^{\frac{1}{2}}\Psi\Vert^{2}}\nonumber \\
 & \leq\Vert\mathcal{N}_{E}^{\frac{1}{2}}\Psi\Vert\left\Vert \mathcal{N}_{E}\Psi\right\Vert \sqrt{\sum_{l\in\mathbb{Z}_{\ast}^{3}}\sum_{p\in S_{l}}\left\Vert K_{l}e_{p_{4}}\right\Vert ^{2}}\sqrt{\sum_{k\in\mathbb{Z}_{\ast}^{3}}\sum_{p\in S_{k}}\max_{q\in L_{k}}\left|\left\langle e_{p_{1}},B_{k}e_{q}\right\rangle \right|^{2}}\\
 & \leq\sqrt{\sum_{k\in\mathbb{Z}_{\ast}^{3}}\sum_{p\in L_{k}}\max_{q\in L_{k}}\left|\left\langle e_{p},B_{k}e_{q}\right\rangle \right|^{2}}\sqrt{\sum_{l\in\mathbb{Z}_{\ast}^{3}}\left\Vert K_{l}\right\Vert _{\mathrm{HS}}^{2}}\,\Vert\mathcal{N}_{E}^{\frac{1}{2}}\Psi\Vert\left\Vert \mathcal{N}_{E}\Psi\right\Vert . \nonumber \tag*{$\square$}
\end{align}

\subsubsection*{Estimation of the Double Commutator Terms}

Finally we have the double commutator terms
\begin{align}
&\sum_{k,l\in\mathbb{Z}_{\ast}^{3}}\sum_{p\in S_{k}\cap S_{l}}\tilde{c}_{p_{2}}^{\ast}\left[b_{k}\left(B_{k}e_{p_{1}}\right),\left[b_{l}\left(K_{l}e_{p_{4}}\right),\tilde{c}_{p_{3}}^{\ast}\right]^{\ast}\right],\nonumber\\
&\sum_{k,l\in\mathbb{Z}_{\ast}^{3}}\sum_{p\in S_{k}\cap S_{l}}\left[b_{l}\left(K_{l}e_{p_{4}}\right),\left[b_{k}\left(B_{k}e_{p_{1}}\right),\tilde{c}_{p_{2}}^{\ast}\right]^{\ast}\right]^{\ast}\tilde{c}_{p_{3}},\nonumber\\
&\sum_{k,l\in\mathbb{Z}_{\ast}^{3}}\sum_{p\in S_{k}\cap S_{l}}\left[b_{l}\left(K_{l}e_{p_{4}}\right),\tilde{c}_{p_{3}}^{\ast}\right]^{\ast}\left[b_{k}\left(B_{k}e_{p_{1}}\right),\tilde{c}_{p_{2}}^{\ast}\right].
\end{align}
An identity for the iterated commutators is obtained by applying the
identity of equation (\ref{eq:bcastCommutator}) to itself: For any
$k,l\in\mathbb{Z}_{\ast}^{3}$, $\varphi\in\ell^{2}(L_{k})$,
$\psi\in\ell^{2}(L_l)$ and $p\in\mathbb{Z}_{\ast}^{3}$
\begin{align}
\left[b_{k}(\varphi),\left[b_{l}\left(\psi\right),\tilde{c}_{p}^{\ast}\right]^{\ast}\right] & =\begin{cases}
-1_{L_l}(p+l)\left\langle e_{p+l},\psi\right\rangle \left[b_{k}(\varphi),\tilde{c}_{p+l}^{\ast}\right] & p\in B_{F}\\
1_{L_l}(p)\left\langle e_{p},\psi\right\rangle \left[b_{k}(\varphi),\tilde{c}_{p-l}^{\ast}\right] & p\in B_{F}^{c}
\end{cases}\label{eq:bbcastastIdentity}\\
 & =\begin{cases}
-1_{L_k}(p+l)1_{L_l}(p+l)\left\langle \varphi,e_{p+l}\right\rangle \left\langle e_{p+l},\psi\right\rangle \tilde{c}_{p+l-k} & p\in B_{F}\\
-1_{L_{k}}\left(p-l+k\right)1_{L_l}(p)\left\langle \varphi,e_{p-l+k}\right\rangle \left\langle e_{p},\psi\right\rangle \tilde{c}_{p-l+k} & p\in B_{F}^{c}
\end{cases}.\nonumber 
\end{align}
The estimates are the following:
\begin{prop}
\label{prop:calE2DoubleCommutatorEstimates}For any collection of
symmetric operators $(B_k)$ and $\Psi\in\mathcal{H}_{N}$
it holds that
\begin{align*}
\sum_{k,l\in\mathbb{Z}_{\ast}^{3}}\sum_{p\in S_{k}\cap S_{l}}\left|\left\langle \Psi,\tilde{c}_{p_{2}}^{\ast}\left[b_{k}\left(B_{k}e_{p_{1}}\right),\left[b_{l}\left(K_{l}e_{p_{4}}\right),\tilde{c}_{p_{3}}^{\ast}\right]^{\ast}\right]\Psi\right\rangle \right|,\\
\sum_{k,l\in\mathbb{Z}_{\ast}^{3}}\sum_{p\in S_{k}\cap S_{l}}\left|\left\langle \Psi,\left[b_{l}\left(K_{l}e_{p_{4}}\right),\left[b_{k}\left(B_{k}e_{p_{1}}\right),\tilde{c}_{p_{2}}^{\ast}\right]^{\ast}\right]^{\ast}\tilde{c}_{p_{3}}\Psi\right\rangle \right|,\\
\sum_{k,l\in\mathbb{Z}_{\ast}^{3}}\sum_{p\in S_{k}\cap S_{l}}\left|\left\langle \Psi,\left[b_{l}\left(K_{l}e_{p_{4}}\right),\tilde{c}_{p_{3}}^{\ast}\right]^{\ast}\left[b_{k}\left(B_{k}e_{p_{1}}\right),\tilde{c}_{p_{2}}^{\ast}\right]\Psi\right\rangle \right|,
\end{align*}
are all bounded by $$Ck_F^{-\frac 1 2}\sqrt{\sum_{k\in\mathbb{Z}_{\ast}^{3}}\max_{p\in L_{k}}\Vert h_{k}^{-\frac{1}{2}}B_{k}e_{p}\Vert^{2}}\left\langle \Psi,\mathcal{N}_{E}\Psi\right\rangle $$
for a constant $C>0$ depending only on $\sum_{k\in\mathbb{Z}_{\ast}^{3}}\hat{V}_{k}^{2}$.
\end{prop}

\textbf{Proof:} For these estimates we consider only the case $S_{k}=L_{k}$
for the sake of clarity, i.e. we let
\begin{equation}
\left(p_{1},p_{2},p_{3},p_{4}\right)=\left(p,-p+l,-p+k,p\right);
\end{equation}
the case $S_{k}=L_{k}-k$ can be handled by similar manipulations.

Using the identity of equation (\ref{eq:bbcastastIdentity}) we start
by estimating (by the bound of Proposition \ref{prop:calE2KSplitEstimate})
\begin{align}
 & \quad\,\sum_{k,l\in\mathbb{Z}_{\ast}^{3}}\sum_{p\in L_{k}\cap L_{l}}\left|\left\langle \Psi,\tilde{c}_{p_{2}}^{\ast}\left[b_{k}\left(B_{k}e_{p_{1}}\right),\left[b_{l}\left(K_{l}e_{p_{4}}\right),\tilde{c}_{p_{3}}^{\ast}\right]^{\ast}\right]\Psi\right\rangle \right|\nonumber \\
 & =\sum_{k,l\in\mathbb{Z}_{\ast}^{3}}\sum_{p\in L_{k}\cap L_{l}}\left|1_{L_{k}}(p_{3}+l)1_{L_{l}}(p_{3}+l)\left\langle B_{k}e_{p_{1}},e_{p_{3}+l}\right\rangle \left\langle e_{p_{3}+l},K_{l}e_{p_{4}}\right\rangle \left\langle \Psi,\tilde{c}_{p_{2}}^{\ast}\tilde{c}_{p_{3}+l-k}\Psi\right\rangle \right|\nonumber \\
 & \leq C\sum_{k,l\in\mathbb{Z}_{\ast}^{3}}\sum_{p\in L_{k}\cap L_{l}}1_{L_{k}}(p_{3}+l)\left|\left\langle B_{k}e_{p_{1}},e_{p_{3}+l}\right\rangle \right|\frac{\hat{V}_{l}k_{F}^{-1}1_{L_{k}}(p_{2}+k)1_{L_{l}}(p_{3}+l)}{\sqrt{\lambda_{k,p_{1}}+\lambda_{k,p_{2}+k}}\sqrt{\lambda_{l,p_{3}+l}+\lambda_{l,p_{4}}}}\left\langle \Psi,\tilde{c}_{p_{2}}^{\ast}\tilde{c}_{p_{2}}\Psi\right\rangle \nonumber \\
 & \leq Ck_{F}^{-1}\sum_{l\in\mathbb{Z}_{\ast}^{3}}\hat{V}_{l}\sum_{p\in L_{l}}\sqrt{\sum_{k\in\mathbb{Z}_{\ast}^{3}}1_{L_k}(p)1_{L_{k}}(p_{3}+l)\left|\left\langle e_{p},h_{k}^{-\frac{1}{2}}B_{k}e_{p_{3}+l}\right\rangle \right|^{2}} \nonumber\\
 &\qquad \cdot \sqrt{\sum_{k\in\mathbb{Z}_{\ast}^{3}}\frac{1_{L_{l}}(p_{3}+l)}{\lambda_{l,p_{3}+l}}}\left\langle \Psi,\tilde{c}_{-p+l}^{\ast}\tilde{c}_{-p+l}\Psi\right\rangle \nonumber \\
 & \leq Ck_{F}^{-\frac{1}{2}}\sum_{l\in\mathbb{Z}_{\ast}^{3}}\hat{V}_{l}\sum_{p\in\left(L_{l}-l\right)}\sqrt{\sum_{k\in\mathbb{Z}_{\ast}^{3}}1_{L_k}(p+l)1_{L_{k}}(p_{3})\left|\left\langle e_{p+l},h_{k}^{-\frac{1}{2}}B_{k}e_{p_{3}}\right\rangle \right|^{2}}\left\langle \Psi,\tilde{c}_{-p}^{\ast}\tilde{c}_{-p}\Psi\right\rangle \\
 & \leq Ck_{F}^{-\frac{1}{2}}\sum_{p\in B_{F}}\sqrt{\sum_{l\in\mathbb{Z}_{\ast}^{3}}\hat{V}_{l}^{2}}\sqrt{\sum_{k,l\in\mathbb{Z}_{\ast}^{3}}1_{L_k}(p+l)1_{L_{k}}(p_{3})\left|\left\langle e_{p+l},h_{k}^{-\frac{1}{2}}B_{k}e_{p_{3}}\right\rangle \right|^{2}}\left\langle \Psi,\tilde{c}_{-p}^{\ast}\tilde{c}_{-p}\Psi\right\rangle \nonumber \\
 & \leq Ck_{F}^{-\frac{1}{2}}\sqrt{\sum_{k\in\mathbb{Z}_{\ast}^{3}}\max_{p\in L_{k}}\Vert h_{k}^{-\frac{1}{2}}B_{k}e_{p}\Vert^{2}}\sqrt{\sum_{l\in\mathbb{Z}_{\ast}^{3}}\hat{V}_{l}^{2}}\left\langle \Psi,\mathcal{N}_{E}\Psi\right\rangle \nonumber 
\end{align}
where we used $\sum_{k\in\mathbb{Z}_{\ast}^{3}}1_{L_{l}}(p_{3}+l)\lambda_{l,p_{3}+l}^{-1}\leq\sum_{q\in L_{l}}\lambda_{l,q}^{-1}\leq Ck_{F}$. From (\ref{eq:bbcastastIdentity}) we have
\begin{align}
&\left[b_{l}\left(K_{l}e_{p_{4}}\right),\left[b_{k}\left(B_{k}e_{p_{1}}\right),\tilde{c}_{p}^{\ast}\right]^{\ast}\right] \nonumber\\
& =-1_{L_{l}}(p_{2}+k)1_{L_{k}}(p_{2}+k)\left\langle K_{l}e_{p_{4}},e_{p_{2}+k}\right\rangle \left\langle e_{p_{2}+k},B_{k}e_{p_{1}}\right\rangle \tilde{c}_{p_{2}+k-l} \nonumber\\
 & =-1_{L_{k}}(p_{2}+k)1_{L_{l}}(p_{3}+l)\left\langle K_{l}e_{p_{4}},e_{p_{3}+l}\right\rangle \left\langle e_{p_{2}+k},B_{k}e_{p_{1}}\right\rangle \tilde{c}_{p_{3}}
\end{align}
so the second term can be similarly estimated as
\begin{align}
 & \qquad\,\sum_{k,l\in\mathbb{Z}_{\ast}^{3}}\sum_{p\in S_{k}\cap S_{l}}\left|\left\langle \Psi,\left[b_{l}\left(K_{l}e_{p_{4}}\right),\left[b_{k}\left(B_{k}e_{p_{1}}\right),\tilde{c}_{p_{2}}^{\ast}\right]^{\ast}\right]^{\ast}\tilde{c}_{p_{3}}\Psi\right\rangle \right|\nonumber \\
 & \leq C\sum_{k,l\in\mathbb{Z}_{\ast}^{3}}\sum_{p\in L_{k}\cap L_{l}}\frac{\hat{V}_{l}k_{F}^{-1}1_{L_{k}}(p_{2}+k)1_{L_{l}}(p_{3}+l)}{\sqrt{\lambda_{k,p_{1}}+\lambda_{k,p_{2}+k}}\sqrt{\lambda_{l,p_{3}+l}+\lambda_{l,p_{4}}}}\left|\left\langle e_{p_{2}+k},B_{k}e_{p_{1}}\right\rangle \right|\left\langle \Psi,\tilde{c}_{p_{3}}^{\ast}\tilde{c}_{p_{3}}\Psi\right\rangle \nonumber \\
 & \leq Ck_{F}^{-1}\sum_{k\in\mathbb{Z}_{\ast}^{3}}\sum_{p\in L_{k}}\sqrt{\sum_{l\in\mathbb{Z}_{\ast}^{3}}1_{L_l}(p)\frac{\hat{V}_{l}^{2}}{\lambda_{l,p_{4}}}}\sqrt{\sum_{l\in\mathbb{Z}_{\ast}^{3}}1_{L_{k}}(p_{2}+k)\left|\left\langle e_{p_{2}+k},h_{k}^{-\frac{1}{2}}B_{k}e_{p_{1}}\right\rangle \right|^{2}}\left\langle \Psi,\tilde{c}_{-p+k}^{\ast}\tilde{c}_{-p+k}\Psi\right\rangle \nonumber \\
 & \leq Ck_{F}^{-1}\sum_{p\in B_{F}}\sum_{k\in\mathbb{Z}_{\ast}^{3}}1_{L_{k}-k}(p)\sqrt{\sum_{l\in\mathbb{Z}_{\ast}^{3}}\hat{V}_{l}^{2}\frac{1_{L_{l}}(p+k)}{\lambda_{l,p+k}}}\Vert h_{k}^{-\frac{1}{2}}B_{k}e_{p+k}\Vert\left\langle \Psi,\tilde{c}_{-p}^{\ast}\tilde{c}_{-p}\Psi\right\rangle \\
 & \leq Ck_{F}^{-1}\sum_{p\in B_{F}}\sqrt{\sum_{l\in\mathbb{Z}_{\ast}^{3}}\hat{V}_{l}^{2}\sum_{k\in\mathbb{Z}_{\ast}^{3}}\frac{1_{L_{l}}(p+k)}{\lambda_{l,p+k}}}\sqrt{\sum_{k\in\mathbb{Z}_{\ast}^{3}}\Vert h_{k}^{-\frac{1}{2}}B_{k}e_{p+k}\Vert^{2}}\left\langle \Psi,\tilde{c}_{-p}^{\ast}\tilde{c}_{-p}\Psi\right\rangle \nonumber \\
 & \leq Ck_{F}^{-\frac{1}{2}}\sqrt{\sum_{k\in\mathbb{Z}_{\ast}^{3}}\max_{p\in L_{k}}\Vert h_{k}^{-\frac{1}{2}}B_{k}e_{p}\Vert^{2}}\sqrt{\sum_{l\in\mathbb{Z}_{\ast}^{3}}\hat{V}_{l}^{2}}\left\langle \Psi,\mathcal{N}_{E}\Psi\right\rangle .\nonumber 
\end{align}
Finally, from  (\ref{eq:Klep4cp3Commutator}) and (\ref{eq:Bkep1cp2Commutator})
we see that $\left[b_{l}\left(K_{l}e_{p_{4}}\right),\tilde{c}_{p_{3}}^{\ast}\right]^{\ast}\left[b_{k}\left(B_{k}e_{p_{1}}\right),\tilde{c}_{p_{2}}^{\ast}\right]$ is equal to 
\begin{align}
1_{L_{k}}(p_{2}+k)1_{L_{l}}(p_{3}+l)\left\langle B_{k}e_{p_{1}},e_{p_{2}+k}\right\rangle \left\langle e_{p_{3}+l},K_{l}e_{p_{4}}\right\rangle \tilde{c}_{p_{3}+l}^{\ast}\tilde{c}_{p_{2}+k},
\end{align}
so we estimate
\begin{align}
 & \qquad\,\sum_{k,l\in\mathbb{Z}_{\ast}^{3}}\sum_{p\in S_{k}\cap S_{l}}\left|\left\langle \Psi,\left[b_{l}\left(K_{l}e_{p_{4}}\right),\tilde{c}_{p_{3}}^{\ast}\right]^{\ast}\left[b_{k}\left(B_{k}e_{p_{1}}\right),\tilde{c}_{p_{2}}^{\ast}\right]\Psi\right\rangle \right|\nonumber \\
 & \leq C\sum_{k,l\in\mathbb{Z}_{\ast}^{3}}\sum_{p\in L_{k}\cap L_{l}}\frac{\hat{V}_{l}k_{F}^{-1}1_{L_{k}}(p_{2}+k)1_{L_{l}}(p_{3}+l)}{\sqrt{\lambda_{k,p_{1}}+\lambda_{k,p_{2}+k}}\sqrt{\lambda_{l,p_{3}+l}+\lambda_{l,p_{4}}}}\left|\left\langle B_{k}e_{p_{1}},e_{p_{2}+k}\right\rangle \right|\left\langle \Psi,\tilde{c}_{p_{3}+l}^{\ast}\tilde{c}_{p_{2}+k}\Psi\right\rangle \nonumber\\
 & \leq Ck_{F}^{-1}\sum_{p\in B_{F}^{c}}\sum_{k,l\in\mathbb{Z}_{\ast}^{3}}1_{L_{k}\cap L_{l}}(p)1_{L_{k}\cap L_{l}}(-p+k+l)\frac{\hat{V}_{l}}{\sqrt{\lambda_{l,p}}}\left|\left\langle e_{p},h_{k}^{-\frac{1}{2}}B_{k}e_{-p+k+l}\right\rangle \right| \nonumber\\
&\qquad \cdot \left\langle \Psi,\tilde{c}_{-p+k+l}^{\ast}\tilde{c}_{-p+k+l}\Psi\right\rangle \nonumber \\
 & =Ck_{F}^{-1}\sum_{p\in B_{F}^{c}}\sum_{k,l\in\mathbb{Z}_{\ast}^{3}}1_{L_{k}\cap L_{l}}(p+k+l)1_{L_{k}\cap L_{l}}(-p)\frac{\hat{V}_{l}}{\sqrt{\lambda_{l,p+k+l}}}\left|\left\langle e_{p+k+l},h_{k}^{-\frac{1}{2}}B_{k}e_{-p}\right\rangle \right|\left\langle \Psi,\tilde{c}_{-p}^{\ast}\tilde{c}_{-p}\Psi\right\rangle \nonumber \\
 & \leq Ck_{F}^{-1}\sum_{p\in B_{F}^{c}}\sqrt{\sum_{k,l\in\mathbb{Z}_{\ast}^{3}}1_{L_{k}}(p+k+l)1_{L_{k}}(-p)\left|\left\langle e_{p+k+l},h_{k}^{-\frac{1}{2}}B_{k}e_{-p}\right\rangle \right|^{2}}\nonumber \\
 & \qquad\cdot\sqrt{\sum_{k,l\in\mathbb{Z}_{\ast}^{3}}\hat{V}_{l}^{2}\frac{1_{L_{l}}(p+k+l)}{\lambda_{l,p+k+l}}}\left\langle \Psi,\tilde{c}_{-p}^{\ast}\tilde{c}_{-p}\Psi\right\rangle \\
 &\leq Ck_{F}^{-\frac{1}{2}}\sqrt{\sum_{k\in\mathbb{Z}_{\ast}^{3}}\max_{p\in L_{k}}\Vert h_{k}^{-\frac{1}{2}}B_{k}e_{p}\Vert^{2}}\sqrt{\sum_{l\in\mathbb{Z}_{\ast}^{3}}\hat{V}_{l}^{2}}\left\langle \Psi,\mathcal{N}_{E}\Psi\right\rangle .\nonumber \tag*{$\square$}
\end{align}

The $\mathcal{E}_{k}^{2}$ bound of Theorem \ref{them:ExchangeTermsEstimates}
now follows:
\begin{prop}
For any $\Psi\in\mathcal{H}_{N}$ and $t\in\left[0,1\right]$ it holds
that
\[
\sum_{k\in\mathbb{Z}_{\ast}^{3}}\left|\left\langle \Psi,\left(\mathcal{E}_{k}^{2}(B_k(t))-\left\langle \psi_{\mathrm{FS}},\mathcal{E}_{k}^{2}(B_k(t))\psi_{\mathrm{FS}}\right\rangle \right)\Psi\right\rangle \right|\leq C\sqrt{\sum_{k\in\mathbb{Z}_{\ast}^{3}}\hat{V}_{k}^{2}\min\left\{ \left|k\right|,k_{F}\right\} }\left\langle \Psi,\mathcal{N}_{E}^{3}\Psi\right\rangle 
\]
for a constant $C>0$ depending only on $\sum_{k\in\mathbb{Z}_{\ast}^{3}}\hat{V}_{k}^{2}$.
\end{prop}

\textbf{Proof:} Clearly
\begin{equation}
\max_{p\in L_{k}}\left\Vert B_{k}e_{p}\right\Vert ^{2}\leq\sum_{p\in L_{k}}\max_{q\in L_{k}}\left|\left\langle e_{p},B_{k}e_{q}\right\rangle \right|^{2},\quad\max_{p\in L_{k}}\Vert h_{k}^{-\frac{1}{2}}B_{k}e_{p}\Vert^{2}\leq\Vert B_{k}h_{k}^{-\frac{1}{2}}\Vert_{\mathrm{HS}}^{2},
\end{equation}
for any $B_{k}$, and as our estimate for $B_{k}(t)$ in
Theorem \ref{them:OneBodyEstimates} is the same as that for $A_{k}(t)$,
the bounds
\[
\sum_{k\in\mathbb{Z}_{\ast}^{3}}\sum_{p\in L_{k}}\max_{q\in L_{k}}\left|\left\langle e_{p},B_{k}e_{q}\right\rangle \right|^{2},\,k_{F}^{-1}\sum_{k\in\mathbb{Z}_{\ast}^{3}}\Vert B_{k}h_{k}^{-\frac{1}{2}}\Vert_{\mathrm{HS}}^{2}\leq C\left(1+\Vert\hat{V}\Vert_{\infty}^{4}\right)\sum_{k\in\mathbb{Z}_{\ast}^{3}}\hat{V}_{k}^{2}\min\left\{ \left|k\right|,k_{F}\right\} 
\]
follow exactly as those of Proposition \ref{prop:ParticularcalE1Estimates}.
Insertion into the Propositions \ref{prop:calE2TopEstimates}, \ref{prop:calE2SingleCommutatorEstimates}
and \ref{prop:calE2DoubleCommutatorEstimates} yields the claim.
$\hfill\square$

\subsection{Analysis of the Exchange Contribution}

Finally we determine the leading order of the exchange contribution.
To begin we derive a general formula for a quantity of the form $\left\langle \psi_{\mathrm{FS}},\mathcal{E}_{k}^{2}(B_{k})\psi_{\mathrm{FS}}\right\rangle $:
We can write
\begin{align}
-2\left\langle \psi_{\mathrm{FS}},\mathcal{E}_{k}^{2}(B_{k})\psi_{\mathrm{FS}}\right\rangle  & =-\sum_{l\in\mathbb{Z}_{\ast}^{3}}\sum_{p\in L_{k}}\sum_{q\in L_{l}}\left\langle \psi_{\mathrm{FS}},b_{k}(B_k e_p)\varepsilon_{-k,-l}(e_{-p};e_{-q})b_{l}^{\ast}(K_{l}e_{q})\psi_{\mathrm{FS}}\right\rangle \nonumber \\
 & =\sum_{l\in\mathbb{Z}_{\ast}^{3}}\sum_{p\in L_{k}\cap L_{l}}\left\langle \psi_{\mathrm{FS}},b_{k}(B_k e_p)\tilde{c}_{-p+l}^{\ast}\tilde{c}_{-p+k}b_{l}^{\ast}\left(K_{l}e_{p}\right)\psi_{\mathrm{FS}}\right\rangle \\
 & \quad +\sum_{l\in\mathbb{Z}_{\ast}^{3}}\sum_{p\in\left(L_{k}-k\right)\cap\left(L_{l}-l\right)}\left\langle \psi_{\mathrm{FS}},b_{k}\left(B_{k}e_{p+k}\right)\tilde{c}_{-p-l}^{\ast}\tilde{c}_{-p-k}b_{l}^{\ast}\left(K_{l}e_{p+l}\right)\psi_{\mathrm{FS}}\right\rangle \nonumber \\
 & =:A+B\nonumber 
\end{align}
where, using equation (\ref{eq:bcastCommutator}) in the form
\begin{equation}
\left[b_{l}\left(\psi\right),\tilde{c}_{p}^{\ast}\right]=\begin{cases}
-\sum_{q\in L_{l}}\delta_{p,q-l}\left\langle \psi,e_{q}\right\rangle \tilde{c}_{q} & p\in B_{F}\\
\sum_{q\in\left(L_{l}-l\right)}\delta_{p,q+l}\left\langle \psi,e_{q+l}\right\rangle \tilde{c}_{q} & p\in B_{F}^{c}
\end{cases},
\end{equation}
the terms $A$ and $B$ are given by
\begin{align}
A & =\sum_{l\in\mathbb{Z}_{\ast}^{3}}\sum_{p\in L_{k}\cap L_{l}}\left\langle \psi_{\mathrm{FS}},\left[b_{k}(B_k e_p),\tilde{c}_{-p+l}^{\ast}\right]\left[b_{l}\left(K_{l}e_{p}\right),\tilde{c}_{-p+k}^{\ast}\right]^{\ast}\psi_{\mathrm{FS}}\right\rangle \\
 & =\sum_{l\in\mathbb{Z}_{\ast}^{3}}\sum_{p\in L_{k}\cap L_{l}}\left\langle \psi_{\mathrm{FS}},\left(\sum_{q\in L_{k}}\delta_{-p+l,q-k}\left\langle B_{k}e_{p},e_{q}\right\rangle \tilde{c}_{q}\right)\left(\sum_{q'\in L_{l}}\delta_{-p+k,q'-l}\left\langle e_{q'},K_{l}e_{p}\right\rangle \tilde{c}_{q'}^{\ast}\right)\psi_{\mathrm{FS}}\right\rangle \nonumber \\
 & =\sum_{l\in\mathbb{Z}_{\ast}^{3}}\sum_{p,q\in L_{k}\cap L_{l}}\delta_{p+q,k+l}\left\langle e_{p},B_{k}e_{q}\right\rangle \left\langle e_{q},K_{l}e_{p}\right\rangle \nonumber 
\end{align}
and similarly
\begin{align}
B & =\sum_{l\in\mathbb{Z}_{\ast}^{3}}\sum_{p\in\left(L_{k}-k\right)\cap\left(L_{l}-l\right)}\left\langle \psi_{\mathrm{FS}},\left[b_{k}\left(B_{k}e_{p+k}\right),\tilde{c}_{-p-l}^{\ast}\right]\left[b_{l}\left(K_{l}e_{p+l}\right),\tilde{c}_{-p-k}^{\ast}\right]^{\ast}\psi_{\mathrm{FS}}\right\rangle \\
 & =\sum_{l\in\mathbb{Z}_{\ast}^{3}}\sum_{p,q\in\left(L_{k}-k\right)\cap\left(L_{l}-l\right)}\delta_{-p-q,k+l}\left\langle e_{p+k},B_{k}e_{q+k}\right\rangle \left\langle e_{q+l},K_{l}e_{p+l}\right\rangle .\nonumber 
\end{align}
Although non-obvious, there holds the identity $A=B$. To see this
we rewrite both terms: First, for $A$, we note that the presence
of the $\delta_{p+q,k+l}$ makes the $L_{l}$ of the summation $p,q\in L_{k}\cap L_{l}$
redundant: For any $p,q\in B_{F}^{c}$ there holds the equivalence
\begin{equation}
p,q\in L_{p+q-k}\Longleftrightarrow p,q\in L_{k}
\end{equation}
by the trivial identities
\begin{equation}
\left|p-k\right|=\left|q-\left(p+q-k\right)\right|,\quad\left|q-k\right|=\left|p-\left(p+q-k\right)\right|,
\end{equation}
so $A$ can be written as
\begin{equation}
A=\sum_{p,q\in L_{k}}\sum_{l\in\mathbb{Z}_{\ast}^{3}}\delta_{p+q,k+l}\left\langle e_{p},B_{k}e_{q}\right\rangle \left\langle e_{q},K_{l}e_{p}\right\rangle =\sum_{p,q\in L_{k}}\left\langle e_{p},B_{k}e_{q}\right\rangle \left\langle e_{q},K_{p+q-k}e_{p}\right\rangle .
\end{equation}
A similar observation applies to $B$: For any $p,q\in B_{F}$ we
likewise have
\begin{equation}
p,q\in\left(L_{-p-q-k}+p+q+k\right)\Longleftrightarrow  p+k,q+k\in L_{p+q+k}\Longleftrightarrow  p,q\in\left(L_{k}-k\right)
\end{equation}
so
\begin{align}
B & =\sum_{p,q\in\left(L_{k}-k\right)}\sum_{l\in\mathbb{Z}_{\ast}^{3}}\delta_{-p-q,k+l}\left\langle e_{p+k},B_{k}e_{q+k}\right\rangle \left\langle e_{q+l},K_{l}e_{p+l}\right\rangle \\
 & =\sum_{p,q\in\left(L_{k}-k\right)}\left\langle e_{p+k},B_{k}e_{q+k}\right\rangle \left\langle e_{-p-k},K_{-p-q-k}e_{-q-k}\right\rangle =\sum_{p,q\in L_{k}}\left\langle e_{p},B_{k}e_{q}\right\rangle \left\langle e_{q},K_{p+q-k}e_{p}\right\rangle \nonumber 
\end{align}
where we lastly used that the kernels $K_{k}$ obey
\begin{equation}
\left\langle e_{-p},K_{-k}e_{-q}\right\rangle =\left\langle e_{p},K_{k}e_{q}\right\rangle =\left\langle e_{q},K_{k}e_{p}\right\rangle ,\quad k\in\mathbb{Z}_{\ast}^{3},\,p,q\in L_{k}.
\end{equation}
In all we thus have the identity
\begin{align}
\left\langle \psi_{\mathrm{FS}},\mathcal{E}_{k}^{2}(B_{k})\psi_{\mathrm{FS}}\right\rangle  & =-\sum_{l\in\mathbb{Z}_{\ast}^{3}}\sum_{p,q\in L_{k}\cap L_{l}}\delta_{p+q,k+l}\left\langle e_{p},B_{k}e_{q}\right\rangle \left\langle e_{q},K_{l}e_{p}\right\rangle \\
 & =-\sum_{p,q\in L_{k}}\left\langle e_{p},B_{k}e_{q}\right\rangle \left\langle e_{q},K_{p+q-k}e_{p}\right\rangle .\nonumber 
\end{align}
Our matrix element estimates of the last section now yield the following:
\begin{prop*}[\ref{prop:LeadingExchangeContribution}]
It holds that
\[
\left|\sum_{k\in\mathbb{Z}_{\ast}^{3}}\int_{0}^{1}\left\langle \psi_{\mathrm{FS}},2\,\mathrm{Re}\left(\mathcal{E}_{k}^{2}(B_k(t))\right)\psi_{\mathrm{FS}}\right\rangle dt-E_{\mathrm{corr},\mathrm{ex}}\right|\leq C\sqrt{\sum_{k\in\mathbb{Z}_{\ast}^{3}}\hat{V}_{k}^{2}\min\left\{ \left|k\right|,k_{F}\right\} }
\]
for a constant $C>0$ depending only on $\sum_{k\in\mathbb{Z}_{\ast}^{3}}\hat{V}_{k}^{2}$,
where
\[
E_{\mathrm{corr},\mathrm{ex}}=\frac{k_{F}^{-2}}{4\left(2\pi\right)^{6}}\sum_{k\in\mathbb{Z}_{\ast}^{3}}\sum_{p,q\in L_{k}}\frac{\hat{V}_{k}\hat{V}_{p+q-k}}{\lambda_{k,p}+\lambda_{k,q}}.
\]
\end{prop*}
\textbf{Proof:} Since all the one-body operators are real-valued we
can drop the $\mathrm{Re}\left(\cdot\right)$ and apply the above
identity for
\begin{align}
 & \qquad\sum_{k\in\mathbb{Z}_{\ast}^{3}}\int_{0}^{1}\left\langle \psi_{\mathrm{FS}},2\,\mathrm{Re}\left(\mathcal{E}_{k}^{2}(B_k(t))\right)\psi_{\mathrm{FS}}\right\rangle dt=\sum_{k\in\mathbb{Z}_{\ast}^{3}}2\left\langle \psi_{\mathrm{FS}},\mathcal{E}_{k}^{2}\left(\int_{0}^{1}B_{k}(t)dt\right)\psi_{\mathrm{FS}}\right\rangle \\
 & =2\sum_{k,l\in\mathbb{Z}_{\ast}^{3}}\sum_{p,q\in L_{k}\cap L_{l}}\delta_{p+q,k+l}\left\langle e_{p},\left(\int_{0}^{1}B_{k}(t)dt\right)e_{q}\right\rangle \left\langle e_{q},\left(-K_{l}\right)e_{p}\right\rangle .\nonumber 
\end{align}
Now, note that $E_{\mathrm{corr},\mathrm{ex}}$ can be written as
\begin{equation}
E_{\mathrm{corr},\mathrm{ex}}=\sum_{k,l\in\mathbb{Z}_{\ast}^{3}}\sum_{p,q\in L_{k}\cap L_{l}}\delta_{p+q,k+l}\frac{\hat{V}_{k}k_{F}^{-1}}{2\left(2\pi\right)^{3}}\frac{\hat{V}_{l}k_{F}^{-1}}{2\left(2\pi\right)^{3}}\frac{1}{\lambda_{l,p}+\lambda_{l,q}}
\end{equation}
since, much as in Proposition \ref{prop:calE1KSplitEstimate}, the
$\delta_{p+q,k+l}$ implies the following identity for the denominators:
\begin{align}
\lambda_{l,p}+\lambda_{l,q} & =\frac{1}{2}\left(\left|p\right|^{2}-\left|p-l\right|^{2}\right)+\frac{1}{2}\left(\left|q\right|^{2}-\left|q-l\right|^{2}\right)\\
 & =\frac{1}{2}\left(\left|p\right|^{2}-\left|q-k\right|^{2}\right)+\frac{1}{2}\left(\left|q\right|^{2}-\left|p-k\right|^{2}\right)=\lambda_{k,p}+\lambda_{k,q}.\nonumber 
\end{align}
In conclusion we thus see that
\begin{align}
 & \qquad\sum_{k\in\mathbb{Z}_{\ast}^{3}}\int_{0}^{1}\left\langle \psi_{\mathrm{FS}},2\,\mathrm{Re}\left(\mathcal{E}_{k}^{2}(B_k(t))\right)\psi_{\mathrm{FS}}\right\rangle dt-E_{\mathrm{corr},\mathrm{ex}}\nonumber \\
 & =2\sum_{k,l\in\mathbb{Z}_{\ast}^{3}}\sum_{p,q\in L_{k}\cap L_{l}}\delta_{p+q,k+l}\left(\left\langle e_{p},\left(\int_{0}^{1}B_{k}(t)dt\right)e_{q}\right\rangle -\frac{\hat{V}_{k}k_{F}^{-1}}{4\left(2\pi\right)^{3}}\right)\left\langle e_{q},\left(-K_{l}\right)e_{p}\right\rangle \\
 & +\sum_{k,l\in\mathbb{Z}_{\ast}^{3}}\sum_{p,q\in L_{k}\cap L_{l}}\delta_{p+q,k+l}\frac{\hat{V}_{k}k_{F}^{-1}}{2\left(2\pi\right)^{3}}\left(\left\langle e_{q},\left(-K_{l}\right)e_{p}\right\rangle -\frac{\hat{V}_{l}k_{F}^{-1}}{2\left(2\pi\right)^{3}}\frac{1}{\lambda_{l,p}+\lambda_{l,q}}\right)=:A+B.\nonumber 
\end{align}
We estimate $A$ and $B$. By the matrix element estimates of Theorem
\ref{them:OneBodyEstimates} we have that (using our freedom to replace
$\lambda_{l,p}+\lambda_{l,q}$ by $\lambda_{k,p}+\lambda_{k,q}$)
\begin{align}
\left|A\right| & \leq C\sum_{k,l\in\mathbb{Z}_{\ast}^{3}}\sum_{p,q\in L_{k}\cap L_{l}}\delta_{p+q,k+l}\left(1+\hat{V}_{k}\right)\hat{V}_{k}^{2}k_{F}^{-1}\frac{\hat{V}_{l}k_{F}^{-1}}{\lambda_{l,p}+\lambda_{l,q}}\nonumber \\
 & \leq Ck_{F}^{-2}\left(1+\Vert\hat{V}\Vert_{\infty}\right)\sum_{k\in\mathbb{Z}_{\ast}^{3}}\hat{V}_{k}^{2}\sum_{p\in L_{k}}\frac{1}{\sqrt{\lambda_{k,p}}}\sum_{q\in L_{k}}\frac{\hat{V}_{p+q-k}}{\sqrt{\lambda_{k,q}}}\\
 & \leq Ck_{F}^{-\frac{3}{2}}\left(1+\Vert\hat{V}\Vert_{\infty}\right)\sqrt{\sum_{l\in\mathbb{Z}_{\ast}^{3}}\hat{V}_{l}^{2}}\sum_{k\in\mathbb{Z}_{\ast}^{3}}\hat{V}_{k}^{2}\sum_{p\in L_{k}}\frac{1}{\sqrt{\lambda_{k,p}}}\nonumber \\
 & \leq C\left(1+\Vert\hat{V}\Vert_{\infty}\right)\sqrt{\sum_{l\in\mathbb{Z}_{\ast}^{3}}\hat{V}_{l}^{2}}\sum_{k\in\mathbb{Z}_{\ast}^{3}}\hat{V}_{k}^{2}\left|k\right|^{\frac{1}{2}}\min\,\{1,k_{F}^{\frac{3}{2}}\left|k\right|^{-\frac{3}{2}}\}\nonumber 
\end{align}
where we applied the inequality $\sum_{q\in L_{k}}\lambda_{k,q}^{-1}\leq Ck_{F}$
and also used that Proposition \ref{prop:RiemannSumEstimates} implies
that
\begin{equation}
\sum_{p\in L_{k}}\frac{1}{\sqrt{\lambda_{k,p}}}\leq Ck_{F}^{\frac{3}{2}}\left|k\right|^{\frac{1}{2}}\min\,\{1,k_{F}^{\frac{3}{2}}\left|k\right|^{-\frac{3}{2}}\}
\end{equation}
for a $C>0$ independent of all quantities. By Cauchy-Schwarz we can
further estimate
\begin{align}
\sum_{k\in\mathbb{Z}_{\ast}^{3}}\hat{V}_{k}^{2}\left|k\right|^{\frac{1}{2}}\min\,\{1,k_{F}^{\frac{3}{2}}\left|k\right|^{-\frac{3}{2}}\} & \leq\sqrt{\sum_{k\in\mathbb{Z}_{\ast}^{3}}\hat{V}_{k}^{2}}\sqrt{\sum_{k\in\mathbb{Z}_{\ast}^{3}}\hat{V}_{k}^{2}\left|k\right|\min\,\{1,k_{F}^{3}\left|k\right|^{-3}\}}\\
 & \leq\sqrt{\sum_{k\in\mathbb{Z}_{\ast}^{3}}\hat{V}_{k}^{2}}\sqrt{\sum_{k\in\mathbb{Z}_{\ast}^{3}}\hat{V}_{k}^{2}\min\left\{ \left|k\right|,k_{F}\right\} }\nonumber 
\end{align}
for the bound of the statement. By similar estimation also
\[
\left|B\right|\leq C\sum_{k,l\in\mathbb{Z}_{\ast}^{3}}\sum_{p,q\in L_{k}\cap L_{l}}\delta_{p+q,k+l}\hat{V}_{k}k_{F}^{-1}\frac{\hat{V}_{l}^{2}k_{F}^{-1}}{\lambda_{l,p}+\lambda_{l,q}}\leq C\sqrt{\sum_{k\in\mathbb{Z}_{\ast}^{3}}\hat{V}_{k}^{2}}\sum_{l\in\mathbb{Z}_{\ast}^{3}}\hat{V}_{l}^{2}\left|l\right|^{\frac{1}{2}}\min\,\{1,k_{F}^{\frac{3}{2}}\left|l\right|^{-\frac{3}{2}}\}
\]
and the claim follows likewise.
$\hfill\square$

\section{\label{sec:EstimationoftheNonBosonizableTerms}Estimation of the
Non-Bosonizable Terms and Gronwall Estimates}

In this section we perform the final work which will allow us to conclude
Theorem \ref{them:MainTheorem}.

The main content of this section lies in the estimation of the \textit{non-bosonizable
terms}, by which we mean the cubic and quartic terms
\begin{align}
\mathcal{C} & =\frac{k_{F}^{-1}}{\left(2\pi\right)^{3}}\sum_{k\in\mathbb{Z}_{\ast}^{3}}\hat{V}_{k}\,\mathrm{Re}\left(\left(B_{k}+B_{-k}^{\ast}\right)^{\ast}D_{k}\right),\\
\mathcal{Q} & =\frac{k_{F}^{-1}}{2\left(2\pi\right)^{3}}\sum_{k\in\mathbb{Z}_{\ast}^{3}}\hat{V}_{k}\left(D_{k}^{\ast}D_{k}-\sum_{p\in L_{k}}\left(c_{p}^{\ast}c_{p}+c_{p-k}c_{p-k}^{\ast}\right)\right).\nonumber 
\end{align}
The cubic terms $\mathcal{C}$ will not present a big obstacle to
us: As was first noted in \cite{BNPSS-20} (in their formulation),
the expectation value of these in fact vanish identically with respect
to the type of trial state we will consider. The bulk of the work
will thus be to estimate the quartic terms. We prove the following
bounds:
\begin{thm}
\label{thm:NBTermEstimates}It holds that $\mathcal{Q}=G+\mathcal{Q}_{\mathrm{LR}}+\mathcal{Q}_{\mathrm{SR}}$
where for any $\Psi\in\mathcal{H}_{N}$
\begin{align*}
\left|\left\langle \Psi,G\Psi\right\rangle \right| & \leq C\sqrt{\sum_{k\in\mathbb{Z}_{\ast}^{3}}\hat{V}_{k}^{2}\min\left\{ \left|k\right|,k_{F}\right\} }\left\langle \Psi,\mathcal{N}_{E}\Psi\right\rangle \\
\left|\left\langle \Psi,\mathcal{Q}_{\mathrm{LR}}\Psi\right\rangle \right| & \leq C\sqrt{\sum_{k\in\mathbb{Z}_{\ast}^{3}}\hat{V}_{k}^{2}\min\left\{ \left|k\right|,k_{F}\right\} }\left\langle \Psi,\mathcal{N}_{E}^{2}\Psi\right\rangle 
\end{align*}
and $e^{\mathcal{K}}\mathcal{Q}_{\mathrm{SR}}e^{-\mathcal{K}}=\mathcal{Q}_{\mathrm{SR}}+\int_{0}^{1}e^{t\mathcal{K}}\left(2\,\mathrm{Re}\left(\mathcal{G}\right)\right)e^{-t\mathcal{K}}dt$
for an operator $\mathcal{G}$ obeying
\[
\left|\left\langle \Psi,\mathcal{G}\Psi\right\rangle \right|\leq C\sqrt{\sum_{k\in\mathbb{Z}_{\ast}^{3}}\hat{V}_{k}^{2}\min\left\{ \left|k\right|,k_{F}\right\} }\left\langle \Psi,\left(\mathcal{N}_{E}^{3}+1\right)\Psi\right\rangle ,
\]
$C>0$ being a constant independent of all quantities.
\end{thm}

With these all the general bounds are established. As all our error
estimates are with respect to $\mathcal{N}_{E}$ and powers thereof,
it then only remains to control the effect which the transformation
$e^{\mathcal{K}}$ has on these. By a standard Gronwall-type argument
this control will follow from the estimate of Proposition \ref{prop:calKTildeBound},
and we then end the paper by concluding Theorem \ref{them:MainTheorem}.

\subsubsection*{Analysis of the Cubic Terms}

Expanding the $\mathrm{Re}\left(\cdot\right)$, the cubic terms are
\begin{equation}
\mathcal{C}=\frac{k_{F}^{-1}}{2\left(2\pi\right)^{3}}\sum_{k\in\mathbb{Z}_{\ast}^{3}}\hat{V}_{k}\left(\left(B_{k}^{\ast}+B_{-k}\right)D_{k}+D_{k}^{\ast}\left(B_{k}+B_{-k}^{\ast}\right)\right).
\end{equation}
The operators $B_{k}$ can be written simply as $B_{k}=\sum_{p\in L_{k}}b_{k,p}$
in terms of the excitation operators $b_{k,p}=c_{p-k}^{\ast}c_{p}$,
whence it is easily seen that
\begin{equation}
\left[\mathcal{N}_{E},B_{k}\right]=-B_{k},\quad\left[\mathcal{N}_{E},B_{k}^{\ast}\right]=B_{k}^{\ast}.
\end{equation}
As a consequence, $B_{k}$ maps the eigenspace $\left\{ \mathcal{N}_{E}=M\right\} $
into $\left\{ \mathcal{N}_{E}=M-1\right\} $ and $B_{k}^{\ast}$ maps
$\left\{ \mathcal{N}_{E}=M\right\} $ into $\left\{ \mathcal{N}_{E}=M+1\right\} $.
Meanwhile, the operators $D_{k}$ preserve the eigenspaces: Writing
$D_{k}=D_{1,k}+D_{2,k}$ for
\begin{align}
D_{1,k} & =\mathrm{d\Gamma}\left(P_{B_{F}}e^{-ik\cdot x}P_{B_{F}}\right)=\sum_{p,q\in B_{F}}\delta_{p,q-k}c_{p}^{\ast}c_{q}=-\sum_{q\in B_{F}\cap\left(B_{F}+k\right)}\tilde{c}_{q}^{\ast}\tilde{c}_{q-k}\\
D_{2,k} & =\mathrm{d\Gamma}\left(P_{B_{F}^{c}}e^{-ik\cdot x}P_{B_{F}^{c}}\right)=\sum_{p,q\in B_{F}^{c}}\delta_{p,q-k}c_{p}^{\ast}c_{q}=\sum_{p\in B_{F}^{c}\cap\left(B_{F}^{c}-k\right)}\tilde{c}_{p}^{\ast}\tilde{c}_{p+k}\nonumber 
\end{align}
these annihilate and create one hole or excitation, respectively,
whence $\left[\mathcal{N}_{E},D_{k}\right]=0=\left[\mathcal{N}_{E},D_{k}^{\ast}\right]$.

It follows that $\mathcal{C}$ maps the eigenspace $\left\{ \mathcal{N}_{E}=M\right\} $
into $\left\{ \mathcal{N}_{E}=M-1\right\} \oplus\left\{ \mathcal{N}_{E}=M+1\right\} $.
Decomposing $\mathcal{H}_{N}$ orthogonally as $\mathcal{H}_{N}=\mathcal{H}_{N}^{\mathrm{even}}\oplus\mathcal{H}_{N}^{\mathrm{odd}}$
for
\begin{equation}
\mathcal{H}_{N}^{\mathrm{even}}=\bigoplus_{m=0}^{\infty}\left\{ \mathcal{N}_{E}=2m\right\} ,\quad\mathcal{H}_{N}^{\mathrm{odd}}=\bigoplus_{m=0}^{\infty}\left\{ \mathcal{N}_{E}=2m+1\right\} ,
\end{equation}
we thus see that $\mathcal{C}$ maps each subspace into the other.
On the other hand, since our transformation kernel $\mathcal{K}$
is of the form
\begin{equation}
\mathcal{K}=\frac{1}{2}\sum_{l\in\mathbb{Z}_{\ast}^{3}}\sum_{p,q\in L_{l}}\left\langle e_{p},K_{l}e_{q}\right\rangle \left(b_{l,p}b_{-l,-q}-b_{-l,-q}^{\ast}b_{l,p}^{\ast}\right)
\end{equation}
we note that $\mathcal{K}$ maps each $\left\{ \mathcal{N}_{E}=M\right\} $
into $\left\{ \mathcal{N}_{E}=M-2\right\} \oplus\left\{ \mathcal{N}_{E}=M+2\right\} $,
hence $\mathcal{K}$ preserves $\mathcal{H}_{N}^{\mathrm{even}}$
and $\mathcal{H}_{N}^{\mathrm{odd}}$, and so too does the transformation
$e^{-\mathcal{K}}$. As any eigenstate $\Psi\in\mathcal{H}_{N}$ of
$\mathcal{N}_{E}$ is contained in either $\mathcal{H}_{N}^{\mathrm{even}}$
or $\mathcal{H}_{N}^{\mathrm{odd}}$, and these are orthogonal, we
conclude the following:
\begin{prop}
\label{prop:VanishingofCubicTerms}For any eigenstate $\Psi$ of $\mathcal{N}_{E}$
it holds that
\[
\left\langle e^{-\mathcal{K}}\Psi,\mathcal{C}e^{-\mathcal{K}}\Psi\right\rangle =0.
\]
\end{prop}

\subsection{Analysis of the Quartic Terms}

Now we consider the quartic terms
\begin{equation}
\mathcal{Q}=\frac{k_{F}^{-1}}{2\left(2\pi\right)^{3}}\sum_{k\in\mathbb{Z}_{\ast}^{3}}\hat{V}_{k}\left(D_{k}^{\ast}D_{k}-\sum_{p\in L_{k}}\left(c_{p}^{\ast}c_{p}+c_{p-k}c_{p-k}^{\ast}\right)\right).
\end{equation}
We begin by rewriting these: Recalling the decomposition $D_{k}=D_{1,k}+D_{2,k}$
above, we calculate
\begin{align}
D_{1,k}^{\ast}D_{1,k} & =\sum_{p,q\in B_{F}\cap\left(B_{F}+k\right)}\tilde{c}_{p-k}^{\ast}\tilde{c}_{p}\tilde{c}_{q}^{\ast}\tilde{c}_{q-k}=\sum_{p,q\in B_{F}\cap\left(B_{F}+k\right)}\tilde{c}_{p-k}^{\ast}\tilde{c}_{q}^{\ast}\tilde{c}_{q-k}\tilde{c}_{p}+\sum_{q\in B_{F}\cap\left(B_{F}+k\right)}\tilde{c}_{q-k}^{\ast}\tilde{c}_{q-k}\nonumber \\
 & =\sum_{p,q\in B_{F}\cap\left(B_{F}+k\right)}\tilde{c}_{p-k}^{\ast}\tilde{c}_{q}^{\ast}\tilde{c}_{q-k}\tilde{c}_{p}+\sum_{p\in B_{F}}1_{B_{F}}(q+k)\tilde{c}_{q}^{\ast}\tilde{c}_{q}
\end{align}
and similarly
\begin{align}
D_{2,k}^{\ast}D_{2,k} & =\sum_{p,q\in B_{F}^{c}\cap\left(B_{F}^{c}-k\right)}\tilde{c}_{p+k}^{\ast}\tilde{c}_{p}\tilde{c}_{q}^{\ast}\tilde{c}_{q+k}=\sum_{p,q\in B_{F}^{c}\cap\left(B_{F}^{c}-k\right)}\tilde{c}_{p+k}^{\ast}\tilde{c}_{q}^{\ast}\tilde{c}_{q+k}\tilde{c}_{p}+\sum_{p\in B_{F}^{c}}1_{B_{F}^{c}}(p-k)\tilde{c}_{p}^{\ast}\tilde{c}_{p}\nonumber \\
 & =\sum_{p,q\in B_{F}^{c}\cap\left(B_{F}^{c}-k\right)}\tilde{c}_{p+k}^{\ast}\tilde{c}_{q}^{\ast}\tilde{c}_{q+k}\tilde{c}_{p}+\mathcal{N}_{E}-\sum_{p\in B_{F}^{c}}1_{B_{F}}(p-k)\tilde{c}_{p}^{\ast}\tilde{c}_{p}.
\end{align}
For any $k\in\mathbb{Z}_{\ast}^{3}$ we can likewise write $\sum_{p\in L_{k}}\left(c_{p}^{\ast}c_{p}+c_{p-k}c_{p-k}^{\ast}\right)$
in the form
\begin{align}
\sum_{p\in L_{k}}\left(c_{p}^{\ast}c_{p}+c_{p-k}c_{p-k}^{\ast}\right) & =\sum_{p\in B_{F}^{c}}1_{B_{F}}(p-k)\tilde{c}_{p}^{\ast}\tilde{c}_{p}+\sum_{q\in B_{F}}1_{B_{F}^{c}}(q+k)\tilde{c}_{q}^{\ast}\tilde{c}_{q}\\
 & =\sum_{p\in B_{F}^{c}}1_{B_{F}}(p-k)\tilde{c}_{p}^{\ast}\tilde{c}_{p}+\mathcal{N}_{E}-\sum_{q\in B_{F}}1_{B_{F}}(q+k)\tilde{c}_{q}^{\ast}\tilde{c}_{q}.\nonumber 
\end{align}
Noting that $D_{1,k}=0$ for $\left|k\right|>2k_{F}$, as then $B_{F}\cap\left(B_{F}+k\right)=\emptyset$,
we thus obtain the decomposition
\begin{equation}
\mathcal{Q}=G+\mathcal{Q}_{\mathrm{LR}}+\mathcal{Q}_{\mathrm{SR}}
\end{equation}
where $G$ is the one-body operator
\begin{equation}
G=\frac{k_{F}^{-1}}{\left(2\pi\right)^{3}}\sum_{k\in\mathbb{Z}_{\ast}^{3}}\hat{V}_{k}\left(\sum_{q\in B_{F}}1_{B_{F}}(q+k)\tilde{c}_{q}^{\ast}\tilde{c}_{q}-\sum_{p\in B_{F}^{c}}1_{B_{F}}(p-k)\tilde{c}_{p}^{\ast}\tilde{c}_{p}\right),
\end{equation}
the \textit{long-range terms} $\mathcal{Q}_{\mathrm{LR}}$ are given
by
\begin{equation}
\mathcal{Q}_{\mathrm{LR}}=\frac{k_{F}^{-1}}{2\left(2\pi\right)^{3}}\sum_{k\in\overline{B}\left(0,2k_{F}\right)\cap\mathbb{Z}_{\ast}^{3}}\hat{V}_{k}\left(\sum_{p,q\in B_{F}\cap\left(B_{F}+k\right)}\tilde{c}_{p-k}^{\ast}\tilde{c}_{q}^{\ast}\tilde{c}_{q-k}\tilde{c}_{p}+D_{1,k}^{\ast}D_{2,k}+D_{2,k}^{\ast}D_{1,k}\right)\label{eq:calQLRDefinition}
\end{equation}
and the \textit{short-range terms} $\mathcal{Q}_{\mathrm{SR}}$ are
\begin{equation}
\mathcal{Q}_{\mathrm{SR}}=\frac{k_{F}^{-1}}{2\left(2\pi\right)^{3}}\sum_{k\in\mathbb{Z}_{\ast}^{3}}\hat{V}_{k}\sum_{p,q\in B_{F}^{c}\cap\left(B_{F}^{c}-k\right)}\tilde{c}_{p+k}^{\ast}\tilde{c}_{q}^{\ast}\tilde{c}_{q+k}\tilde{c}_{p}.
\end{equation}

\subsubsection*{Estimation of $G$ and $\mathcal{Q}_{\mathrm{LR}}$}

$G$ and the long-range terms are easily controlled: First, interchanging
the summations we can write $G$ as
\begin{equation}
G=\frac{k_{F}^{-1}}{\left(2\pi\right)^{3}}\sum_{q\in B_{F}}\left(\sum_{k\in\left(B_{F}-q\right)\cap\mathbb{Z}_{\ast}^{3}}\hat{V}_{k}\right)\tilde{c}_{q}^{\ast}\tilde{c}_{q}-\frac{k_{F}^{-1}}{\left(2\pi\right)^{3}}\sum_{p\in B_{F}^{c}}\left(\sum_{k\in\left(B_{F}+p\right)\cap\mathbb{Z}_{\ast}^{3}}\hat{V}_{k}\right)\tilde{c}_{p}^{\ast}\tilde{c}_{p}
\end{equation}
from which it is obvious that $G$ obeys
\begin{equation}
\pm G\leq\max_{p\in\mathbb{Z}_{\ast}^{3}}\left(\frac{k_{F}^{-1}}{\left(2\pi\right)^{3}}\sum_{k\in\left(B_{F}+p\right)\cap\mathbb{Z}_{\ast}^{3}}\hat{V}_{k}\right)\mathcal{N}_{E}.
\end{equation}
This implies the following:
\begin{prop}
For any $\Psi\in\mathcal{H}_{N}$ it holds that
\[
\left|\left\langle \Psi,G\Psi\right\rangle \right|\leq C\sqrt{\sum_{k\in\mathbb{Z}_{\ast}^{3}}\hat{V}_{k}^{2}\min\left\{ \left|k\right|,k_{F}\right\} }\left\langle \Psi,\mathcal{N}_{E}\Psi\right\rangle 
\]
for a constant $C>0$ independent of all quantities.
\end{prop}

\textbf{Proof:} For any $p\in\mathbb{Z}^{3}$ we estimate by Cauchy-Schwarz
\begin{align}
\sum_{k\in\left(B_{F}+p\right)\cap\mathbb{Z}_{\ast}^{3}}\hat{V}_{k} & \leq\sqrt{\sum_{k\in\left(B_{F}+p\right)\cap\mathbb{Z}_{\ast}^{3}}\hat{V}_{k}^{2}\min\left\{ \left|k\right|,k_{F}\right\} }\sqrt{\sum_{k\in\left(B_{F}+p\right)\cap\mathbb{Z}_{\ast}^{3}}\min\left\{ \left|k\right|,k_{F}\right\} ^{-1}}\label{eq:VkSumBound}\\
 & \leq\sqrt{\sum_{k\in\mathbb{Z}_{\ast}^{3}}\hat{V}_{k}^{2}\min\left\{ \left|k\right|,k_{F}\right\} }\sqrt{\sum_{k\in B_{F}\backslash\left\{ 0\right\} }\left|k\right|^{-1}+k_{F}^{-1}}\nonumber 
\end{align}
where we lastly used that $k\mapsto\min\left\{ \left|k\right|,k_{F}\right\} ^{-1}$
is radially decreasing and that $\left(B_{F}+p\right)\cap\mathbb{Z}_{\ast}^{3}$
contains at most $\left|B_{F}\right|$ points. As it is well-known
that $\sum_{k\in\overline{B}\left(0,R\right)\backslash\left\{ 0\right\} }\left|k\right|^{-1}\leq CR^{2}$
as $R\rightarrow\infty$ the bound follows.
$\hfill\square$

\medskip

$\mathcal{Q}_{\mathrm{LR}}$ can be handled in a similar manner:
\begin{prop}
For any $\Psi\in\mathcal{H}_{N}$ it holds that
\[
\left|\left\langle \Psi,\mathcal{Q}_{\mathrm{LR}}\Psi\right\rangle \right|\leq C\sqrt{\sum_{k\in\mathbb{Z}_{\ast}^{3}}\hat{V}_{k}^{2}\min\left\{ \left|k\right|,k_{F}\right\} }\left\langle \Psi,\mathcal{N}_{E}^{2}\Psi\right\rangle 
\]
for a constant $C>0$ independent of all quantities.
\end{prop}

\textbf{Proof:} Consider the first term in the parenthesis of (\ref{eq:calQLRDefinition}): For any $k\in\mathbb{Z}_{\ast}^{3}$
we can estimate
\begin{align}
 & \sum_{p,q\in B_{F}\cap\left(B_{F}+k\right)}\left|\left\langle \Psi,\tilde{c}_{p-k}^{\ast}\tilde{c}_{q}^{\ast}\tilde{c}_{q-k}\tilde{c}_{p}\Psi\right\rangle \right|\leq\sum_{p,q\in B_{F}\cap\left(B_{F}+k\right)}\left\Vert \tilde{c}_{q}\tilde{c}_{p-k}\Psi\right\Vert \left\Vert \tilde{c}_{q-k}\tilde{c}_{p}\Psi\right\Vert \\
 & \leq\sqrt{\sum_{p,q\in B_{F}\cap\left(B_{F}+k\right)}\left\Vert \tilde{c}_{q}\tilde{c}_{p-k}\Psi\right\Vert ^{2}}\sqrt{\sum_{p,q\in B_{F}\cap\left(B_{F}+k\right)}\left\Vert \tilde{c}_{q-k}\tilde{c}_{p}\Psi\right\Vert ^{2}}\leq\left\langle \Psi,\mathcal{N}_{E}^{2}\Psi\right\rangle .\nonumber 
\end{align}
As e.g.
\[
D_{1,k}^{\ast}D_{2,k}=\sum_{p\in B_{F}^{c}\cap\left(B_{F}^{c}-k\right)}\sum_{q\in B_{F}\cap\left(B_{F}+k\right)}\tilde{c}_{q-k}^{\ast}\tilde{c}_{q}\tilde{c}_{p}^{\ast}\tilde{c}_{p+k}=\sum_{p\in B_{F}^{c}\cap\left(B_{F}^{c}-k\right)}\sum_{q\in B_{F}\cap\left(B_{F}+k\right)}\tilde{c}_{p}^{\ast}\tilde{c}_{q-k}^{\ast}\tilde{c}_{q}\tilde{c}_{p+k}
\]
the terms $D_{1,k}^{\ast}D_{2,k}$ and $D_{2,k}^{\ast}D_{1,k}$ can
be handled similarly, whence
\[
\left|\left\langle \Psi,\mathcal{Q}_{\mathrm{LR}}\Psi\right\rangle \right|\leq\frac{3k_{F}^{-1}}{2\left(2\pi\right)^{3}}\left(\sum_{k\in\overline{B}\left(0,2k_{F}\right)\cap\mathbb{Z}_{\ast}^{3}}\hat{V}_{k}\right)\left\langle \Psi,\mathcal{N}_{E}^{2}\Psi\right\rangle \leq C\sqrt{\sum_{k\in\mathbb{Z}_{\ast}^{3}}\hat{V}_{k}^{2}\min\left\{ \left|k\right|,k_{F}\right\} }\left\langle \Psi,\mathcal{N}_{E}^{2}\Psi\right\rangle 
\]
where $\sum_{k\in\overline{B}\left(0,2k_{F}\right)\cap\mathbb{Z}_{\ast}^{3}}\hat{V}_{k}$
was bounded as in equation (\ref{eq:VkSumBound}).
$\hfill\square$

\subsubsection*{Analysis of $\mathcal{Q}_{\mathrm{SR}}$}

Lastly we come to
\begin{equation}
\mathcal{Q}_{\mathrm{SR}}=\frac{k_{F}^{-1}}{2\left(2\pi\right)^{3}}\sum_{k\in\mathbb{Z}_{\ast}^{3}}\hat{V}_{k}\sum_{p,q\in B_{F}^{c}\cap\left(B_{F}^{c}-k\right)}\tilde{c}_{p+k}^{\ast}\tilde{c}_{q}^{\ast}\tilde{c}_{q+k}\tilde{c}_{p}.
\end{equation}
Recall that the transformation $\mathcal{K}$ can be written as $\mathcal{K}=\tilde{\mathcal{K}}-\tilde{\mathcal{K}}^{\ast}$
for
\begin{equation}
\tilde{\mathcal{K}}=\frac{1}{2}\sum_{l\in\mathbb{Z}_{\ast}^{3}}\sum_{p,q\in L_{l}}\left\langle e_{p},K_{l}e_{q}\right\rangle b_{l,p}b_{-l,-q}=\frac{1}{2}\sum_{l\in\mathbb{Z}_{\ast}^{3}}\sum_{q\in L_{l}}b_{l}(K_{l}e_{q})b_{-l,-q}.
\end{equation}
To determine $e^{\mathcal{K}}\mathcal{Q}_{\mathrm{SR}}e^{-\mathcal{K}}$
we will need the commutator $\left[\mathcal{K},\mathcal{Q}_{\mathrm{SR}}\right]=2\,\mathrm{Re}\left(\left[\tilde{\mathcal{K}},\mathcal{Q}_{\mathrm{SR}}\right]\right)$.
Noting that for any $p\in B_{F}^{c}$ and $l\in\mathbb{Z}_{\ast}^{3}$,
$q\in L_{l}$, we have
\begin{equation}
\left[b_{l,q},\tilde{c}_{p}^{\ast}\right]=\left[c_{q-l}^{\ast}c_{q},c_{p}^{\ast}\right]=\delta_{p,q}c_{q-l}^{\ast}=\delta_{p,q}\tilde{c}_{q-l},
\end{equation}
we deduce (with the help of Lemma \ref{lemma:TraceFormLemma}) that
\begin{align}
\left[\tilde{\mathcal{K}},\tilde{c}_{p}^{\ast}\right] & =\frac{1}{2}\sum_{l\in\mathbb{Z}_{\ast}^{3}}\sum_{q\in L_{l}}\left(b_{l}(K_{l}e_{q})\left[b_{-l,-q},\tilde{c}_{p}^{\ast}\right]+\left[b_{l}(K_{l}e_{q}),\tilde{c}_{p}^{\ast}\right]b_{-l,-q}\right)\nonumber \\
 & =\frac{1}{2}\sum_{l\in\mathbb{Z}_{\ast}^{3}}\sum_{q\in L_{l}}\left(b_{l}(K_{l}e_{q})\left[b_{-l,-q},\tilde{c}_{p}^{\ast}\right]+\left[b_{l,q},\tilde{c}_{p}^{\ast}\right]b_{-l}(K_{-l}e_{-q})\right)\label{eq:calKtildecpastCommutator}\\
 & =\frac{1}{2}\sum_{l\in\mathbb{Z}_{\ast}^{3}}\sum_{q\in L_{l}}\left(b_{l}(K_{l}e_{q})\delta_{p,-q}\tilde{c}_{-q+l}+\delta_{p,q}\tilde{c}_{q-l}b_{-l}(K_{-l}e_{-q})\right)\nonumber \\
 & =\sum_{l\in\mathbb{Z}_{\ast}^{3}}\sum_{q\in L_{l}}\delta_{p,-q}b_{l}(K_{l}e_{q})\tilde{c}_{-q+l}=\sum_{l\in\mathbb{Z}_{\ast}^{3}}1_{L_{l}}(-p)b_{l}\left(K_{l}e_{-p}\right)\tilde{c}_{p+l}.\nonumber 
\end{align}
Using this we conclude the following:
\begin{prop}
It holds that $e^{\mathcal{K}}\mathcal{Q}_{\mathrm{SR}}e^{-\mathcal{K}}=\mathcal{Q}_{\mathrm{SR}}+\int_{0}^{1}e^{t\mathcal{K}}\left(2\,\mathrm{Re}\left(\mathcal{G}\right)\right)e^{-t\mathcal{K}}dt$
for
\begin{align*}
\mathcal{G} & =\frac{k_{F}^{-1}}{\left(2\pi\right)^{3}}\sum_{k,l\in\mathbb{Z}_{\ast}^{3}}\hat{V}_{k}\sum_{p,q\in B_{F}^{c}\cap\left(B_{F}^{c}+k\right)}1_{L_l}(q)\tilde{c}_{p}^{\ast}b_{l}(K_{l}e_{q})\tilde{c}_{-q+l}\tilde{c}_{-q+k}\tilde{c}_{p-k}\\
 & +\frac{k_{F}^{-1}}{2\left(2\pi\right)^{3}}\sum_{k,l\in\mathbb{Z}_{\ast}^{3}}\hat{V}_{k}\sum_{p,q\in B_{F}^{c}\cap\left(B_{F}^{c}+k\right)}1_{L_l}(p)1_{L_l}(q)\left\langle K_{l}e_{q},e_{p}\right\rangle \tilde{c}_{p-l}\tilde{c}_{-q+l}\tilde{c}_{-q+k}\tilde{c}_{p-k}.
\end{align*}
\end{prop}

\textbf{Proof:} By the fundamental theorem of calculus
\begin{equation}
e^{\mathcal{K}}\mathcal{Q}_{\mathrm{SR}}e^{-\mathcal{K}}=\mathcal{Q}_{\mathrm{SR}}+\int_{0}^{1}e^{t\mathcal{K}}\left[\mathcal{K},\mathcal{Q}_{\mathrm{SR}}\right]e^{-t\mathcal{K}}dt
\end{equation}
and as noted $\left[\mathcal{K},\mathcal{Q}_{\mathrm{SR}}\right]=2\,\mathrm{Re}\left(\left[\tilde{\mathcal{K}},\mathcal{Q}_{\mathrm{SR}}\right]\right)$.
Using equation (\ref{eq:calKtildecpastCommutator}) we compute that
$\mathcal{G}:=\left[\tilde{\mathcal{K}},\mathcal{Q}_{\mathrm{SR}}\right]$
is given by
\begin{align}
\mathcal{G} & =\frac{k_{F}^{-1}}{2\left(2\pi\right)^{3}}\sum_{k\in\mathbb{Z}_{\ast}^{3}}\hat{V}_{k}\sum_{p\in B_{F}^{c}\cap\left(B_{F}^{c}+k\right)}\sum_{q\in B_{F}^{c}\cap\left(B_{F}^{c}-k\right)}\left(\tilde{c}_{p}^{\ast}\left[\tilde{\mathcal{K}},\tilde{c}_{q}^{\ast}\right]+\left[\tilde{\mathcal{K}},\tilde{c}_{p}^{\ast}\right]\tilde{c}_{q}^{\ast}\right)\tilde{c}_{q+k}\tilde{c}_{p-k}\nonumber \\
 & =\frac{k_{F}^{-1}}{2\left(2\pi\right)^{3}}\sum_{k,l\in\mathbb{Z}_{\ast}^{3}}\hat{V}_{k}\sum_{p\in B_{F}^{c}\cap\left(B_{F}^{c}+k\right)}\sum_{q\in B_{F}^{c}\cap\left(B_{F}^{c}-k\right)}1_{L_{l}}(-q)\tilde{c}_{p}^{\ast}b_{l}\left(K_{l}e_{-q}\right)\tilde{c}_{q+l}\tilde{c}_{q+k}\tilde{c}_{p-k}\nonumber \\
 & +\frac{k_{F}^{-1}}{2\left(2\pi\right)^{3}}\sum_{k,l\in\mathbb{Z}_{\ast}^{3}}\hat{V}_{k}\sum_{p\in B_{F}^{c}\cap\left(B_{F}^{c}+k\right)}\sum_{q\in B_{F}^{c}\cap\left(B_{F}^{c}-k\right)}1_{L_{l}}(-p)b_{l}\left(K_{l}e_{-p}\right)\tilde{c}_{p+l}\tilde{c}_{q}^{\ast}\tilde{c}_{q+k}\tilde{c}_{p-k}\\
 & =\frac{k_{F}^{-1}}{2\left(2\pi\right)^{3}}\sum_{k,l\in\mathbb{Z}_{\ast}^{3}}\hat{V}_{k}\sum_{p\in B_{F}^{c}\cap\left(B_{F}^{c}+k\right)}\sum_{q\in B_{F}^{c}\cap\left(B_{F}^{c}-k\right)}1_{L_{l}}(-q)\left\{ b_{l}\left(K_{l}e_{-q}\right),\tilde{c}_{p}^{\ast}\right\} \tilde{c}_{q+l}\tilde{c}_{q+k}\tilde{c}_{p-k}\nonumber \\
 & =\frac{k_{F}^{-1}}{2\left(2\pi\right)^{3}}\sum_{k,l\in\mathbb{Z}_{\ast}^{3}}\hat{V}_{k}\sum_{p,q\in B_{F}^{c}\cap\left(B_{F}^{c}+k\right)}1_{L_l}(q)\left\{ b_{l}(K_{l}e_{q}),\tilde{c}_{p}^{\ast}\right\} \tilde{c}_{-q+l}\tilde{c}_{-q+k}\tilde{c}_{p-k},\nonumber 
\end{align}
where we for the third inequality substituted $p\rightarrow  q$
and $k\rightarrow-k$ in the second sum. By the identity of equation
(\ref{eq:bcastCommutator}) the anti-commutator is given by
\begin{equation}
\left\{ b_{l}(K_{l}e_{q}),\tilde{c}_{p}^{\ast}\right\} =2\,\tilde{c}_{p}^{\ast}b_{l}(K_{l}e_{q})+1_{L_l}(p)\left\langle K_{l}e_{q},e_{p}\right\rangle \tilde{c}_{p-l}
\end{equation}
which is inserted into the previous equation for the claim.
$\hfill\square$

\medskip

We bound the $\mathcal{G}$ operator as follows:
\begin{prop}
For any $\Psi\in\mathcal{H}_{N}$ it holds that
\[
\left|\left\langle \Psi,\mathcal{G}\Psi\right\rangle \right|\leq C\sqrt{\sum_{k\in\mathbb{Z}_{\ast}^{3}}\hat{V}_{k}^{2}\min\left\{ \left|k\right|,k_{F}\right\} }\left\langle \Psi,\left(\mathcal{N}_{E}^{3}+1\right)\Psi\right\rangle 
\]
for a constant $C>0$ depending only on $\sum_{k\in\mathbb{Z}_{\ast}^{3}}\hat{V}_{k}^{2}$.
\end{prop}

\textbf{Proof:} Using Proposition \ref{prop:bbastEstimates} we estimate
the sum of the first term of $\mathcal{G}$ as
\begin{align}
 & \quad\,\sum_{k,l\in\mathbb{Z}_{\ast}^{3}}\hat{V}_{k}\sum_{p,q\in B_{F}^{c}\cap\left(B_{F}^{c}+k\right)}1_{L_l}(q)\left|\left\langle \Psi,\tilde{c}_{p}^{\ast}b_{l}(K_{l}e_{q})\tilde{c}_{-q+l}\tilde{c}_{-q+k}\tilde{c}_{p-k}\Psi\right\rangle \right|\nonumber \\
 & \leq\sum_{k,l\in\mathbb{Z}_{\ast}^{3}}\hat{V}_{k}\sum_{p,q\in B_{F}^{c}\cap\left(B_{F}^{c}+k\right)}1_{L_l}(q)\left\Vert b_{l}^{\ast}(K_{l}e_{q})\tilde{c}_{p}\Psi\right\Vert \left\Vert \tilde{c}_{-q+l}\tilde{c}_{-q+k}\tilde{c}_{p-k}\Psi\right\Vert \nonumber \\
 & \leq\sum_{k,l\in\mathbb{Z}_{\ast}^{3}}\hat{V}_{k}\sum_{p,q\in B_{F}^{c}\cap\left(B_{F}^{c}+k\right)}1_{L_l}(q)\left\Vert K_{l}e_{q}\right\Vert \Vert\tilde{c}_{p}\left(\mathcal{N}_{E}+1\right)^{\frac{1}{2}}\Psi\Vert\left\Vert \tilde{c}_{p-k}\tilde{c}_{-q+l}\tilde{c}_{-q+k}\Psi\right\Vert \nonumber \\
 & \leq\left\Vert \left(\mathcal{N}_{E}+1\right)\Psi\right\Vert \sum_{l\in\mathbb{Z}_{\ast}^{3}}\sum_{q\in L_{l}}\left\Vert K_{l}e_{q}\right\Vert \sum_{k\in\mathbb{Z}_{\ast}^{3}}1_{B_{F}^{c}+k}(q)\hat{V}_{k}\Vert\tilde{c}_{-q+k}\tilde{c}_{-q+l}\mathcal{N}_{E}^{\frac{1}{2}}\Psi\Vert\\
 & \leq\sqrt{\sum_{k\in\mathbb{Z}_{\ast}^{3}}\hat{V}_{k}^{2}}\left\Vert \left(\mathcal{N}_{E}+1\right)\Psi\right\Vert \sum_{l\in\mathbb{Z}_{\ast}^{3}}\sum_{q\in L_{l}}\left\Vert K_{l}e_{q}\right\Vert \left\Vert \tilde{c}_{-q+l}\mathcal{N}_{E}\Psi\right\Vert \nonumber \\
 & \leq\sqrt{\sum_{k\in\mathbb{Z}_{\ast}^{3}}\hat{V}_{k}^{2}}\left(\sum_{l\in\mathbb{Z}_{\ast}^{3}}\left\Vert K_{l}\right\Vert _{\mathrm{HS}}\right)\left\Vert \left(\mathcal{N}_{E}+1\right)\Psi\right\Vert \Vert\mathcal{N}_{E}^{\frac{3}{2}}\Psi\Vert.\nonumber 
\end{align}
Now, the $\left\Vert K_{k}\right\Vert _{\mathrm{HS}}$ estimate of
Theorem \ref{them:OneBodyEstimates} and Cauchy-Schwarz lets us estimate
\[
\sum_{k\in\mathbb{Z}_{\ast}^{3}}\left\Vert K_{k}\right\Vert _{\mathrm{HS}}\leq C\sum_{k\in\mathbb{Z}_{\ast}^{3}}\hat{V}_{k}\min\,\{1,k_{F}^{2}\left|k\right|^{-2}\}\leq C\sqrt{\sum_{k\in\mathbb{Z}_{\ast}^{3}}\frac{\min\,\{1,k_{F}^{4}\left|k\right|^{-4}\}}{\min\left\{ \left|k\right|,k_{F}\right\} }}\sqrt{\sum_{k\in\mathbb{Z}_{\ast}^{3}}\hat{V}_{k}^{2}\min\left\{ \left|k\right|,k_{F}\right\} },
\]
and
\begin{equation}
\sum_{k\in\mathbb{Z}_{\ast}^{3}}\frac{\min\,\{1,k_{F}^{4}\left|k\right|^{-4}\}}{\min\left\{ \left|k\right|,k_{F}\right\} }=\sum_{k\in B_{F}\backslash\left\{ 0\right\} }\frac{1}{\left|k\right|}+k_{F}^{3}\sum_{k\in\mathbb{Z}_{\ast}^{3}\backslash B_{F}}\frac{1}{\left|k\right|^{4}}\leq Ck_{F}^{2}
\end{equation}
for a constant $C>0$ independent of all quantities, so in all the
first term of $\mathcal{G}$ obeys
\begin{align}
 & \;\frac{k_{F}^{-1}}{2\left(2\pi\right)^{3}}\sum_{k,l\in\mathbb{Z}_{\ast}^{3}}\hat{V}_{k}\sum_{p,q\in B_{F}^{c}\cap\left(B_{F}^{c}+k\right)}1_{L_l}(q)\left|\left\langle \Psi,\tilde{c}_{p}^{\ast}b_{l}(K_{l}e_{q})\tilde{c}_{-q+l}\tilde{c}_{-q+k}\tilde{c}_{p-k}\Psi\right\rangle \right|\\
 & \leq C\sqrt{\sum_{k\in\mathbb{Z}_{\ast}^{3}}\hat{V}_{k}^{2}}\sqrt{\sum_{k\in\mathbb{Z}_{\ast}^{3}}\hat{V}_{k}^{2}\min\left\{ \left|k\right|,k_{F}\right\} }\left\Vert \left(\mathcal{N}_{E}+1\right)\Psi\right\Vert \Vert\mathcal{N}_{E}^{\frac{3}{2}}\Psi\Vert.\nonumber 
\end{align}
Similarly, for the second term (using simply that $\left\Vert \tilde{c}_{p-l}\right\Vert _{\mathrm{Op}}=1$
at the beginning)
\begin{align}
 & \quad\sum_{k,l\in\mathbb{Z}_{\ast}^{3}}\hat{V}_{k}\sum_{p,q\in B_{F}^{c}\cap\left(B_{F}^{c}+k\right)}1_{L_l}(p)1_{L_l}(q)\left|\left\langle K_{l}e_{q},e_{p}\right\rangle \left\langle \Psi,\tilde{c}_{p-l}\tilde{c}_{-q+l}\tilde{c}_{-q+k}\tilde{c}_{p-k}\Psi\right\rangle \right|\nonumber \\
 & \leq\left\Vert \Psi\right\Vert \sum_{k,l\in\mathbb{Z}_{\ast}^{3}}\hat{V}_{k}\sum_{p,q\in B_{F}^{c}\cap\left(B_{F}^{c}+k\right)}1_{L_l}(p)1_{L_l}(q)\left|\left\langle K_{l}e_{q},e_{p}\right\rangle \right|\left\Vert \tilde{c}_{p-k}\tilde{c}_{-q+l}\tilde{c}_{-q+k}\Psi\right\Vert \\
 & \leq\left\Vert \Psi\right\Vert \sum_{l\in\mathbb{Z}_{\ast}^{3}}\sum_{q\in L_{l}}\left\Vert K_{l}e_{q}\right\Vert \sum_{k\in\mathbb{Z}_{\ast}^{3}}1_{B_{F}^{c}+k}(q)\hat{V}_{k}\Vert\tilde{c}_{-q+k}\tilde{c}_{-q+l}\mathcal{N}_{E}^{\frac{1}{2}}\Psi\Vert\nonumber \\
 & \leq\sqrt{\sum_{k\in\mathbb{Z}_{\ast}^{3}}\hat{V}_{k}^{2}}\left(\sum_{l\in\mathbb{Z}_{\ast}^{3}}\left\Vert K_{l}\right\Vert _{\mathrm{HS}}\right)\left\Vert \Psi\right\Vert \Vert\mathcal{N}_{E}^{\frac{3}{2}}\Psi\Vert.\nonumber \tag*{$\square$}
\end{align}

\subsection{Gronwall Estimates}

We now establish control over the operators $e^{\mathcal{K}}\mathcal{N}_{E}^{m}e^{-\mathcal{K}}$
for $m=1,2,3$. Consider first the mapping $t\mapsto e^{t\mathcal{K}}\mathcal{N}_{E}e^{-t\mathcal{K}}$:
Noting that for any $\Psi\in\mathcal{H}_{N}$
\begin{equation}
\frac{d}{dt}\left\langle \Psi,e^{t\mathcal{K}}\left(\mathcal{N}_{E}+1\right)e^{-t\mathcal{K}}\Psi\right\rangle =\left\langle \Psi,e^{-t\mathcal{K}}\left[\mathcal{K},\mathcal{N}_{E}\right]e^{-t\mathcal{K}}\Psi\right\rangle ,
\end{equation}
Gronwall's lemma implies that to bound $e^{t\mathcal{K}}\left(\mathcal{N}_{E}+1\right)e^{-t\mathcal{K}}$
it suffices to control $\left[\mathcal{K},\mathcal{N}_{E}\right]$
with respect to $\mathcal{N}_{E}+1$ itself. We determine the commutator:
As $\mathcal{K}=\mathcal{\tilde{K}}-\mathcal{\tilde{K}}^{\ast}$ for
\begin{equation}
\tilde{\mathcal{K}}=\frac{1}{2}\sum_{l\in\mathbb{Z}_{\ast}^{3}}\sum_{p,q\in L_{l}}\left\langle e_{p},K_{l}e_{q}\right\rangle b_{l,p}b_{-l,-q}
\end{equation}
and $\left[b_{l,p},\mathcal{N}_{E}\right]=b_{l,p}$ it holds that
$\left[\tilde{\mathcal{K}},\mathcal{N}_{E}\right]=2\,\tilde{\mathcal{K}}$,
whence
\begin{equation}
\left[\mathcal{K},\mathcal{N}_{E}\right]=2\,\mathrm{Re}\left(\left[\tilde{\mathcal{K}},\mathcal{N}_{E}\right]\right)=2\,\tilde{\mathcal{K}}+2\,\tilde{\mathcal{K}}^{\ast}.
\end{equation}
The estimate of Proposition \ref{prop:calKTildeBound} immediately
yields that
\begin{equation}
\pm\left[\mathcal{K},\mathcal{N}_{E}\right]\leq C\left(\mathcal{N}_{E}+1\right)\label{eq:calKNeCommutatorBound}
\end{equation}
for a constant $C>0$ depending only on $\sum_{k\in\mathbb{Z}_{\ast}^{3}}\hat{V}_{k}^{2}$,
whence by Gronwall's lemma
\begin{equation}
\left\langle \Psi,e^{t\mathcal{K}}\left(\mathcal{N}_{E}+1\right)e^{-t\mathcal{K}}\Psi\right\rangle \leq e^{C\left|t\right|}\left\langle \Psi,\left(\mathcal{N}_{E}+1\right)\Psi\right\rangle \leq C'\left\langle \Psi,\left(\mathcal{N}_{E}+1\right)\Psi\right\rangle ,\quad\left|t\right|\leq1.
\end{equation}
This proves the bound for $\mathcal{N}_{E}$; for $\mathcal{N}_{E}^{2}$
we will as in \cite{CHN-21} apply the following lemma:
\begin{lem}
\label{lemma:CommutatorRootEstimate}Let $A,B,Z$ be given with $A>0$,
$Z\geq0$ and $\left[A,Z\right]=0$. Then if $\pm\left[A,\left[A,B\right]\right]\leq Z$
it holds that
\[
\pm[A^{\frac{1}{2}},[A^{\frac{1}{2}},B]]\leq\frac{1}{4}A^{-1}Z.
\]
\end{lem}

The estimates are as follows:
\begin{prop}
\label{prop:GronwallEstimates}For any $\Psi\in\mathcal{H}_{N}$ and
$\left|t\right|\leq1$ it holds that
\[
\left\langle e^{-t\mathcal{K}}\Psi,\left(\mathcal{N}_{E}^{m}+1\right)e^{-t\mathcal{K}}\Psi\right\rangle \leq C\left\langle \Psi,\left(\mathcal{N}_{E}^{m}+1\right)\Psi\right\rangle ,\quad m=1,2,3,
\]
for a constant $C>0$ depending only on $\sum_{k\in\mathbb{Z}_{\ast}^{3}}\hat{V}_{k}^{2}$.
\end{prop}

\textbf{Proof:} The case of $m=1$ was proved above. For $m=2$ it
suffices to control $\left[\mathcal{K},\mathcal{N}_{E}^{2}\right]$
in terms of $\mathcal{N}_{E}^{2}+1$; by the identity $\left\{ A,B\right\} =A^{\frac{1}{2}}BA^{\frac{1}{2}}+[A^{\frac{1}{2}},[A^{\frac{1}{2}},B]]$
we can write
\begin{align}
\left[\mathcal{K},\mathcal{N}_{E}^{2}\right] & =\left\{ \mathcal{N}_{E},\left[\mathcal{K},\mathcal{N}_{E}\right]\right\} =\left\{ \mathcal{N}_{E}+1,\left[\mathcal{K},\mathcal{N}_{E}\right]\right\} -2\left[\mathcal{K},\mathcal{N}_{E}\right]\\
 & =\left(\mathcal{N}_{E}+1\right)^{\frac{1}{2}}\left[\mathcal{K},\mathcal{N}_{E}\right]\left(\mathcal{N}_{E}+1\right)^{\frac{1}{2}}+[\left(\mathcal{N}_{E}+1\right)^{\frac{1}{2}},[\left(\mathcal{N}_{E}+1\right)^{\frac{1}{2}},\left[\mathcal{K},\mathcal{N}_{E}\right]]]-2\left[\mathcal{K},\mathcal{N}_{E}\right]\nonumber 
\end{align}
and note that the commutator $\left[\tilde{\mathcal{K}},\mathcal{N}_{E}\right]=2\,\tilde{\mathcal{K}}$
also implies that
\begin{equation}
\left[\mathcal{N}_{E},\left[\mathcal{N}_{E},\left[\mathcal{K},\mathcal{N}_{E}\right]\right]\right]=4\left[\mathcal{K},\mathcal{N}_{E}\right],
\end{equation}
so by Lemma \ref{lemma:CommutatorRootEstimate} and equation (\ref{eq:calKNeCommutatorBound})
\begin{equation}
\pm\left[\mathcal{K},\mathcal{N}_{E}^{2}\right]\leq C\left(\left(\mathcal{N}_{E}+1\right)^{2}+1+\left(\mathcal{N}_{E}+1\right)\right)\leq C'\left(\mathcal{N}_{E}^{2}+1\right).
\end{equation}
Similarly, for $\mathcal{N}_{E}^{3}$,
\begin{equation}
\left[\mathcal{K},\mathcal{N}_{E}^{3}\right]=3\,\mathcal{N}_{E}\left[\mathcal{K},\mathcal{N}_{E}\right]\mathcal{N}_{E}+\left[\mathcal{N}_{E},\left[\mathcal{N}_{E},\left[\mathcal{K},\mathcal{N}_{E}\right]\right]\right]=3\,\mathcal{N}_{E}\left[\mathcal{K},\mathcal{N}_{E}\right]\mathcal{N}_{E}+4\left[\mathcal{K},\mathcal{N}_{E}\right]
\end{equation}
implies that
\begin{equation}
\pm\left[\mathcal{K},\mathcal{N}_{E}^{3}\right]\leq C\left(\mathcal{N}_{E}\left(\mathcal{N}_{E}+1\right)\mathcal{N}_{E}+\left(\mathcal{N}_{E}+1\right)\right)\leq C'\left(\mathcal{N}_{E}^{3}+1\right)
\end{equation}
hence the $m=3$ bound.
$\hfill\square$

\subsubsection*{Conclusion of Theorem \ref{them:MainTheorem}}

We can now conclude:
\begin{thm*}[\ref{them:MainTheorem}]
It holds that
\[
\inf\sigma\left(H_{N}\right)\leq E_{F}+E_{\mathrm{corr},\mathrm{bos}}+E_{\mathrm{corr},\mathrm{ex}}+C\sqrt{\sum_{k\in\mathbb{Z}_{\ast}^{3}}\hat{V}_{k}^{2}\min\left\{ \left|k\right|,k_{F}\right\} },\quad k_{F}\rightarrow\infty,
\]
for a constant $C>0$ depending only on $\sum_{k\in\mathbb{Z}_{\ast}^{3}}\hat{V}_{k}^{2}$.
\end{thm*}
\textbf{Proof:} By the variational principle applied to the trial
state $e^{-\mathcal{K}}\psi_{\mathrm{FS}}$ we have by Proposition
\ref{prop:LocalizationofHN} and the Theorems \ref{thm:DiagonalizationoftheBosonizableTerms},
\ref{them:OneBodyEstimates} and \ref{thm:NBTermEstimates} that
\begin{align}
 & \inf\sigma\left(H_{N}\right)\leq E_{F}+\left\langle \psi_{\mathrm{FS}},e^{\mathcal{K}}\left(H_{\mathrm{kin}}^{\prime}+\sum_{k\in\mathbb{Z}_{\ast}^{3}}\frac{\hat{V}_{k}k_{F}^{-1}}{2\left(2\pi\right)^{3}}\left(2B_{k}^{\ast}B_{k}+B_{k}B_{-k}+B_{-k}^{\ast}B_{k}^{\ast}\right)\right)e^{-\mathcal{K}}\psi_{\mathrm{FS}}\right\rangle \nonumber \\
 & \qquad\qquad\qquad\qquad\qquad\qquad\qquad\qquad\qquad\quad+\left\langle \psi_{\mathrm{FS}},e^{\mathcal{K}}\mathcal{C}e^{-\mathcal{K}}\psi_{\mathrm{FS}}\right\rangle +\left\langle \psi_{\mathrm{FS}},e^{\mathcal{K}}\mathcal{Q}e^{-\mathcal{K}}\psi_{\mathrm{FS}}\right\rangle \nonumber \\
 & =E_{F}+E_{\mathrm{corr},\mathrm{bos}}+\left\langle \psi_{\mathrm{FS}},H_{\mathrm{kin}}^{\prime}\psi_{\mathrm{FS}}\right\rangle +2\sum_{k\in\mathbb{Z}_{\ast}^{3}}\left\langle \psi_{\mathrm{FS}},Q_{1}^{k}\left(e^{-K_{k}}h_{k}e^{-K_{k}}-h_{k}\right)\psi_{\mathrm{FS}}\right\rangle \\
 & +\sum_{k\in\mathbb{Z}_{\ast}^{3}}\int_{0}^{1}\left\langle e^{-\left(1-t\right)\mathcal{K}}\psi_{\mathrm{FS}},\left(\varepsilon_{k}(\left\{ K_{k},B_{k}(t)\right\})+2\,\mathrm{Re}\left(\mathcal{E}_{k}^{1}(A_k(t))\right)+2\,\mathrm{Re}\left(\mathcal{E}_{k}^{2}(B_k(t))\right)\right)e^{-\left(1-t\right)\mathcal{K}}\psi_{\mathrm{FS}}\right\rangle dt\nonumber \\
 & +\left\langle e^{\mathcal{K}}\psi_{\mathrm{FS}},\left(G+\mathcal{Q}_{\mathrm{LR}}\right)e^{-\mathcal{K}}\psi_{\mathrm{FS}}\right\rangle +\left\langle \psi_{\mathrm{FS}},\mathcal{Q}_{\mathrm{SR}}\psi_{\mathrm{FS}}\right\rangle +\int_{0}^{1}\left\langle e^{-t\mathcal{K}}\psi_{\mathrm{FS}},\left(2\,\mathrm{Re}\left(\mathcal{G}\right)\right)e^{-t\mathcal{K}}\psi_{\mathrm{FS}}\right\rangle dt\nonumber \\
 & =E_{F}+E_{\mathrm{corr},\mathrm{bos}}+E_{\mathrm{corr},\mathrm{ex}}+\epsilon_{1}+\epsilon_{2}+\epsilon_{3},\nonumber 
\end{align}
where we also used that
\begin{equation}
H_{\mathrm{kin}}^{\prime}\psi_{\mathrm{FS}}=Q_{1}^{k}(A)\psi_{\mathrm{FS}}=\mathcal{Q}_{\mathrm{SR}}\psi_{\mathrm{FS}}=0
\end{equation}
and that $\left\langle \psi_{\mathrm{FS}},e^{\mathcal{K}}\mathcal{C}e^{-\mathcal{K}}\psi_{\mathrm{FS}}\right\rangle =0$
by Proposition \ref{prop:VanishingofCubicTerms}. The errors $\epsilon_{1}$,
$\epsilon_{2}$ and $\epsilon_{3}$ obey
\begin{equation}
\epsilon_{1}=\sum_{k\in\mathbb{Z}_{\ast}^{3}}\int_{0}^{1}\left\langle \psi_{\mathrm{FS}},2\,\mathrm{Re}\left(\mathcal{E}_{k}^{2}(B_k(t))\right)\psi_{\mathrm{FS}}\right\rangle dt-E_{\mathrm{corr},\mathrm{ex}}\leq C\sum_{k\in\mathbb{Z}_{\ast}^{3}}\sqrt{\sum_{k\in\mathbb{Z}_{\ast}^{3}}\hat{V}_{k}^{2}\min\left\{ \left|k\right|,k_{F}\right\} }
\end{equation}
by Proposition \ref{prop:LeadingExchangeContribution},
\begin{align}
\epsilon_{2} & =\sum_{k\in\mathbb{Z}_{\ast}^{3}}\int_{0}^{1}\left\langle e^{-\left(1-t\right)\mathcal{K}}\psi_{\mathrm{FS}},\left(\varepsilon_{k}(\left\{ K_{k},B_{k}(t)\right\})+2\,\mathrm{Re}\left(\mathcal{E}_{k}^{1}(A_k(t))\right)\right)e^{-\left(1-t\right)\mathcal{K}}\psi_{\mathrm{FS}}\right\rangle dt\nonumber \\
 & +\sum_{k\in\mathbb{Z}_{\ast}^{3}}\int_{0}^{1}\left\langle e^{-\left(1-t\right)\mathcal{K}}\psi_{\mathrm{FS}},\left(2\,\mathrm{Re}\left(\mathcal{E}_{k}^{2}(B_k(t))-\left\langle \psi_{F},\mathcal{E}_{k}^{2}(B_k(t))\psi_{F}\right\rangle \right)\right)e^{-\left(1-t\right)\mathcal{K}}\psi_{\mathrm{FS}}\right\rangle dt\\
 & \leq Ck_{F}^{-1}+C\sqrt{\sum_{k\in\mathbb{Z}_{\ast}^{3}}\hat{V}_{k}^{2}\min\left\{ \left|k\right|,k_{F}\right\} }\leq C'\sqrt{\sum_{k\in\mathbb{Z}_{\ast}^{3}}\hat{V}_{k}^{2}\min\left\{ \left|k\right|,k_{F}\right\} }\nonumber 
\end{align}
by Theorem \ref{them:ExchangeTermsEstimates}, and
\begin{align}
\epsilon_{3} & =\left\langle e^{-\mathcal{K}}\psi_{\mathrm{FS}},\left(G+\mathcal{Q}_{\mathrm{LR}}\right)e^{-\mathcal{K}}\psi_{\mathrm{FS}}\right\rangle +\int_{0}^{1}\left\langle e^{-t\mathcal{K}}\psi_{\mathrm{FS}},\left(2\,\mathrm{Re}\left(\mathcal{G}\right)\right)e^{-t\mathcal{K}}\psi_{\mathrm{FS}}\right\rangle dt\\
 & \leq C\sqrt{\sum_{k\in\mathbb{Z}_{\ast}^{3}}\hat{V}_{k}^{2}\min\left\{ \left|k\right|,k_{F}\right\} }\nonumber 
\end{align}
by Theorem \ref{thm:NBTermEstimates}, where we for the last error
terms also used that
\begin{equation}
\left\langle e^{-t\mathcal{K}}\psi_{\mathrm{FS}},\left(\mathcal{N}_{E}^{m}+1\right)e^{-t\mathcal{K}}\psi_{\mathrm{FS}}\right\rangle \leq C,\quad\left|t\right|\leq1,\,m=1,2,3,
\end{equation}
as follows by Proposition \ref{prop:GronwallEstimates}.
$\hfill\square$

\appendix

\section{\label{sec:DiagonalizationoftheBosonizableTerms}Diagonalization
of the Bosonizable Terms}

In this section we derive the identity
of Theorem \ref{thm:DiagonalizationoftheBosonizableTerms}. This is
to a degree equivalent with the contents of Section 5 of \cite{CHN-21},
but for the reader's convenience, and since the notation used in the
papers differ, we include a brief derivation in this appendix.


To determine the action of $e^{\mathcal{K}}$, 
we must first compute several commutators involving $\mathcal{K}$.
To simplify the calculations we will make repeated use of the following
result (\cite[Lemma 3.2]{CHN-21}). 
\begin{lem}
\label{lemma:TraceFormLemma}Let $\left(V,\left\langle \cdot,\cdot\right\rangle \right)$
be an $n$-dimensional Hilbert space and let $q:V\times V\rightarrow W$
be a sesquilinear mapping into a vector space $W$. Let $\left(e_{i}\right)_{i=1}^{N}$
be an orthonormal basis for $V$. Then for any linear operators $S,T:V\rightarrow V$
it holds that
\[
\sum_{i=1}^{n}q\left(Se_{i},Te_{i}\right)=\sum_{i=1}^{n}q\left(ST^{\ast}e_{i},e_{i}\right).
\]
\end{lem}

The lemma is easily proved by orthonormal expansion. In our case, 
where we regard $\ell^{2}(L_{k})$ as real vector
spaces, sesquilinearity is simply bilinearity.
Moreover, the operators $K_{k}$ satisfy 
%
\begin{equation}
I_{k}K_{k}=K_{-k}I_{k}
\end{equation}
where $I_{k}:\ell^{2}(L_{k})\rightarrow\ell^{2}\left(L_{-k}\right)$
denotes the unitary mapping defined by $I_{k}e_{p}=e_{-p}$, $p\in L_{k}$.
Thus Lemma \ref{lemma:TraceFormLemma} allows us to move operators from one argument
to another (when summed), as e.g.
\begin{align}
\sum_{q\in L_{l}}b_{l}(K_{l}e_{q})b_{-l,-q}  
=\sum_{q\in L_{l}}b_{l}(K_{l}e_{q})b_{-l}\left(I_{l}e_{q}\right)
  =\sum_{q\in L_{l}}b_{l}\left(e_{q}\right)b_{-l}\left(I_{l}K_{l}^{\ast}e_{q}\right)
 =\sum_{q\in L_{l}}b_{l,q}b_{-l}(K_{-l}e_{-q}).
\end{align}

We start by computing the commutator of $\mathcal{K}$ with an excitation
operator:
\begin{prop}
\label{prop:calKExcitationCommutator}For any $k\in\mathbb{Z}_{\ast}^{3}$
and $\varphi\in\ell^{2}(L_{k})$ it holds that
\begin{align*}
\left[\mathcal{K},b_{k}(\varphi)\right]  =b_{-k}^{\ast}\left(I_{k}K_{k}\varphi\right)+\mathcal{E}_{k}(\varphi),\quad \left[\mathcal{K},b_{k}^{\ast}(\varphi)\right] & =b_{-k}\left(I_{k}K_{k}\varphi\right)+\mathcal{E}_{k}(\varphi)^{\ast}
\end{align*}
where
\[
\mathcal{E}_{k}(\varphi)=\frac{1}{2}\sum_{l\in\mathbb{Z}_{\ast}^{3}}\sum_{q\in L_{l}}\left\{ \varepsilon_{k,l}\left(\varphi;e_{q}\right),b_{-l}^{\ast}(K_{-l}e_{-q})\right\} .
\]
\end{prop}

\textbf{Proof:} It suffices to determine $\left[\mathcal{K},b_{k}(\varphi)\right]$.
Using Lemma \ref{lemma:TraceFormLemma} and Lemma \ref{lemma:QuasiBosonicCommutationRelations} we calculate that
\begin{align}
\left[\mathcal{K},b_{k}(\varphi)\right] 
&=\frac{1}{2}\sum_{l\in\mathbb{Z}_{\ast}^{3}}\sum_{q\in L_{l}}\left\{ \left[b_{k}(\varphi),b_{-l}^{\ast}\left(e_{-q}\right)\right],b_{l}^{\ast}(K_{l}e_{q})\right\} \nonumber\\
& =\frac{1}{2}\sum_{l\in\mathbb{Z}_{\ast}^{3}}\sum_{q\in L_{l}}\left\{ \delta_{k,-l}\left\langle \varphi,e_{-q}\right\rangle +\varepsilon_{k,-l}\left(\varphi;e_{-q}\right),b_{l}^{\ast}(K_{l}e_{q})\right\} \nonumber \\
 & =b_{-k}^{\ast}\left(I_{k}K_{k}\varphi\right)+\mathcal{E}_{k}(\varphi) 
\end{align}
where in the last identity we recognized $K_{-k}\sum_{q\in L_{-k}}\left\langle \varphi,e_{-q}\right\rangle e_{q}=K_{-k}I_{k}\varphi=I_{k}K_{k}\varphi.
$
$\hfill\square$

\medskip

Using this relation we can now determine the commutators with $Q_{1}^{k}$
terms:
\begin{prop}
\label{prop:calKQ1Commutator}For any $k\in\mathbb{Z}_{\ast}^{3}$
and symmetric operators $A_{\pm k}:\ell^{2}(L_{\pm k})\rightarrow\ell^{2}(L_{\pm k})$
such that $I_{k}A_{k}=A_{-k}I_{k}$, it holds that
\[
\left[\mathcal{K},2\,Q_{1}^{k}(A_{k})+2\,Q_{1}^{-k}(A_{-k})\right]=Q_{2}^{k}(\left\{ K_{k},A_{k}\right\})+2\,\mathrm{Re}\left(\mathcal{E}_{k}^{1}(A_{k})\right)+\left(k\rightarrow-k\right)
\]
where
\[
\mathcal{E}_{k}^{1}(A_{k})=\sum_{l\in\mathbb{Z}_{\ast}^{3}}\sum_{p\in L_{k}}\sum_{q\in L_{l}}b_{k}^{\ast}(A_k e_p)\left\{ \varepsilon_{k,l}(e_{p};e_{q}),b_{-l}^{\ast}(K_{-l}e_{-q})\right\} .
\]
\end{prop}

\textbf{Proof:} Using Proposition \ref{prop:calKExcitationCommutator}
(and Lemma \ref{lemma:TraceFormLemma} together with symmetry of $A_{k}$)
we find that
\begin{align}
\left[\mathcal{K},Q_{1}^{k}(A_{k})\right] 
&=\sum_{p\in L_{k}}\left(b_{k}^{\ast}(A_k e_p)\left[\mathcal{K},b_{k}\left(e_{p}\right)\right]+\left[\mathcal{K},b_{k}^{\ast}(A_k e_p)\right]b_{k}\left(e_{p}\right)\right)\nonumber \\
 & =\sum_{p\in L_{k}}\left(b_{-k,-p}^{\ast}b_{k}^{\ast}\left(A_{k}K_{k}e_{p}\right)+b_{k}\left(A_{k}K_{k}e_{p}\right)b_{-k,-p}\right)+2\,\mathrm{Re}\left(\sum_{p\in L_{k}}b_{k}^{\ast}(A_k e_p)\mathcal{E}_{k}\left(e_{p}\right)\right)\nonumber \\
 & 
 =Q_{2}^{k}(A_{k}K_{k})+2\,\mathrm{Re}\left(\sum_{p\in L_{k}}b_{k}^{\ast}(A_k e_p)\mathcal{E}_{k}\left(e_{p}\right)\right). 
\end{align}
The assumption that $I_{k}A_{k}=A_{-k}I_{k}$ yields
$
Q_{2}^{k}(A_{k}K_{k})=Q_{2}^{-k}(K_{-k}A_{-k}).
$ Summing over both $k$ and $-k$, we obtain the desired identity.
$\hfill\square$

\medskip

To state the commutator of $\mathcal{K}$ with $Q_{2}^{k}$ terms
we note the identity
\begin{align} \label{eq:ekk-intro}
\sum_{p\in L_{k}}b_{k}\left(e_{p}\right)b_{k}^{\ast}(A_k e_p) 
 =Q_{1}^{k}(A_{k})+\mathrm{tr}(A_{k})+\varepsilon_{k}(A_{k}) 
\end{align}
where we introduced the convenient notation 
\begin{align}
\varepsilon_{k}(A_{k}) & =\sum_{p\in L_{k}}\varepsilon_{k,k}\left(e_{p};A_{k}e_{p}\right)
  =-\sum_{p\in L_{k}}\left\langle e_{p},A_{k}e_{p}\right\rangle \left(c_{p}^{\ast}c_{p}+c_{p-k}c_{p-k}^{\ast}\right). 
\end{align}
The commutator is then given as follows:
\begin{prop}
\label{prop:calKQ2Commutator}For any $k\in\mathbb{Z}_{\ast}^{3}$
and symmetric operators $B_{\pm k}:\ell^{2}(L_{\pm k})\rightarrow\ell^{2}(L_{\pm k})$
such that $I_{k}B_{k}=B_{-k}I_{k}$, it holds that
\begin{align*}
\left[\mathcal{K},Q_{2}^{k}(B_{k})+Q_{2}^{-k}(B_{-k})\right] & =2\,Q_{1}^{k}(\left\{ K_{k},B_{k}\right\})+\mathrm{tr}(\left\{ K_{k},B_{k}\right\})+\varepsilon_{k}(\left\{ K_{k},B_{k}\right\})\\
 & \quad +2\,\mathrm{Re}\left(\mathcal{E}_{k}^{2}(B_{k})\right)+\left(k\rightarrow-k\right)
\end{align*}
where
\[
\mathcal{E}_{k}^{2}(B_{k})=\frac{1}{2}\sum_{l\in\mathbb{Z}_{\ast}^{3}}\sum_{p\in L_{k}}\sum_{q\in L_{l}}\left\{ b_{k}(B_k e_p),\left\{ \varepsilon_{-k,-l}(e_{-p};e_{-q}),b_{l}^{\ast}(K_{l}e_{q})\right\} \right\} .
\]
\end{prop}

\textbf{Proof:} Writing $Q_{2}^{k}(B_{k})=2\,\mathrm{Re}\left(\sum_{p\in L_{k}}b_{k}(B_k e_p)b_{-k}\left(e_{-p}\right)\right)$ and using Proposition \ref{prop:calKExcitationCommutator} we get
\begin{align}
\left[\mathcal{K},Q_{2}^{k}(B_{k})\right] 
& =2\,\mathrm{Re}\left(\sum_{p\in L_{k}}\left(b_{k}(B_k e_p)\left[\mathcal{K},b_{-k}\left(e_{-p}\right)\right]+\left[\mathcal{K},b_{k}(B_k e_p)\right]b_{-k}\left(e_{-p}\right)\right)\right)\nonumber \\
 & =2\,\mathrm{Re}\left(\sum_{p\in L_{k}}\left(b_{k,p}b_{k}^{\ast}\left(K_{k}B_{k}e_{p}\right)+b_{-k}^{\ast}\left(K_{-k}B_{-k}e_{-p}\right)b_{-k,-p}\right)\right)\nonumber \\
 &\quad +2\,\mathrm{Re}\left(\sum_{p\in L_{k}}\left(b_{k}(B_k e_p)\mathcal{E}_{-k}\left(e_{-p}\right)+\mathcal{E}_{k}\left(e_{p}\right)b_{-k}\left(B_{-k}e_{-p}\right)\right)\right)= {\rm (I)} + {\rm (II)}.\nonumber 
\end{align}
For (I), the first term on the right-hand side, using \eqref{eq:ekk-intro} we find that 
\begin{align}
 {\rm (I)} =Q_{1}^{k}(\left\{ K_{k},B_{k}\right\})+\mathrm{tr}(\left\{ K_{k},B_{k}\right\})+\varepsilon_{k}(\left\{ K_{k},B_{k}\right\})+Q_{1}^{-k}(\left\{ K_{-k},B_{-k}\right\}).
\end{align}
Summing over $k$ and $-k$ and using $\mathcal{E}_{k}^{2}(B_{k})=\sum_{p\in L_{k}}\left\{ b_{k}(B_k e_p),\mathcal{E}_{-k}\left(e_{-p}\right)\right\} $ for {\rm (II)}, we obtain the  desired identity.
%
$\hfill\square$

\medskip

Finally we calculate the commutator with $H_{\mathrm{kin}}^{\prime}$:
\begin{prop}
\label{prop:calKHkinCommutator}It holds that
\[
\left[\mathcal{K},H_{\mathrm{kin}}^{\prime}\right]=\sum_{k\in\mathbb{Z}_{\ast}^{3}}Q_{2}^{k}(\left\{ K_{k},h_{k}\right\}).
\]
\end{prop}

\textbf{Proof:} By equation (\ref{eq:HkinbkpCommutator}) we have
\begin{equation}
\left[H_{\mathrm{kin}}^{\prime},b_{k}(\varphi)\right]=-2\,b_{k}\left(h_{k}\varphi\right),\quad\left[H_{\mathrm{kin}}^{\prime},b_{k}^{\ast}(\varphi)\right]=2\,b_{k}^{\ast}\left(h_{k}\varphi\right),
\end{equation}
so using that $I_{k}h_{k}=h_{-k}I_{k}$ we find
\begin{align}
&\left[\mathcal{K},H_{\mathrm{kin}}^{\prime}\right]  =\frac{1}{2}\sum_{k\in\mathbb{Z}_{\ast}^{3}}\sum_{q\in L_{k}}\left(\left[b_{k}\left(K_{k}e_{q}\right)b_{-k}\left(e_{-q}\right),H_{\mathrm{kin}}^{\prime}\right]-\left[b_{-k}^{\ast}\left(e_{-q}\right)b_{k}^{\ast}\left(K_{k}e_{q}\right),H_{\mathrm{kin}}^{\prime}\right]\right) \\
 & =\sum_{k\in\mathbb{Z}_{\ast}^{3}}\sum_{q\in L_{k}}\left(b_{k}\left(\left\{ K_{k},h_{k}\right\} e_{q}\right)b_{-k}\left(e_{-q}\right)+b_{-k}^{\ast}\left(e_{-q}\right)b_{k}^{\ast}\left(\left\{ K_{k},h_{k}\right\} e_{q}\right)\right) =\sum_{k\in\mathbb{Z}_{\ast}^{3}}Q_{2}^{k}(\left\{ K_{k},h_{k}\right\}).\nonumber \tag*{$\square$}
\end{align}

%
Now we can now determine the action
of $e^{\mathcal{K}}$ on quadratic operators:
\begin{prop}
\label{prop:TransformationofQuadraticTerms}For any $k\in\mathbb{Z}_{\ast}^{3}$
and symmetric operators $T_{\pm k}:\ell^{2}(L_{\pm k})\rightarrow\ell^{2}(L_{\pm k})$
such that $I_{k}T_{k}=T_{-k}I_{k}$ it holds that
\begin{align*}
 & \;\,e^{\mathcal{K}}\left(2\,Q_{1}^{k}(T_{k})+2\,Q_{1}^{-k}(T_{-k})\right)e^{-\mathcal{K}}=\mathrm{tr}\left(T_{k}^{1}(1)-T_{k}\right)+2\,Q_{1}^{k}(T_{k}^{1}(1))+Q_{2}^{k}\left(T_{k}^{2}(1)\right)\\
 & +\int_{0}^{1}e^{\left(1-t\right)\mathcal{K}}\left(\varepsilon_{k}\left(\left\{ K_{k},T_{k}^{2}(t)\right\} \right)+2\,\mathrm{Re}\left(\mathcal{E}_{k}^{1}(T_{k}^{1}(t))\right)+2\,\mathrm{Re}\left(\mathcal{E}_{k}^{2}\left(T_{k}^{2}(t)\right)\right)\right)e^{-\left(1-t\right)\mathcal{K}}dt+\left(k\rightarrow-k\right)
\end{align*}
and
\begin{align*}
 & \;\,e^{\mathcal{K}}\left(Q_{2}^{k}(T_{k})+Q_{2}^{-k}(T_{-k})\right)e^{-\mathcal{K}}=\mathrm{tr}\left(T_{k}^{2}(1)\right)+2\,Q_{1}^{k}\left(T_{k}^{2}(1)\right)+Q_{2}^{k}(T_{k}^{1}(1))\\
 & +\int_{0}^{1}e^{\left(1-t\right)\mathcal{K}}\left(\varepsilon_{k}(\left\{ K_{k},T_{k}^{1}(t)\right\})+2\,\mathrm{Re}\left(\mathcal{E}_{k}^{1}\left(T_{k}^{2}(t)\right)\right)+2\,\mathrm{Re}\left(\mathcal{E}_{k}^{2}(T_{k}^{1}(t))\right)\right)e^{-\left(1-t\right)\mathcal{K}}dt+\left(k\rightarrow-k\right)
\end{align*}
where for $t\in\left[0,1\right]$, 
\begin{align*}
T_{k}^{1}(t)  =\frac{1}{2}\left(e^{tK_{k}}T_{k}e^{tK_{k}}+e^{-tK_{k}}T_{k}e^{-tK_{k}}\right), \quad T_{k}^{2}(t) =\frac{1}{2}\left(e^{tK_{k}}T_{k}e^{tK_{k}}-e^{-tK_{k}}T_{k}e^{-tK_{k}}\right).
\end{align*}
\end{prop}

\textbf{Proof:} We prove the first identity, the second following
by a similar argument. Note that the operators $A_{k}(t)=T_k^1(t)$, $B_k(t)=T_k^2(t)$ satisfy 
\begin{equation} \label{eq:diff-eq-At-Bt}
A_{k}^{\prime}(t)=\left\{ K_{k},B_{k}(t)\right\} ,\quad B_{k}^{\prime}(t)=\left\{ K_{k},A_{k}(t)\right\} ,\quad A_{k}\left(0\right)=T_{k},\quad B_{k}\left(0\right)=0.
\end{equation}
%
By Propositions \ref{prop:calKQ1Commutator} and \ref{prop:calKQ2Commutator} we get 
\begin{align}
 & \,\frac{d}{dt}e^{-t\mathcal{K}}\left(2\,Q_{1}^{k}(A_k(t))+Q_{2}^{k}(B_k(t))\right)e^{t\mathcal{K}}+\left(k\rightarrow-k\right)\nonumber \\
 & =e^{-t\mathcal{K}}\left(2\,Q_{1}^{k}\left(A_{k}^{\prime}(t)\right)+Q_{2}^{k}\left(B_{k}^{\prime}(t)\right)-\left[\mathcal{K},2\,Q_{1}^{k}(A_k(t))+Q_{2}^{k}(B_k(t))\right]\right)e^{t\mathcal{K}}+\left(k\rightarrow-k\right)\nonumber \\
 & =-\mathrm{tr}(\left\{ K_{k},B_{k}(t)\right\})+e^{-t\mathcal{K}}\left(2\,Q_{1}^{k}\left(A_{k}^{\prime}(t)-\left\{ K_{k},B_{k}(t)\right\} \right)+Q_{2}^{k}\left(B_{k}^{\prime}(t)-\left\{ K_{k},A_{k}(t)\right\} \right)\right)e^{t\mathcal{K}}\nonumber \\
 &\quad  -e^{-t\mathcal{K}}\left(\varepsilon_{k}(\left\{ K_{k},B_{k}(t)\right\})+2\,\mathrm{Re}\left(\mathcal{E}_{k}^{1}(A_k(t))\right)+2\,\mathrm{Re}\left(\mathcal{E}_{k}^{2}(B_k(t))\right)\right)e^{t\mathcal{K}}+\left(k\rightarrow-k\right).\nonumber 
\end{align}
The second term on the right-hand side vanishes due to \eqref{eq:diff-eq-At-Bt}. Specifying also the initial conditions in \eqref{eq:diff-eq-At-Bt} we conclude by the fundamental theorem of calculus, 
\begin{align}
 & \;\,e^{\mathcal{K}}\left(2\,Q_{1}^{k}(T_{k})+2\,Q_{1}^{-k}(T_{-k})\right)e^{-\mathcal{K}}=\mathrm{tr}\left(A_{k}(1)-T_{k}\right)+2\,Q_{1}^{k}(A_{k}(1))+Q_{2}^{k}(B_{k}(1))\\
 & +\int_{0}^{1}e^{\left(1-t\right)\mathcal{K}}\left(\varepsilon_{k}(\left\{ K_{k},B_{k}(t)\right\})+2\,\mathrm{Re}\left(\mathcal{E}_{k}^{1}(A_k(t))\right)+2\,\mathrm{Re}\left(\mathcal{E}_{k}^{2}(B_k(t))\right)\right)e^{-\left(1-t\right)\mathcal{K}}dt+\left(k\rightarrow-k\right)\nonumber 
\end{align}
where we also used that by the assumptions on $A_{k}(t)$
and $B_{k}(t)$
\begin{equation}
\int_{0}^{1}\mathrm{tr}(\left\{ K_{k},B_{k}(t)\right\})dt=\mathrm{tr}\left(\int_{0}^{1}A_{k}^{\prime}(t)dt\right)=\mathrm{tr}\left(A_{k}(1)-T_{k}\right).
\end{equation}
The proof of Proposition \ref{prop:TransformationofQuadraticTerms} is complete. 
$\hfill\square$

\medskip

From this we can also easily deduce the action of $e^{\mathcal{K}}$
on $H_{\mathrm{kin}}^{\prime}$:
\begin{prop}
It holds that
\begin{align*}
e^{\mathcal{K}}H_{\mathrm{kin}}^{\prime}e^{-\mathcal{K}} & =\sum_{k\in\mathbb{Z}_{\ast}^{3}}\mathrm{tr}\left(h_{k}^{1}(1)-h_{k}\right)+H_{\mathrm{kin}}^{\prime}+\sum_{k\in\mathbb{Z}_{\ast}^{3}}\left(2\,Q_{1}^{k}\left(h_{k}^{1}(1)-h_{k}\right)+Q_{2}^{k}\left(h_{k}^{2}(1)\right)\right)\\
 & +\sum_{k\in\mathbb{Z}_{\ast}^{3}}\int_{0}^{1}e^{\left(1-t\right)\mathcal{K}}\left(\varepsilon_{k}\left(\left\{ K_{k},h_{k}^{2}(t)\right\} \right)+\mathcal{E}_{k}^{1}\left(h_{k}^{1}(t)-h_{k}\right)+\mathcal{E}_{k}^{2}\left(h_{k}^{2}(t)\right)\right)e^{-\left(1-t\right)\mathcal{K}}dt
\end{align*}
where for $t\in\left[0,1\right]$,
\begin{align*}
h_{k}^{1}(t)  =\frac{1}{2}\left(e^{tK_{k}}h_{k}e^{tK_{k}}+e^{-tK_{k}}h_{k}e^{-tK_{k}}\right),\quad h_{k}^{2}(t)  =\frac{1}{2}\left(e^{tK_{k}}h_{k}e^{tK_{k}}-e^{-tK_{k}}h_{k}e^{-tK_{k}}\right).
\end{align*}
\end{prop}

\textbf{Proof:} By the Propositions \ref{prop:calKQ1Commutator} and
\ref{prop:calKHkinCommutator} we see that
\begin{equation}
\left[\mathcal{K},H_{\mathrm{kin}}^{\prime}-\sum_{k\in\mathbb{Z}_{\ast}^{3}}2\,Q_{1}^{k}(h_{k})\right]=-\sum_{k\in\mathbb{Z}_{\ast}^{3}}2\,\mathrm{Re}\left(\mathcal{E}_{k}^{1}(h_{k})\right)
\end{equation}
whence by the fundamental theorem of calculus
\begin{equation}
e^{\mathcal{K}}\left(H_{\mathrm{kin}}^{\prime}-\sum_{k\in\mathbb{Z}_{\ast}^{3}}2\,Q_{1}^{k}(h_{k})\right)e^{-\mathcal{K}}=H_{\mathrm{kin}}^{\prime}-\sum_{k\in\mathbb{Z}_{\ast}^{3}}2\,Q_{1}^{k}(h_{k})-\sum_{k\in\mathbb{Z}_{\ast}^{3}}\int_{0}^{1}e^{t\mathcal{K}}\left(2\,\mathrm{Re}\left(\mathcal{E}_{k}^{1}(h_{k})\right)\right)e^{-t\mathcal{K}}dt.
\end{equation}
Applying Proposition \ref{prop:TransformationofQuadraticTerms} now
yields the claim.
$\hfill\square$

We are now equipped to conclude Theorem \ref{thm:DiagonalizationoftheBosonizableTerms}. By the two previous propositions, we see that
\begin{align}
e^{\mathcal{K}}H_{\mathrm{eff}}e^{-\mathcal{K}} & =e^{\mathcal{K}}\left(H_{\mathrm{kin}}^{\prime}+\sum_{k\in\mathbb{Z}_{\ast}^{3}}\left(2\,Q_{1}^{k}(P_{k})+Q_{2}^{k}(P_{k})\right)\right)e^{-\mathcal{K}}\nonumber \\
 & =\sum_{k\in\mathbb{Z}_{\ast}^{3}}\mathrm{tr}(A_{k}(1)-P_{k})+H_{\mathrm{kin}}^{\prime}+\sum_{k\in\mathbb{Z}_{\ast}^{3}}\left(2\,Q_{1}^{k}(A_{k}(1))+Q_{2}^{k}(B_{k}(1))\right)\\
 & +\sum_{k\in\mathbb{Z}_{\ast}^{3}}\int_{0}^{1}e^{\left(1-t\right)\mathcal{K}}\left(\varepsilon_{k}(\left\{ K_{k},B_{k}(t)\right\})+\mathcal{E}_{k}^{1}(A_k(t))+\mathcal{E}_{k}^{2}(B_k(t))\right)e^{-\left(1-t\right)\mathcal{K}}dt\nonumber 
\end{align}
where the operators $A_{k}(t),B_{k}(t):\ell^{2}(L_{k})\rightarrow\ell^{2}(L_{k})$
are given by 
\begin{align}
A_{k}(t) & =h_{k}^{1}(t)+P_{k}^{1}(t)+P_{k}^{2}(t)-h_{k}=\frac{1}{2}\left(e^{tK_{k}}\left(h_{k}+2P_{k}\right)e^{tK_{k}}+e^{-tK_{k}}h_{k}e^{-tK_{k}}\right)-h_{k}\\
B_{k}(t) & =h_{k}^{2}(t)+P_{k}^{1}(t)+P_{k}^{2}(t)=\frac{1}{2}\left(e^{tK_{k}}\left(h_{k}+2P_{k}\right)e^{tK_{k}}-e^{-tK_{k}}h_{k}e^{-tK_{k}}\right).\nonumber 
\end{align}
Now we choose $K_{k}$ such that $B_{k}(1)=0$. This amounts to the diagonalization condition
\begin{equation} \label{eq:diag-cond}
e^{K_{k}}\left(h_{k}+2P_{k}\right)e^{K_{k}}=e^{-K_{k}}h_{k}e^{-K_{k}},
\end{equation}
of which the solution is given in \eqref{eq:KkDefinition}.
Since \eqref{eq:diag-cond} is fulfilled, it follows
that also $A_{k}(1)=e^{-K_{k}}h_{k}e^{-K_{k}}-h_{k}$, and so the identity in Theorem \ref{thm:DiagonalizationoftheBosonizableTerms} follows
provided we can show that
\begin{equation} \label{eq:final-trace-identity}
\sum_{k\in\mathbb{Z}_{\ast}^{3}}\mathrm{tr}\left(e^{-K_{k}}h_{k}e^{-K_{k}}-h_{k}-P_{k}\right)=E_{\mathrm{corr},\mathrm{bos}}.
\end{equation}

%

To establish this final identity we will use the following integral
representation of the square root of a one-dimensional perturbation,
first used in \cite{BNPSS-20}:
\begin{lem}
\label{lemma:SquareRootofAOneDimensionalPerturbation}Let $A:V\rightarrow V$
be a positive self-adjoint operator. Then for any $w\in V$ and $g\in\mathbb{R}$
such that $A+gP_{w}>0$ it holds that
\begin{align*}
\left(A+gP_{w}\right)^{\frac{1}{2}}&=A^{\frac{1}{2}}+\frac{2g}{\pi}\int_{0}^{\infty}\frac{t^{2}}{1+g\left\langle w,\left(A+t^{2}\right)^{-1}w\right\rangle }P_{\left(A+t^{2}\right)^{-1}w}dt, \\
\mathrm{tr}\left(\left(A+gP_{w}\right)^{\frac{1}{2}}\right)&=\mathrm{tr}\left(A^{\frac{1}{2}}\right)+\frac{1}{\pi}\int_{0}^{\infty}\log\left(1+g\left\langle w,\left(A+t^{2}\right)^{-1}w\right\rangle \right)dt.
\end{align*}
\end{lem}

%
The trace identity \eqref{eq:final-trace-identity} now follows (note that this is essentially Proposition
7.6 of \cite{CHN-21}):
\begin{prop}
\label{prop:GeneralBosonizableContribution} Let $F\left(x\right)=\log\left(1+x\right)-x.$ For any $k\in\mathbb{Z}_{\ast}^{3}$
it holds that
\[
\mathrm{tr}\left(e^{-K_{k}}h_{k}e^{-K_{k}}-h_{k}-P_{k}\right)=\frac{1}{\pi}\int_{0}^{\infty}F\left(\frac{\hat{V}_{k}k_{F}^{-1}}{\left(2\pi\right)^{3}}\sum_{p\in L_{k}}\frac{\lambda_{k,p}}{\lambda_{k,p}^2+t^{2}}\right)dt. 
\]
\end{prop}


\textbf{Proof:} By cyclicity of the trace and Lemma \ref{lemma:SquareRootofAOneDimensionalPerturbation}, $\mathrm{tr}\left(e^{-K_{k}}h_{k}e^{-K_{k}}-h_{k}\right)$ is equal to 
\begin{align}
  \mathrm{tr}\left(\left(h_{k}^{2}+2P_{h_{k}^{\frac{1}{2}}v_{k}}\right)^{\frac{1}{2}}\right) =\frac{1}{\pi}\int_{0}^{\infty}\log\left(1+2\left\langle v_{k},h_{k}\left(h_{k}^{2}+t^{2}\right)^{-1}v_{k}\right\rangle \right)dt.
\end{align}
The claim follows by inserting the definition of $h_{k}$ and $v_{k}$, and noting also that  \begin{equation}
\frac{2}{\pi}\int_{0}^{\infty}\left\langle v_{k},h_{k}\left(h_{k}^{2}+t^{2}\right)^{-1}v_{k}\right\rangle dt 
=\frac{\hat{V}_{k}k_{F}^{-1}}{2\left(2\pi\right)^{3}}\left|L_{k}\right|=\left\Vert v_{k}\right\Vert ^{2} = {\rm tr}(P_k)
\end{equation}
where we used the integral identity $\int_{0}^{\infty}a/(a^2+t^2)dt=\pi/2$ for every $a>0$. 
$\hfill\square$

\end{document}